\documentclass[12pt, notitlepage]{thesis}

\usepackage{hyperref}
\usepackage{caption}
\usepackage{subcaption}
\usepackage{verbatim}
\usepackage{booktabs}
\usepackage{adjustbox}
\usepackage{deluxetable}
\usepackage{comment}
\usepackage{graphicx}
\usepackage{grffile}
\usepackage{lscape}
\usepackage{indentfirst}
\usepackage{latexsym}
\usepackage{amssymb}
\usepackage{amsthm}
\usepackage{amsmath}
\usepackage[titletoc]{appendix}
\usepackage{multirow}
\usepackage{tabls}
\usepackage{wrapfig}
\usepackage{slashbox}
\usepackage{longtable}
\usepackage{rotating}
\usepackage{psfig}
\usepackage{url}
\usepackage{natbib}
\usepackage{braket}
\usepackage{setspace}
\usepackage{pdflscape}
\usepackage{geometry}
\renewcommand{\baselinestretch}{2}
\newcommand{\aap}{{A\&A}}

\newcommand{\apjl}{ApJL}

\newcommand{\apjs}{ApJS}

\newcommand\blfootnote[1]{%
  \begingroup
  \renewcommand\thefootnote{}\footnote{#1}%
  \addtocounter{footnote}{-1}%
  \endgroup
}

\def\arcdeg{\hbox{$^\circ$}}
\def\slantfrac#1#2{\hbox{$\,^#1\!/_#2$}}

\begin{document}

\renewcommand{\baselinestretch}{1}
\begin{titlepage}
    \centering
        \vspace*{1cm}
        \LARGE
        \textbf{Modeling the Galactic Compact Binary Neutron Star Population and Studying the Double Pulsar System}
        
        \vspace{0.5cm}
        \large
        \textbf{Nihan~Pol}
        
        \vspace{2cm}
        \normalsize
        Dissertation Submitted to\\
		The Eberly College of Arts and Sciences\\ 
		at West Virginia University\\ 
		in partial fulfillment of the requirements\\ 
		for the degree of\\
        \vspace{1.5cm}
        Doctor of Philosophy\\
		in\\
		Physics\\
		\vspace{1.5cm}
    Maura McLaughlin, Ph.D., Chair

	Sarah Burke-Spolaor, Ph.D.

	Daniel Pisano, Ph.D
	
	Paul Cassak, Ph.D.
	
	David Mebane, Ph.D.
    
        \vspace{0.8cm}
        
        Morgantown, West Virginia, USA

		2020
        \vfill

Copyright 2020 Nihan~Pol

\end{titlepage}

\begin{abstract}
\begin{center}
\vspace{0.5cm}
\textbf{Modeling the Galactic Compact Binary Neutron Star Population and Studying the Double Pulsar System}

\vspace{0.5cm}

Nihan Pol
\end{center}
\vspace{0.5cm}

Binary neutron star (BNS) systems consisting of at least one neutron star provide an avenue for testing a broad range of physical phenomena ranging from tests of General Relativity to probing magnetospheric physics to understanding the behavior of matter in the densest environments in the Universe. Ultra-compact BNS systems with orbital periods less than few tens of minutes emit gravitational waves with frequencies $\sim$mHz and are detectable by the planned space-based Laser Interferometer Space Antenna (LISA), while merging BNS systems produce a chirping gravitational wave signal that can be detected by the ground-based Laser Interferometer Gravitational-Wave Observatory (LIGO). Thus, BNS systems are the most promising sources for the burgeoning field of multi-messenger astrophysics.

In this thesis, we estimate the population of different classes of BNS systems that are visible to gravitational-wave observatories. Given that no ultra-compact BNS systems have been discovered in pulsar radio surveys, we place a 95\% confidence upper limit of $\sim$850 and $\sim$1100 ultra-compact neutron star--white dwarf and double neutron star (DNS) systems beaming towards the Earth, respectively. We show that among all of the current radio pulsar surveys, the ones at the Arecibo radio telescope have the best chance of detecting an ultra-compact BNS system. We also show that adopting a survey integration time of $t_{\rm int} \sim 1$~min will maximize the signal-to-noise ratio, and thus, the probability of detecting an ultra-compact BNS system.

Similarly, we use the sample of nine observed DNS systems to derive a Galactic DNS merger rate of $\mathcal{R}_{\rm MW} = 37^{+24}_{-11}$~Myr$^{-1}$, where the errors represent 90\% confidence intervals. Extrapolating this rate to the observable volume for LIGO, we derive a merger detection rate of $\mathcal{R} = 1.9^{+1.2}_{-0.6} \times \left(D_{\rm r}/100 \ \rm Mpc \right)^3 \rm yr^{-1}$, where $D_{\rm r}$ is the range distance for LIGO. This rate is consistent with that derived using the DNS mergers observed by LIGO.

Finally, to illustrate the unique opportunities for science presented by compact DNS systems, we study the J0737--3039 DNS system, also known as the Double Pulsar system. This is the only known DNS system where both of the neutron stars have been observed as pulsars.
We measure the sense of rotation of the older millisecond pulsar, pulsar A, in the DNS J0737--3039 system and find that it rotates prograde with respect to its orbit. This is the first direct measurement of the sense of rotation of a pulsar and a direct confirmation of the rotating lighthouse model for pulsars. This result confirms that the spin angular momentum vector is closely aligned with the orbital angular momentum, suggesting that kick of the supernova producing the second born pulsar J0737--3039B was small.

\end{abstract}
\pagestyle{empty}

\pagestyle{plain}
\pagenumbering{roman}
\setcounter{page}{3}

\begin{center}
\large
Acknowledgements
\end{center}
\normalsize


I would like to thank Maura McLaughlin for all of her guidance over the past five years. Everything I learned about doing research was by watching and interacting with her over these past five years. She also afforded me an incredible amount of patience and freedom as I explored multiple areas of research and for that, I will be eternally grateful. I could not imagine a better advisor. 

I would also like to thank Duncan Lorimer for introducing me to the statistical wizardry which ended up becoming a central part of my research. I would also like to thank Sarah Burke-Spolaor who introduced me to the art of radio interferometry and taught me how null results can still be scientifically interesting and important.
I would also like to thank all of my colleagues and friends in NANOGrav for welcoming me into the collaboration with open arms and allowing me to learn and contribute at my own speed.

Thank you also to all my friends at WVU. You made getting through my courses at WVU so much more fun than it would have otherwise been, while all the time we spent together coming as a welcome reprieve from my academic life. I will cherish all the memories we made during my time here at WVU.

Finally, I would like to thank my family for their support, advice and guidance as I made my journey across the world to start my PhD in an unknown land. I would like to thank my grandfather, Madhukar Pol, for sitting down a 10-year old and explaining non-Euclidean geometries to him, which sparked his interest in Astrophysics. I would like to thank my parents, Satish and Sae Pol, for always being there for me whenever I needed them and for sending delicious food to me from India whenever I asked for it. This whole journey and everything I have accomplished would not be possible without you. 

\small\normalsize
\tableofcontents 
\newpage
\listoftables 
\newpage
\listoffigures 
\newpage

\newpage
\setlength{\parskip}{0em}
\renewcommand{\baselinestretch}{2}
\small\normalsize

\setcounter{page}{1}
\pagenumbering{arabic}
\pagestyle{plain}
\newpage
\renewcommand{\thechapter}{1}

\chapter{Introduction}
\label{chap:intro}
    
    Neutron stars (NSs) are the leftover cores of stars with masses greater than $\sim$10~$M_{\odot}$ which collapse under their own gravitational force. They are also some of the densest objects in the universe, with a mass similar to that of the Sun packed into a sphere with radius $\sim$10~km. They are predominantly visible as pulsars, which are highly magnetized rotating NSs. Similar to a lighthouse, pulsar emission can be thought of as a beam of radiation that periodically sweeps across the line-of-sight to Earth. Pulsars can been observed across the electromagnetic spectrum, with a majority of them visible in the radio band. Radio pulsars have been observed to have rotational periods ranging from $\sim$1~ms all the way to few tens of seconds \citep{psrcat}.
    
    With the advent of gravitational-wave (GW) astronomy, NSs are expected to be one the most promising sources of GW emission, especially if they are in a binary system with another neutron star, white dwarf star or black hole. Isolated NSs/pulsars are expected to emit GWs if they are not perfectly symmetric about their rotation axis, i.e. if they have a deformity (such as a ``mountain") on their surface. On the other hand, the inspiraling of binary NS (BNS) systems, which can consist of a NS in orbit around a white dwarf star (a NS--WD system), another NS (a double neutron star, or DNS, system), or a black hole (a NS--BH system), also results in the emission of GWs from these systems. The Laser Interferometer Gravitational-Wave Observatory (LIGO) has already detected two DNS mergers in 2017 \citep{THE_DNS_merger} and 2020 \citep{dns_merger_2}.
    NSs and pulsars are thus one of the few objects in the Universe that can be explored through a ``multi-messenger" lens, i.e. combining observations from both the electromagnetic and GW spectra.
    Thus, it is important to study these NS and pulsar systems, especially those in binary systems, to pave the way for science with new and planned GW observatories. Apart from being excellent sources for multi-messenger science, BNS systems can provide a lot of unique science using only the electromagnetic band, such as tests of General Relativity (GR), studies of magnetospheric physics and binary stellar evolution. 
    
    In this chapter, we give an introduction to the formation of binary NS systems, followed by how we search for pulsars in these BNS systems in the electromagnetic band. We then provide a brief description of how these searches could benefit current and future GW observatories in their search and analysis of the BNS systems. Finally, we briefly describe the Double Pulsar system, a unique BNS system where both the NSs have been observed as pulsars, and describe the ground-breaking science opportunities possible with such systems.
    
    \section{Formation of binary neutron star systems} \label{bns_formation}
        
\begin{figure}
            \centering
            \includegraphics[width = \textwidth]{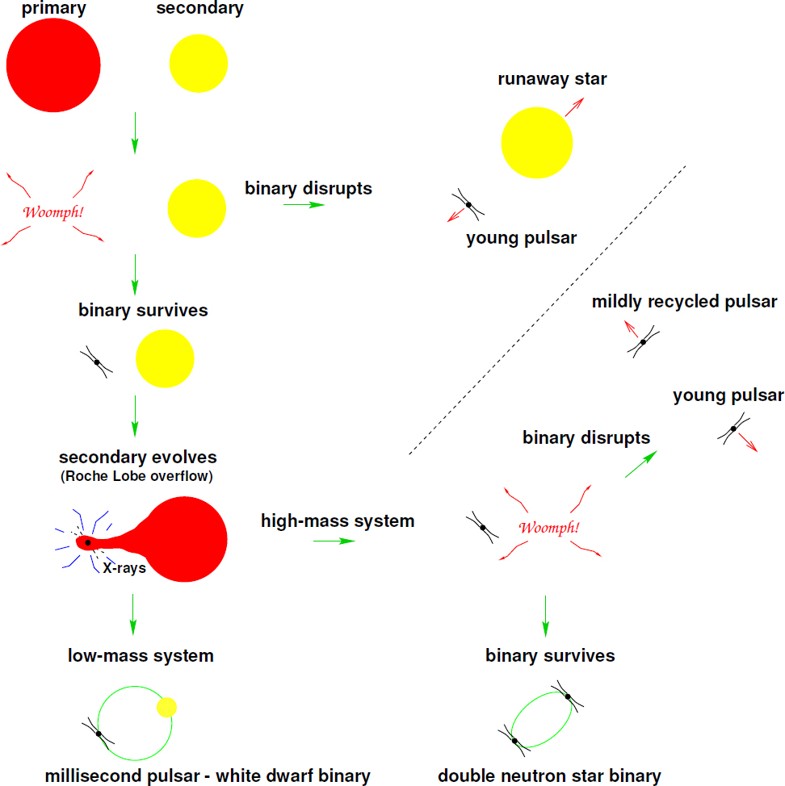}
            \caption[Illustration of the formation of a double neutron star system.]{Illustration of the formation of a double neutron star system. The two neutron stars in the double neutron star system will eventually merge due to decay of the orbit due to emission of gravitational waves resulting in the cataclysmic merger of the two neutron stars, accompanied by its own burst of gravitational wave radiation. This image is reproduced with no changes from \citet{lorimer_msp_review} under the terms of the \href{https://creativecommons.org/licenses/by/4.0/}{Creative Commons Attribution 4.0 International License}.}
            \label{full_bns_evn}
        \end{figure}

        As stated earlier, NSs are the stellar-core remnants of stars with mass greater than 10~$M_{\odot}$ that collapse under their own gravitational field. To form a BNS system requires at least one such massive star in orbit around another star of smaller mass for a NS--WD system, or a similarly massive or heavier star for a DNS or NS--BH system. The formation process for a DNS system is illustrated in Fig.~\ref{full_bns_evn}.
        
        As described by \citet{Tauris_bns_formation}, we start with two stars with mass $\gtrsim$10~$M_{\odot}$ (i.e. OB-type stars) which have just entered the main sequence, called zero age main sequence (ZAMS) stars and are in orbit around each other. Once the heavier, or primary, star exhausts the supply of hydrogen in its core and moves off the main sequence, it begins to burn helium in its core and expands past its Roche Lobe radius. Once this Roche Lobe overflow (RLO) begins, the companion, or secondary, star begins to accrete the mass from the primary star until the primary star is completely stripped of its outer hydrogen envelope. At the end of this process, the core of the primary star is left-over as a Helium star (He-star), i.e. a star that has been stripped of a majority of its hydrogen envelope.
        
        At the end of its thermonuclear evolution, the primary star undergoes a Type Ib/c supernova (SN) explosion and forms a NS. Whether the system survives this explosion depends on the specifics of the mass transfer in the RLO phase and the kick imparted to the NS in the SN explosion itself. If the NS accretes material from the secondary star, which is still on (or close to) the main sequence, the system is visible in the X-ray band as a high-mass X-ray binary (HMXB) system. During this process, the NS is spun up to higher rotational velocities. Such NSs, especially when they are observed as pulsars (see Sec.~\ref{observing_pulsars}) are referred to as ``recycled NSs/pulsars".
        
        Once the secondary star begins to move off the main sequence, its outer shell expands and engulfs the companion NS, forming a common envelope (CE) around the two objects. Depending on the accretion of matter onto the NS, there is the possibility that the NS might collapse into a black hole if it accretes enough material from the common envelope. There is also the possibility that the NS will merge with the Helium core of the secondary star to form a Thorne-Zytkow object (TZO), which eventually results in the formation of a single NS or BH. If the system is able to avoid any of these situations, the outer layers of the envelope are blown away leaving behind a NS orbiting a He-star.
        
        The dynamical friction in the CE phase causes a loss of angular momentum of the system which results in the NS--He-star system having a more compact orbit. Depending on the separation between the two, it is possible to have another round of mass transfer between the He-star and NS (Case BB RLO), which results in further spin-up of the NS. Eventually the secondary star undergoes its own supernova explosion, which is a Type Ib/c if there is not a second round of mass transfer or an ultra-stripped SN if there is. Again, the system can be disrupted depending on the separation between the two stars and the kick imparted during the second SN. If the system survives the second SN explosion, that results in the formation of a DNS system, where the NS formed from the primary star is a recycled NS, while the NS formed from the secondary star is a normal, young NS.
        
        The process for the formation of a NS--WD system is similar to the one described above, though one of the stars has mass $\lesssim$10~$M_{\odot}$. However, depending on the initial mass of the two stars and their separation, the order of formation of the two compact objects in the system can vary \citep{nswd_evolution}. For example, for slightly heavier and more compact progenitor systems, the WD tends to be born before the NS, while for lighter and wider progenitor systems, the NS is born first \citep[see][for an overview of different formation channels for NS--WD systems]{nswd_evolution}. Similarly, to form a NS--BH system will require one of the stars to have mass $\gtrsim$45~$M_{\odot}$, while the rest of the evolution will be broadly similar. Again, similar to NS--WD systems, there can be variations in which compact object is formed first \citep[see, for example,][]{zwart_bns_evolution, nsbh_evolution}. It is also possible to form a NS--BH system by the collapse of the NS to a BH in the binary system due to accretion of matter from the companion resulting in the NS's mass exceeding the Tolman-Oppenheimer-Volkoff limit \citep{nsbh_evolution}.
        
    \section{Observing neutron stars} \label{observing_pulsars}
        
        \subsection{Pulsar emission mechanism}
            
            As mentioned earlier, the majority of the neutron star discoveries have been through their radio emission, i.e. as pulsars. This is also true for NSs in binary systems. Thus, to understand how we can search for BNS systems using radio telescopes, we need to understand the pulsar emission mechanism.
            
            \begin{figure}
                \centering
                \includegraphics[width = \textwidth]{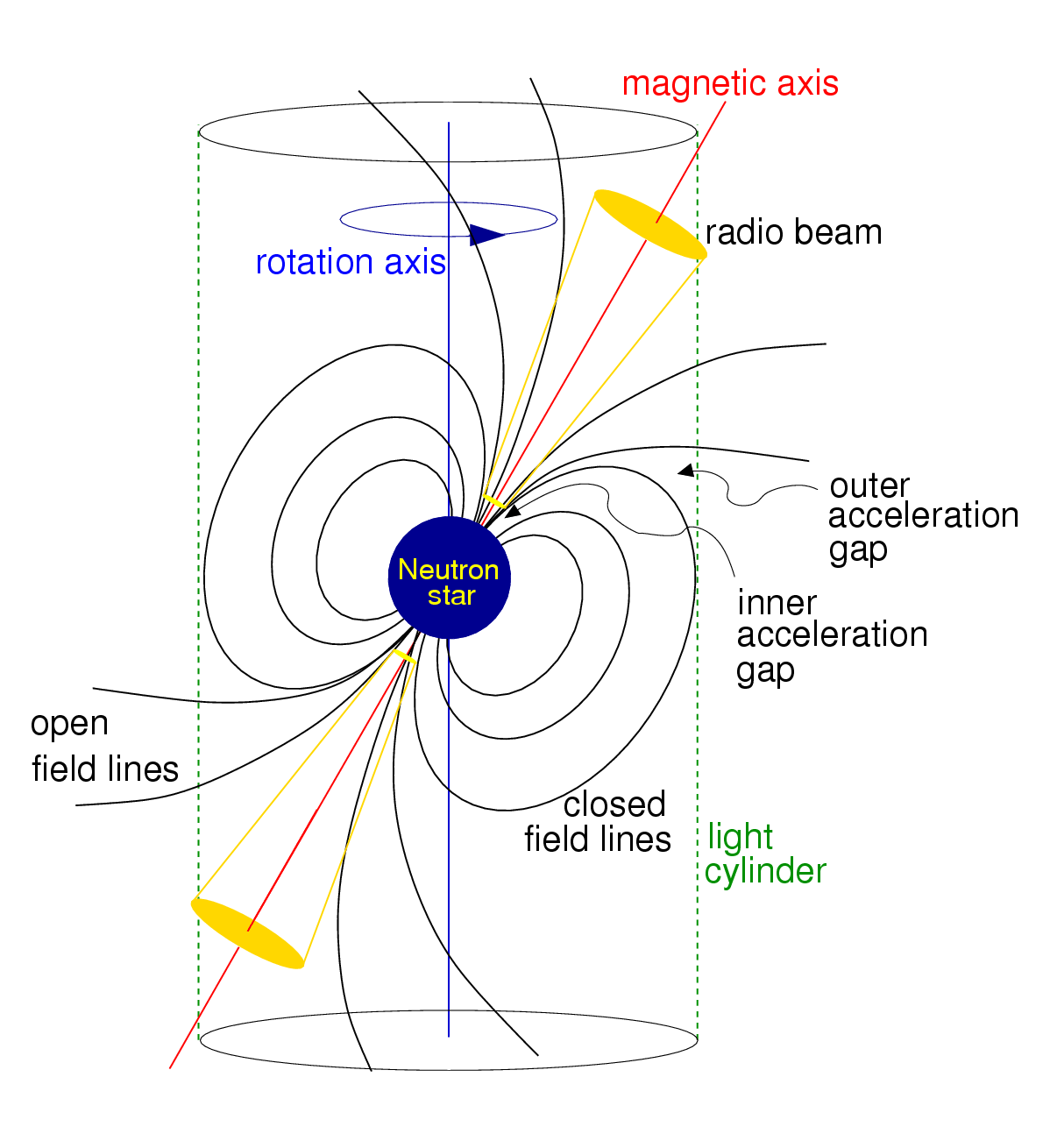}
                \caption[Magnetic dipole model of pulsars.]{The magnetic dipole model of a pulsar. The image is from the Handbook of Pulsar Astronomy \citep[ISBN: 9780521828239,][]{lorimer_kramer} and has been reproduced with permission of Cambridge University Press through PLSclear.}
                \label{psr_emission}
            \end{figure}
            
            The pulsar emission mechanism can be understood using a dipole model \citep[see Chapter 3 and references therein from][]{lorimer_kramer}, as shown in Fig.~\ref{psr_emission}. The high rotational velocity combined with the strong magnetic fields in pulsars results in the presence of a strong electric field at the surface of the NS. The electric field is strong enough to extract charged particles (mostly electrons) from the magnetic poles of the NS, which then travel along the magnetic field lines of the NS. The NS is thus enveloped in this plasma, which is called the magnetosphere of the pulsar.
            
            As these electrons are accelerated along the magnetic field lines near the polar caps, they emit curvature radiation. The high-energy photons produced in this curvature radiation interact with the magnetic field and lower energy photons to produce electron-positron pairs, which radiate even more high-energy photons. This results in a cascade process generating bunches of charged particles that emit coherently (i.e. in phase) at radio wavelengths. The net effect of this process is that a strong beam of radio emission is generated above the magnetic poles of the NS. If the magnetic axis of the pulsar is offset from the rotation axis, then this beam of radio emission will result in a rotating lighthouse effect if the emission beam crosses the line-of-sight to the Earth. Just like a lighthouse, this radio emission from pulsars is typically highly periodic. The periodicity of the pulsar emission can be exploited to our advantage when we search for pulsars. Due to the magnetic dipole radiation carrying away the rotational energy of the pulsar, the spin period of the pulsar is observed to increase as a function of time. This increase in the the spin period is quantified through the measurement of the spin period-derivative, i.e. the change in the observed spin period of the pulsar.
            
        \subsection{Propagation effects}
            
            The radio emission from the pulsar is affected by the ionized component of the interstellar medium (ISM) that lies between the pulsar and Earth. The ionized ISM affects the pulsar emission observed at Earth in three main ways: (a) dispersion; (b) scattering; (c) scintillation. The most important of these with respect to searching for pulsars and BNS systems is the effect of dispersion and scattering.
            
            \subsubsection{Dispersion}
                    
                \begin{figure}
                    \centering
                    \includegraphics[width = \textwidth]{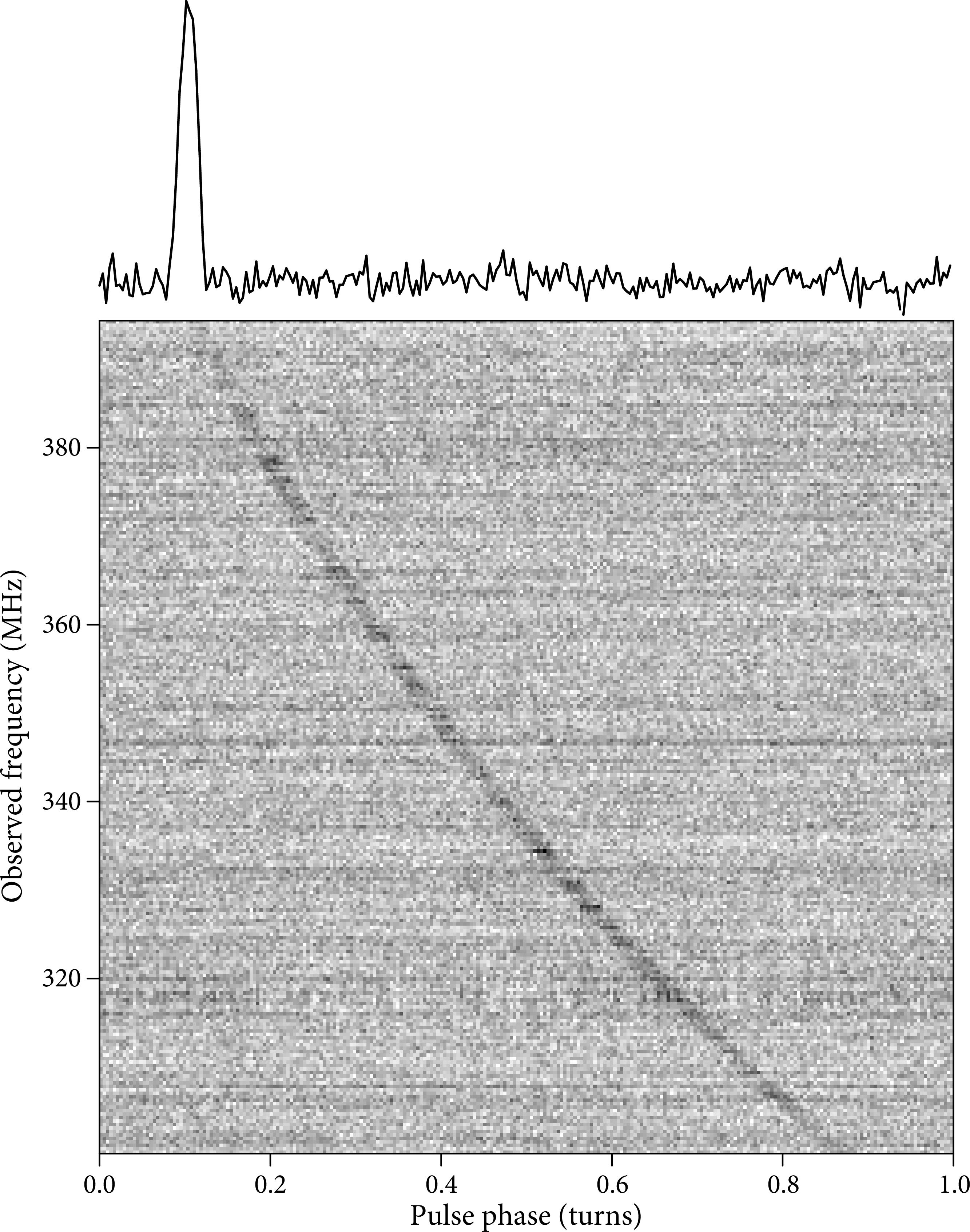}
                    \caption[Dispersion effect of the ionized interstellar medium.]{The $f^{-2}$ dispersed emission from PSR J1400+50 in the bottom panel, while the top panel shows the dispersion corrected pulse profile which is integrated across frequency. Image republished with permission of Princeton University Press, from \citet{ERA}; permission conveyed through Copyright Clearance Center, Inc.}
                    \label{DM_example}
                \end{figure}
                
                Radio waves propagating through a plasma (i.e. the ionized ISM) experience a frequency-dependent refractive index, where high-frequency corresponds to a higher index of refraction. Since the group velocity is proportional to the index of refraction, the radio emission at higher frequencies will arrive at the Earth earlier relative to emission at lower frequencies. This ``dispersive delay", $t_{\rm d}$, can be quantified as \citep{lorimer_kramer},
                \begin{equation}
                    \displaystyle t_{\rm d} = \mathcal{D} \times \frac{\rm DM}{f^2}
                    \label{DM_delay}
                \end{equation}
                where $f$ is the observing frequency, $\mathcal{D} = 4.15 \times 10^3$~MHz$^2$~pc$^{-1}$~cm$^{3}$~s is the dispersion constant and
                DM is the ``dispersion measure" which quantifies the column density of the ionized ISM along the line of sight,
                \begin{equation}
                    \displaystyle {\rm DM} = \int_{0}^{d} n_{\rm e} {\rm d}l \ [{\rm pc \, cm^{-3}}]
                    \label{dm_eqn}
                \end{equation}
                where $n_{\rm e}$ is the free electron density along the line of sight, d$l$, to the pulsar at a distance $d$. Using Eq.~\ref{DM_delay}, the time delay between two frequencies, $f_1 < f_2$ (both in MHz), is given by,
                \begin{equation}
                    \displaystyle \Delta t = 4.15 \times 10^6 \ {\rm ms} \times (f_1^{-2} - f_2^{-2}) \times {\rm DM}
                    \label{DM_delay_2}
                \end{equation}
                
                An example of dispersive smearing is shown for PSR J1400+50 in Fig.~\ref{DM_example}. As we can see, emission at higher frequencies arrives earlier than the emission at lower frequencies. If this dispersive delay is not corrected, it will lead to a significantly lower signal-to-noise (S/N) ratio when the data is integrated across frequency, inhibiting the detection of any pulsar. However, if we correct for the dispersive delay, we can recover the signal from the pulsar, as shown in the top panel of Fig.~\ref{DM_example}, which is easier to detect. This process of correcting the observed dispersion delay is called ``de-dispersion".
                
            \subsubsection{Scattering}
                
\begin{figure}
                    \centering
                    \includegraphics[width = \columnwidth]{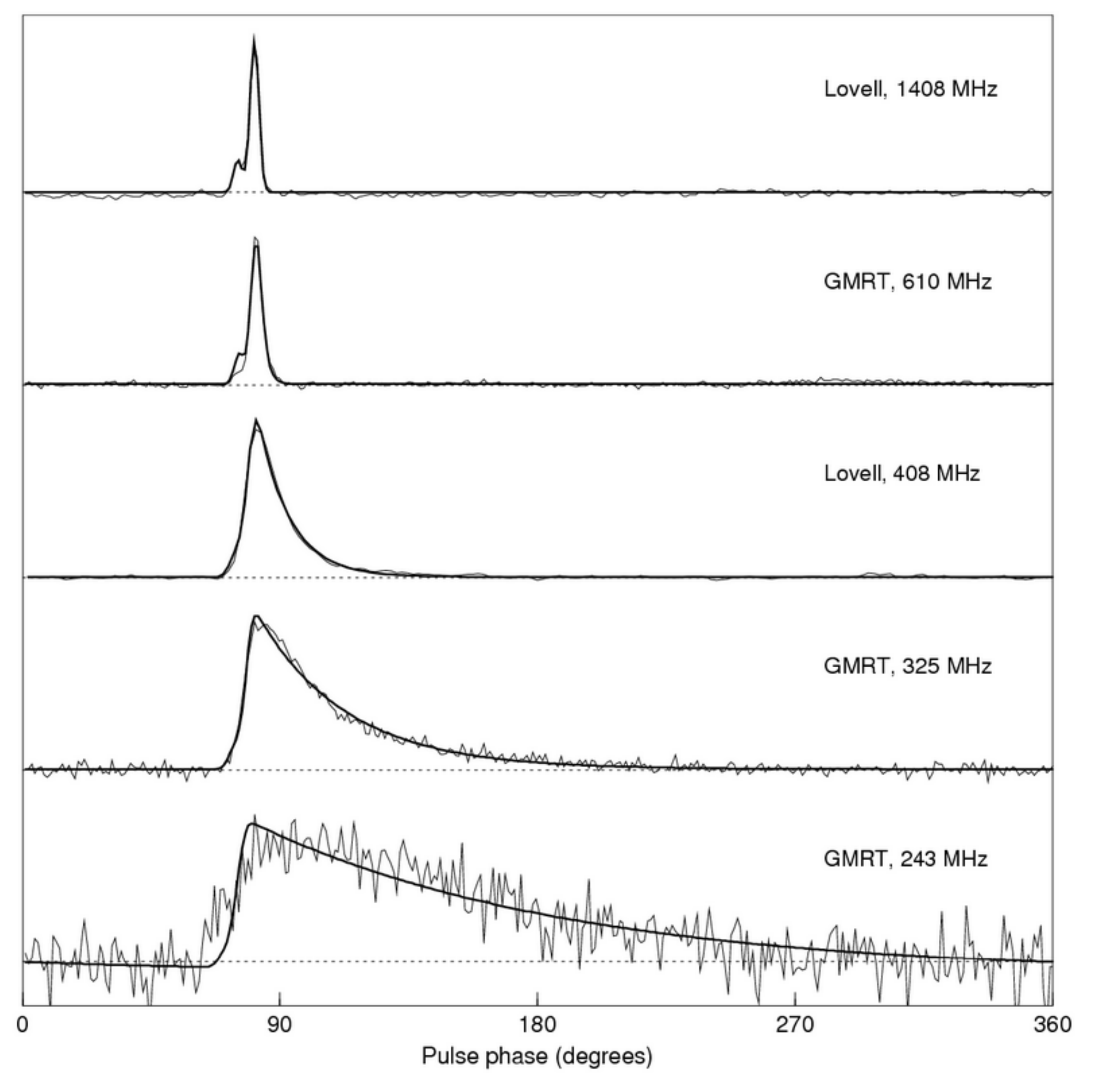}
                    \caption[Effect of interstellar scattering on pulsar emission profile.]{Illustration of the effect of interstellar scattering on the pulsar emission profile. From top to bottom, the pulse profile for B1831--03 is shown at decreasing radio frequency. As described in Sec.~\ref{observing_pulsars}, the amount of scattering has a larger effect at lower radio frequencies. The image is from the Handbook of Pulsar Astronomy \citep[ISBN: 9780521828239,][]{lorimer_kramer} and has been reproduced with permission of Cambridge University Press through PLSclear.}
                    \label{scattering}
                \end{figure}

                Scattering of the pulsar emission is a result of the inhomogeneities in the ionized ISM resulting in multiple ray paths from the pulsar to the Earth. As a result, the emission arriving at the Earth from these scattered paths will arrive later than that from the direct path. This difference in arrival times of the pulsar emission results in a broadening of the intrinsic pulsar emission profile with an exponential tail, as shown in Fig.~\ref{scattering}. 
                
                The broadening can be modeled as the convolution of the intrinsic pulsar profile with an exponential function with a scale height defined by the scattering timescale, $\tau_{\rm s}$. The scattering timescale, and thus the pulse broadening, depend on the observing frequency as $f^{-4}$, i.e. the pulse broadening is more severe at lower radio frequencies. This pulse broadening results in a reduction in the observed S/N ratio for the pulsar and can inhibit their detection. Using a higher radio frequency for observing pulsars can mitigate the effect of scattering to some extent, thereby increasing the S/N, and thus the detection probability of the pulsar.

        \subsection{Pulsar emission spectrum}
            
            As can be seen in Fig.~\ref{DM_example}, pulsar emission is broadband, i.e. it is visible across a wide range of radio frequencies. The flux, $S(f)$, for most of the pulsars can be modeled as a power-law,
            \begin{equation}
                \displaystyle S(f) \propto f^{\alpha}
                \label{psr_spectrum}
            \end{equation}
            where $\alpha$ is the spectral index. \citet{Bates_si_2013} used population synthesis to simulate the observed pulsar population and found that the spectral index can be described by a Gaussian distribution with a mean value of $\hat{\alpha} = -1.4$ and standard deviation $\sigma_{\alpha} = 0.96$. However, there are also a few pulsars where their spectrum is observed to be better fit by broken power-law or turnover models \citep{Bates_si_2013, gps_pulsars}, though the cause of these different spectral characteristics is not yet fully understood.
            
            The power-law nature of the pulsar emission spectrum implies that pulsars will be brighter at lower radio frequencies. However, scattering affects the pulsar emission much more strongly as we move towards lower frequencies. Thus, it is necessary to find an optimum radio frequency for observing and searching for pulsars. Most pulsar surveys adopt a center frequency of $\sim$1.4~GHz to search for pulsars as the effects of scattering are almost negligible for pulsars with low to mid-DM values, while the power-law spectrum implies the pulsar will be bright enough to be detected in these surveys.
    
    \section{Searching for binary neutron star systems} \label{search_BNS_systems}
        
        \subsection{Pulsar surveys and sensitivity}
            
            There have been numerous all-sky pulsar surveys conducted using different radio telescopes around the world. The largest pulsar surveys have been conducted at the Parkes radio telescope in Australia (the Parkes Multi-beam survey \citep[PMSURV,][]{PMSURV} and the High Time Resolution Universe survey \citep[HTRU,][]{htru_low_mid}), the Arecibo radio telescope in Puerto Rico (the Arecibo drift-scan survey \citep[AODRIFT,][]{aodrift_1} and the PALFA survey \citep{PALFA}), USA, and the Green Bank telescope in West Virginia, USA (the Green Bank North Celestial Cap survey \citep[GBNCC,][]{gbncc}).
            
            \begin{figure*}
                \centering
                \begin{subfigure}[b]{0.45\textwidth}
                \centering
                \includegraphics[width = \textwidth]{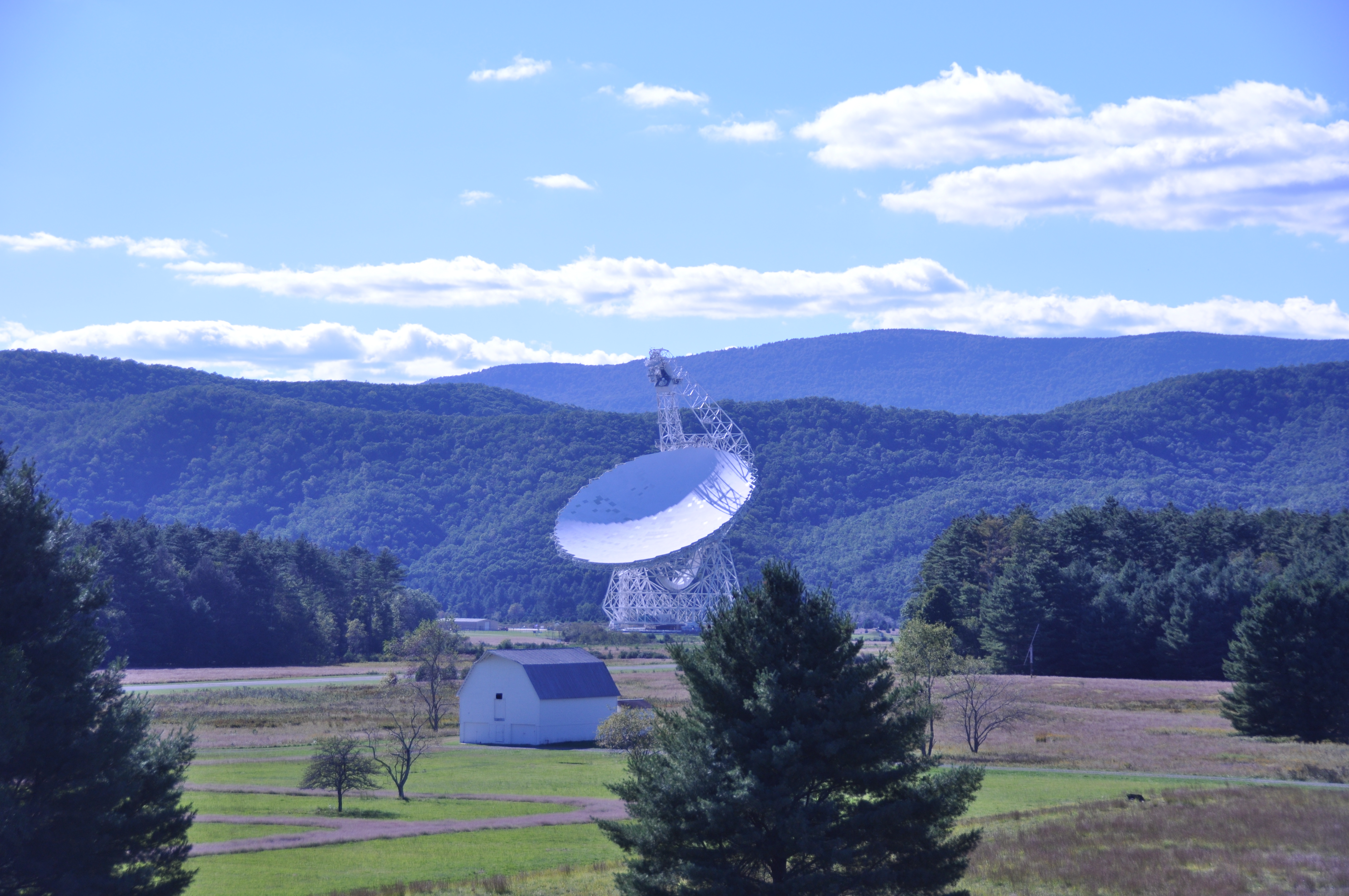}
                \caption{The 100-m diameter Green Bank telescope in Green Bank, West Virginia, USA. Image is reproduced with permission from Green Bank Observatory (GBO), Associated Universities, Inc. (AUI), and the National Science Foundation (NSF), and is reproduced here under the Creative Commons Attribution 3.0 Unported license.}
                \end{subfigure}
                \begin{subfigure}[b]{0.45\textwidth}
                \centering
                \includegraphics[width = \textwidth]{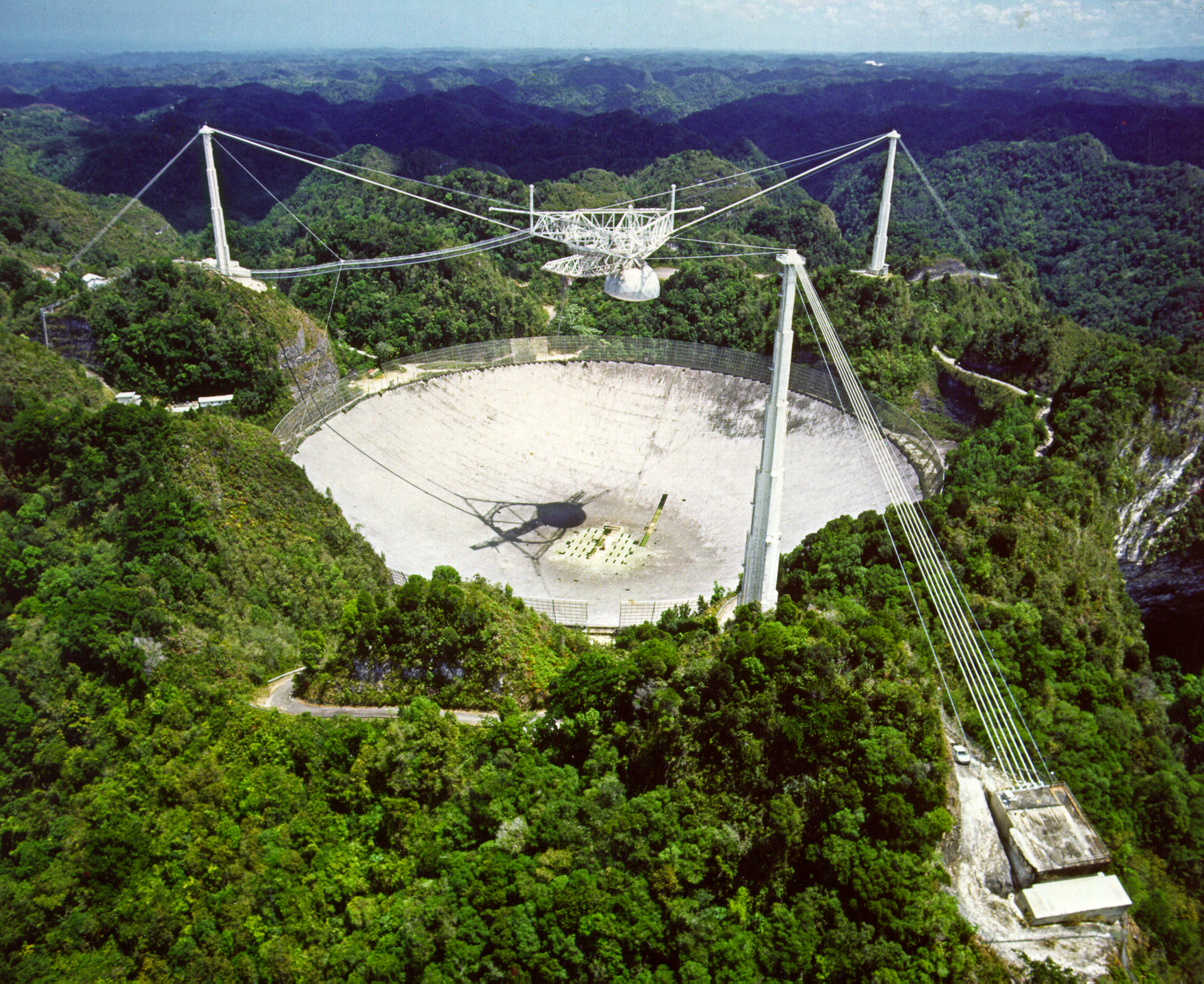}
                \caption{The 300-m diameter Arecibo radio telescope in Puerto Rico, USA. Courtesy of the NAIC - Arecibo Observatory, a facility of the NSF.}
                \end{subfigure}
                \begin{subfigure}[b]{0.45\textwidth}
                \centering
                \includegraphics[width = \textwidth]{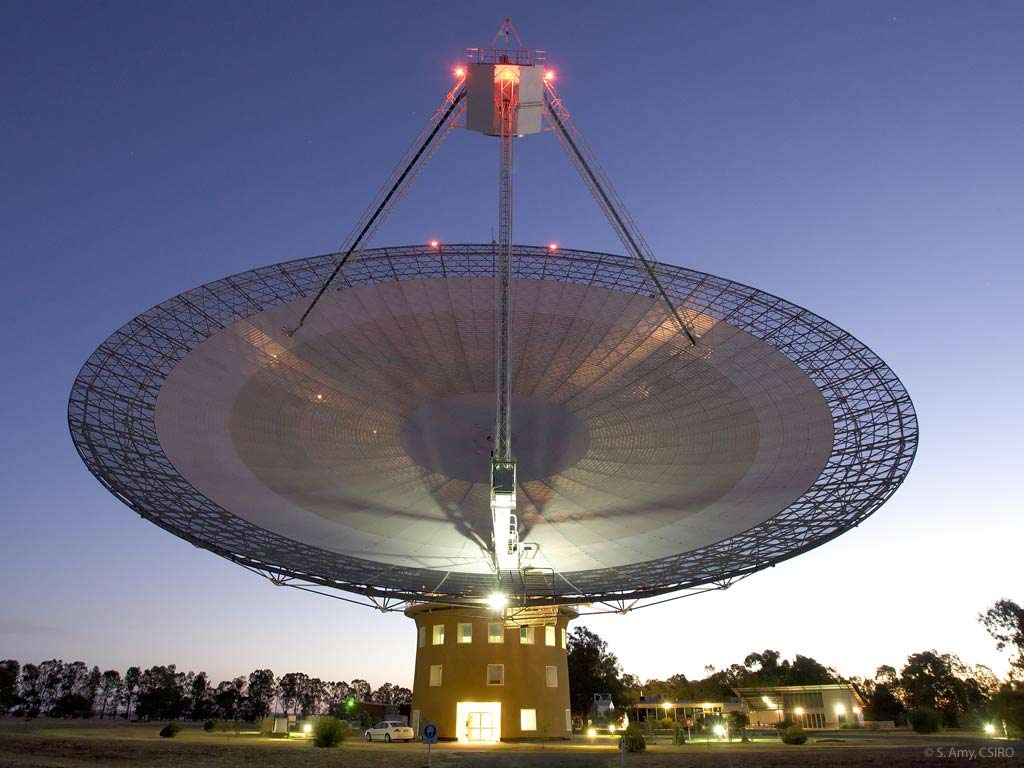}
                \caption{The 64-m diameter Parkes radio telescope in New South Wales, Australia. Copyright CSIRO Australia (2020).}
                \end{subfigure}
                \caption[Examples of radio telescopes around the world.]{Examples of radio telescopes around the world.}
                \label{radio_telescopes}
            \end{figure*}
            
            Given the differences in the telescope and backend setups as well as their different operating frequencies, the sensitivity for each of the pulsar surveys is different. The sensitivity for a pulsar survey can be calculated in terms of the minimum flux density, $S_{\rm min}$, that a pulsar must have in order to be detected with a threshold signal-to-noise (S/N)$_{\rm min}$ ratio \citep{lorimer_kramer},
            \begin{equation}
                \displaystyle S_{\rm min} = \beta \frac{{\rm (S/N)_{\rm min}} T_{\rm sys}}{G \sqrt{n_{\rm p}t_{\rm int} \Delta f}} \sqrt{\frac{W}{P - W}}.
                \label{intro_radiometer_eq}
            \end{equation}
            Here, $\beta$ is a ``correction factor" which accounts for loss in sensitivity due to system imperfections, $T_{\rm sys}$ is the system temperature (including the sky temperature), $G$ is the telescope gain, $n_{\rm p}$ is the number of polarizations summed over in the survey, $t_{\rm int}$ is the integration or observation time, $\Delta f$ is the bandwidth of the receiver, $P$ is the period of the pulsar, and $W$ is the effective pulse-width.
            
            Each of the factors going into Eq.~\ref{intro_radiometer_eq} are different for different surveys. For example, the integration time used for PMSURV is 2100~s, while for PALFA, it is only 268~s. All of these factors result in different sensitivities for the different surveys and thus, different amounts of success in their search for pulsars. Depending on the setup for each survey, they will have their own minimum S/N threshold, but in general, for a convincing detection, a pulsar candidate needs to have a S/N $\geq 8$. Additionally, given their different geographic locations, each telescope will have a different field-of-view of the sky, which may or may not overlap with that of other radio telescopes.
            
        \subsection{Overview of pulsar search techniques}
            
            Given that pulsars are exceptionally regular rotators, their emission is typically highly periodic. As a result, while searching for pulsars, the problem reduces to finding a periodic signal in the data collected by radio telescopes which has been dispersed by an amount quantified by the DM for that pulsar. Given the periodic nature of pulsar emission, Fourier domain searches have been a popular way of searching for pulsars. There also exist time-domain methods for searching for pulsars, but apart from a few exceptions, these methods are not well suited to the discovery of BNS systems. Thus, in this section, we focus on an overview of Fourier-domain based search techniques and refer the reader to \citet{lorimer_kramer} for discussions of other search techniques.
            
            The data collected by radio telescopes when searching for pulsars can be thought of as a three-dimensional array consisting of the time stamp of each sample along one dimension, the frequency that sample was observed at along the other dimension, and the intensity at that time stamp and frequency (for example, see bottom panel of Fig.~\ref{DM_example}). The first step in searching for a pulsar is de-dispersing the data using a trial DM. The resultant data is then usually integrated across frequency to obtain a time-series at the trial DM value. Next, we compute a Fast Fourier Transform (FFT) on this time series, which can then be used to search for significant signals using either the amplitude or power spectrum in the Fourier domain. This process is then repeated for the next trial DM. 
            
            At the correct DM, the signal from the pulsar in the Fourier domain will be sharply peaked at frequencies $f_{\rm n} = n / P_0$, where $P_0$ is the spin period of the pulsar and $n = 1, 2, 3, ...$ represent the harmonics of the signal. Since the pulsar signal is not a pure sinusoid, the power in the Fourier domain will be spread over multiple harmonics. This spread in power across harmonics can be leveraged to increase the S/N ratio of the detection for pulsars by summing over the harmonics (see \citet{lorimer_kramer} for more details).
            
            Once a candidate for a pulsar has been identified with some period, $P$, and DM, the time-series is ``folded" in order to produce a folded or integrated pulse profile. ``Folding" is done by dividing the time series into chunks of data whose length is equal to the period of pulsar and then averaging the data across the individual chunks. Examination of this integrated pulse profile and its structure as a function of frequency is then used to confirm whether the candidate is likely to be a real pulsar. This step of folding the time-series data is necessary because there are numerous sources of radio frequency interference (RFI) which tend to be periodic, and thus, indistinguishable from a pulsar signal in the Fourier domain. Looking at the integrated pulse profile both in the time and frequency as described above provides the ability to distinguish sources of RFI from real pulsar signals.
            
        \subsection{Search techniques for binary neutron star systems}
                
            The technique described above is not optimized for detecting BNS systems, especially those with compact orbits \citep{og_odf}. Due to the orbital motion of the pulsar, the power in each harmonic in the Fourier domain is smeared across adjacent frequency bins. This results in an overall reduction in the S/N with which the pulsar can be detected, thus reducing the overall sensitivity of the survey to a BNS system. The amount of smearing also depends on the integration time of the survey, with longer integration times corresponding to greater smearing and thus a larger reduction in the observed S/N for a pulsar in a BNS system.
            
            \begin{figure}
                \centering
                \includegraphics[width = \columnwidth]{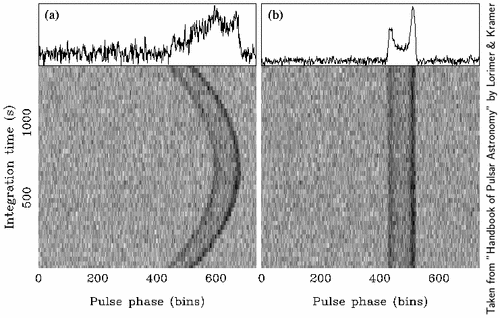}
                \caption[Effect of a pulsar's motion in a binary system on the observed emission from that pulsar.]{An example of the effect of a pulsar's orbital motion on the observed, de-dispersed pulse profile. The image shows the pulse profile from PSR B1913+16, also known as the Hulse-Taylor binary. In the left-hand panel, no corrections to account for the pulsar's orbital motion have been applied, while the right-hand panel shows the effect of the resampled time-series using Eq.~\ref{accel_time_resamp}. Correcting for the pulsar's orbital motion leads to a much stronger detection of the pulsar. The image is from the Handbook of Pulsar Astronomy \citep[ISBN: 9780521828239,][]{lorimer_kramer} and has been reproduced with permission of Cambridge University Press through PLSclear.}
                \label{accel_search_eg}
            \end{figure}
            
            This effect can be mitigated by using a reference frame in which the pulsar in the BNS system is stationary. We can resample the time-series using the Doppler formula \citep{lorimer_kramer},
            \begin{equation}
                \displaystyle \tau = \tau_0 \, \left( 1 + \frac{V(t)}{c} \right)
                \label{accel_time_resamp}
            \end{equation}
            where $V(t)$ is the velocity of the pulsar at time, $t$, and $\tau_0$ is a normalization constant. The amplitude at the resampled time $\tau$ is calculated from the interpolated value at the corresponding time stamp in the original time-series. For a BNS system whose orbital parameters are known, we can directly calculate the correct $V(t)$ and completely remove the effect of the pulsar's orbital motion, as shown in Fig.~\ref{accel_search_eg}.
            
            However, when searching for a BNS system, the orbital parameters of the binary are not known a priori. It is possible to derive the velocity $V(t)$ using Kepler's laws of motion, but that would introduce five additional parameters (orbital period, eccentricity, epoch and angle of periastron passage, and length of the projected semi-major axis) to the search algorithm making the search process computationally expensive. An alternative technique instead assumes a constant acceleration (referred to as ``acceleration search") or a constant jerk (i.e. derivative of the acceleration, referred to as ``jerk search") for the pulsar in the binary and calculates the velocity, $V(t)$, under this assumption. In these types of searches, it is only necessary to search over a single additional parameter (acceleration) for acceleration searches or two additional parameters (acceleration and jerk) in jerk searches. Due to the higher computational cost of the latter search technique, this type of search was not widely implemented in pulsar surveys until only recently. In the following, we describe the methodology behind the acceleration search technique though the methodology is similar for jerk searches as well (see \citet{jerk_search_implementation} for more details on jerk searches).
            
            When doing an acceleration search, the pulsar is assumed to have constant acceleration for the length of the observation. Thus, the velocity of the pulsar can be written as $V(t) = at$, where $a$ is the acceleration of the pulsar. Next, for every trial DM, a range of acceleration values, $a$, are used to resample the time-series. For each trial value of DM and $a$, we take the Fourier transform of the time-series and search for significant signals in the Fourier domain. For the example shown in Fig.~\ref{accel_search_eg}, the right-hand panel uses an acceleration of $a = -16$~m~s$^{-2}$ to correct the observed smearing of the pulsar emission. Jerk searches are implemented in a similar manner, with the only difference being that the velocity is modeled using the jerk, $j$, i.e. $V(t) = at + jt^2$, and thus requires a search over the jerk parameter in addition to the acceleration. 
            
            The repeated Fourier transforms required in these methods can be avoided by doing the entire search in the Fourier domain itself. The effect of resampling the time-series using Eq.~\ref{accel_time_resamp} can be mimicked by using finite impulse response filters in the Fourier domain to achieve the same effect of removing the dispersion of power from the signal harmonics \citep{ransom_accel_fourier_domain}. This results in a significant improvement in the computational efficiency of the search process and as a result, most modern versions of acceleration and jerk searches use the Fourier domain implementation of these methods, such as, for example, in the {\sc PRESTO} \citep{ransom_accel_fourier_domain} software package). A majority of all known DNS and NS--WD systems have been discovered by using this method of acceleration search.
            
        \subsection{Sample of binary neutron star systems in the Galaxy}
            
            \begin{figure}
                \centering
                \includegraphics[width = \textwidth]{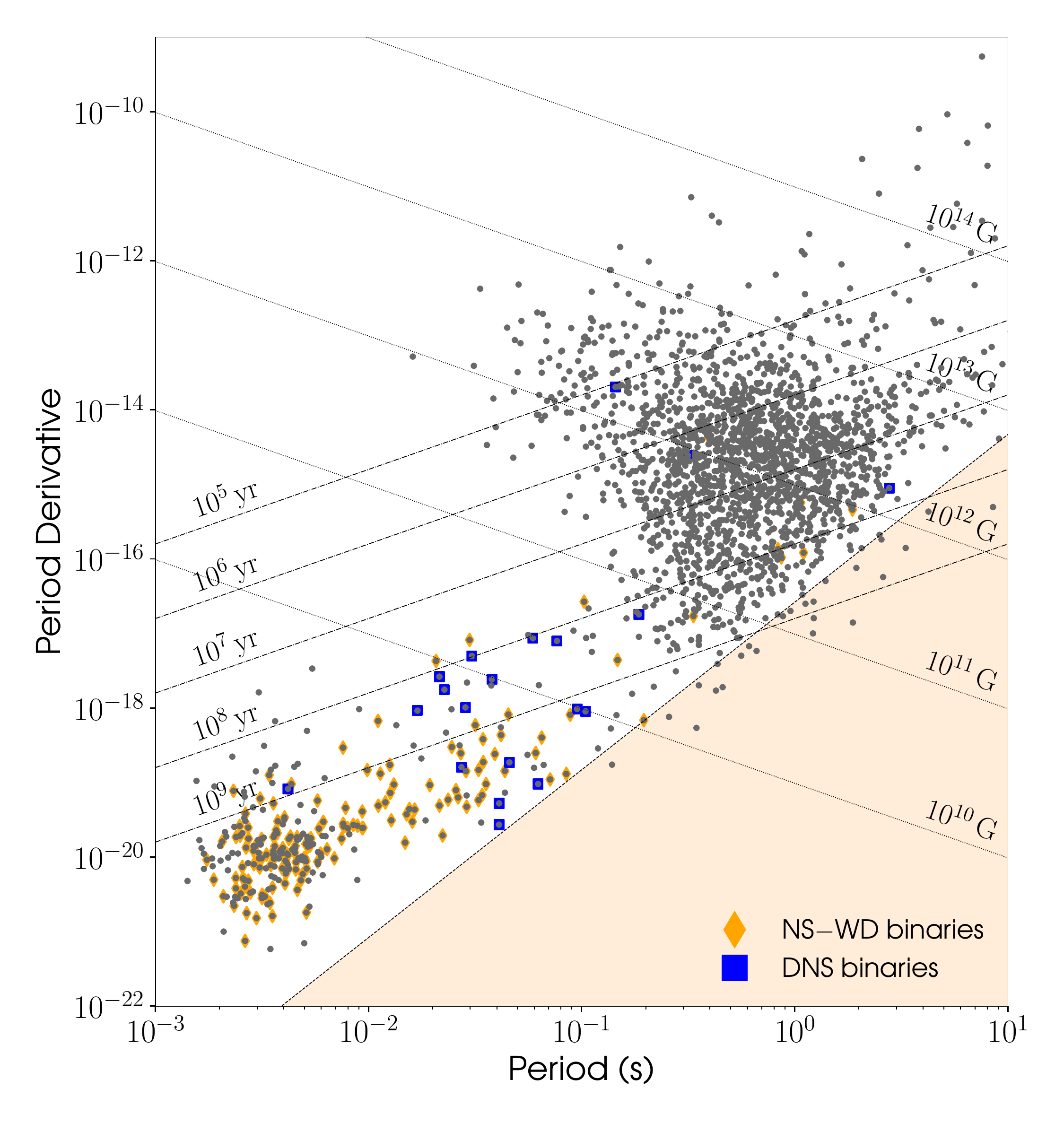}
                \caption[$P-\dot{P}$ diagram showing BNS systems relative to general pulsar population.]{The $P-\dot{P}$ diagram, showing the location of the known BNS systems relative to the general pulsar population. The horizontal axis shows the period in seconds, while the vertical axis shows the period derivative. As described in Sec.~\ref{search_BNS_systems}, pulsars in BNS systems have much smaller periods and period derivatives than the general pulsar population. This figure was generated using data from the ATNF Pulsar catalog \citep{psrcat} and the {\sc psrqpy} software package \citep{psrqpy}.}
                \label{binary_ppdot}
            \end{figure}
            
            To date, we have discovered 20 DNS systems and 185 NS--WD systems, out of $\sim$2800 pulsars in the Milky Way \citep{psrcat}. We have yet to detect a NS--BH system. The period and period-derivative of the known NS--WD and DNS systems are shown in comparison to the canonical pulsar population in Fig.~\ref{binary_ppdot} \citep{psrcat}. As we can see, pulsars in NS--WD systems have some of the smallest spin periods and spin period-derivatives of the known pulsar population. As explained in Sec.~\ref{bns_formation}, this is a result of the large amount of time spent accreting material from the companion by the NS in a NS--WD system. The same argument also explains why the DNS systems have, on average, larger spin periods and period-derivatives, i.e. the first-born NS spends relatively less time accreting material from the companion before the companion collapses into a NS. Since we are observing the first-born NS as the pulsar in most of the BNS systems, these pulsars are older and have smaller magnetic field strengths than the general pulsar population. 
            
            The formation process of BNS systems also leaves an imprint on the observed orbital properties of the BNS system. As shown by \citet{ozel_ns_mass_dist}, DNS systems have a narrow mass distribution peaking at 1.33~$M_{\odot}$ with a dispersion of $\sigma = 0.05$~$M_{\odot}$, while those found in NS--WD systems have a broader mass distribution centered at 1.48~$M_{\odot}$ with a dispersion of $\sigma = 0.2$~$M_{\odot}$. This is indicative of an extended period of mass accretion in the latter type of system.
            The same extended period of mass distribution results in a greater circularizing of the orbit of NS--WD systems relative to DNS systems. As a result, majority of the observed NS--WD systems have eccentricities $e < 10^{-2}$, while the DNS systems have eccentricities ranging from $0.06-0.82$. The observed population of NS--WD systems have orbital periods ranging from $0.08 \, {\rm days} \lesssim P_{\rm b} \lesssim 1200$~days, while the DNS systems have orbital periods $0.08 \, {\rm days} \lesssim P_{\rm b} \lesssim 45$~days \citep[from the pulsar catalog,][]{psrcat}. The difference in the upper limit on the orbital periods of these two types of systems can be explained by the loss of angular momentum of the system during multiple stages of mass transfer in the formation of DNS systems, though a larger sample of observed DNS systems will be necessary to conclusively prove this hypothesis.
            
    \section{The Double Pulsar system}
        
        The J0737--3039 system, also known as the Double Pulsar, is a unique DNS system in which both of the NSs have been observed as pulsars. This system was discovered using the Parkes radio telescope in Australia. One of the pulsars in the system, J0737--3039A (hereafter referred to as ``A"), has a spin period of 22.7~ms \citep{0737A_disc}, while the other pulsar, J0737--3039B (hereafter referred to as ``B") has a spin period of 2.7~s \citep{Lyne_Bdiscovery_2004}.
        Through timing of this DNS system (see \citet{lorimer_kramer} for a review of pulsar timing), pulsar A was found to be slowing down in its spin period due to magnetic dipole-braking at a rate $\dot{P} = 1.8 \times 10^{-18}$~s~s$^{-1}$, while pulsar B was found to be slowing down at a much faster rate of $\dot{P} = 0.9 \times 10^{-15}$~s~s$^{-1}$. The spin-down rate combined with their spin periods tells us that pulsar A is the older, first-formed pulsar in the system that was likely spun up to periods on the order of milliseconds during the accretion phases in the formation of the DNS system (see Sec.~\ref{bns_formation}). On the other hand, pulsar B is the younger, second-formed pulsar in the system, which explains its slower spin period and higher spin-down rate.
        
        In addition to the spin parameters, the timing of the Double Pulsar also revealed that the orbital period of the system was 2.4~hours, with a semi-major axis of $\sim$1.25$R_{\odot}$ and a mild eccentricity of 0.088. The compact configuration of this system combined with both NSs being observed as pulsars offers a unique opportunity to probe a wide range of physical phenomena, ranging from tests of General Relativity \citep{Kramer_GRtest_2006} to studying magnetospheric physics \citep{Maura_mod_2004, david_lyutikov_modelling} to understanding binary stellar evolution \citep{Stairs_0737formation_2006, Ferdman_snevidence_2013}. This system also offers the best opportunity for measuring, for the first time, the moment-of-inertia of a NS \citep{Kramer_testofgravity_2009}, specifically for pulsar A, which in turn will allow us to place some of the most stringent constraints on the NS equation-of-state which will reveal the behavior of matter at densities that are impossible to recreate on Earth. In this section, we provide an overview of one of the unique science opportunities offered by the Double Pulsar system that we will build upon in Sec.~\ref{chap:sense_rotn_A} and refer the reader to \citet{dpsr_review} for a full description of the science with the Double Pulsar.
        
        \begin{figure}
            \centering
            \includegraphics[width = \textwidth]{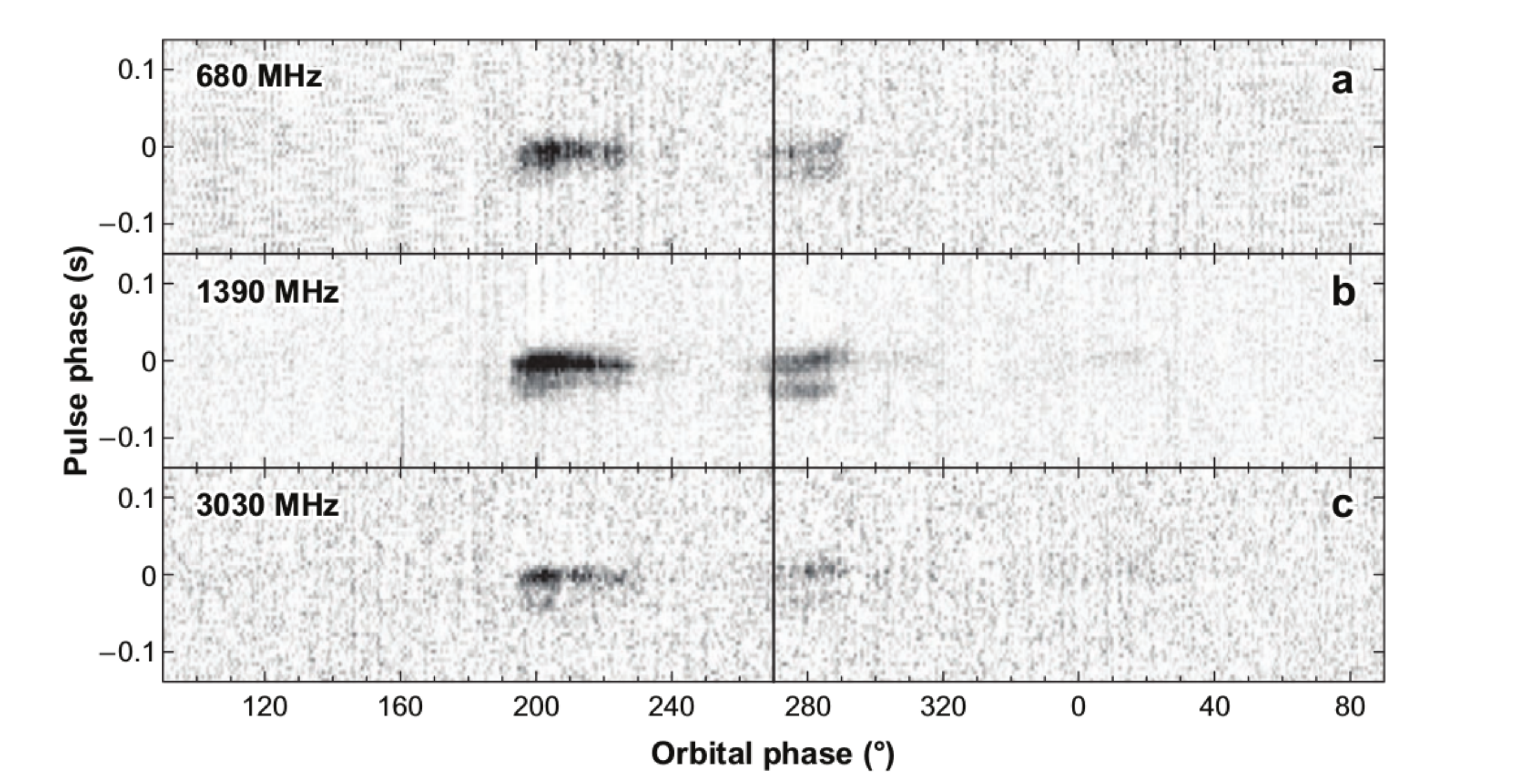}
            \caption[Emission from pulsar B in the Double Pulsar as a function of orbital phase.]{Intensity of pulsar B's emission as a function of orbital phase relative to the ascending node at three different frequencies. Pulsar B is bright in only two parts of its orbit. Image republished with permission of Annual Reviews, Inc., from \citet{dpsr_review}; permission conveyed through Copyright Clearance Center, Inc.}
            \label{B_emission_orbphase}
        \end{figure}
        
        We can calculate the spin-down energy loss, $\dot{E}$, as \citep{lorimer_kramer},
        \begin{equation}
            \displaystyle \dot{E} = 4 \pi^2I \frac{\dot{P}}{P^3}
            \label{spin_down_energy_eq}
        \end{equation}
        where $I$ is the moment of inertia of the pulsar, and $P$ and $\dot{P}$ are the spin period and spin-down rate of the pulsar respectively. As described in \citet{Lyne_Bdiscovery_2004}, the energy emitted by pulsar A is approximately two orders of magnitude greater than that for pulsar B. Correspondingly, the energy emitted by A, particularly in the pulsar wind from A, impinges on and deforms the magnetosphere of pulsar B \citep{Maura_mod_2004, david_lyutikov_modelling}. As shown in Fig.~\ref{B_emission_orbphase}, this results in the emission from pulsar B being visible in only two parts of its orbit when this interaction pushes the emission from B into the line of sight to Earth.
        
        \begin{figure}
            \centering
            \includegraphics[width = \textwidth]{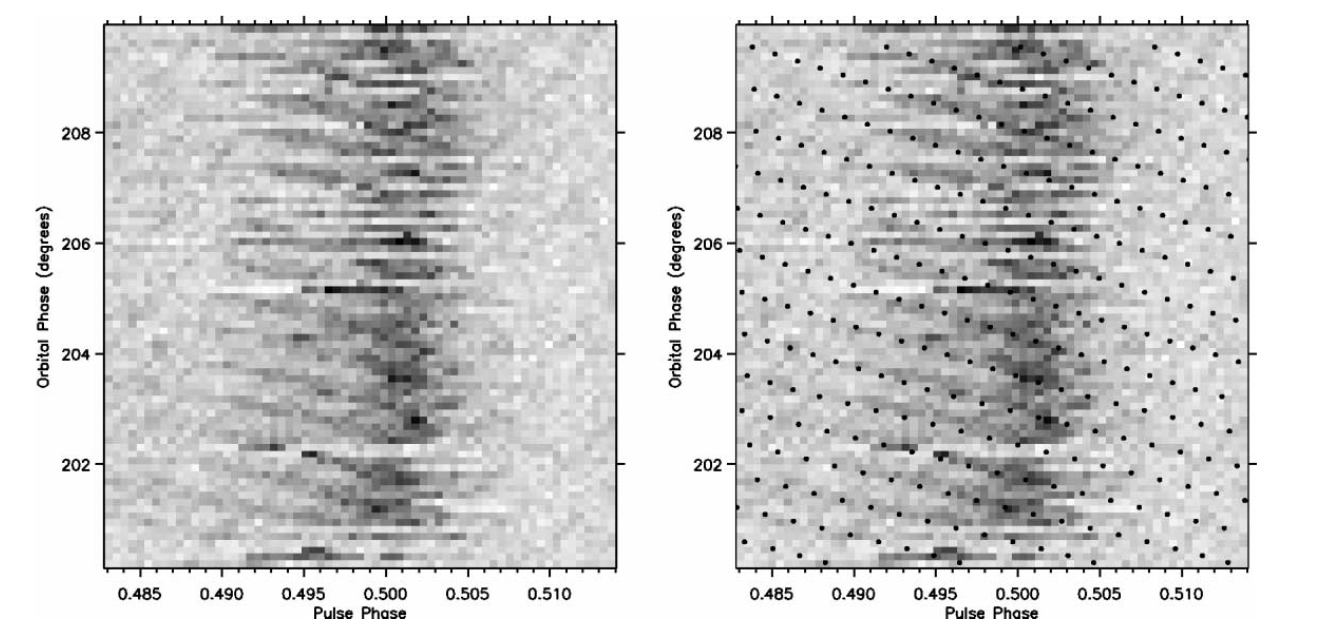}
            \caption[Single pulse emission from pulsar B in the Double Pulsar showing the drifting features due to pulsar A stimulating emission in B's magnetosphere.]{Single pulse emission from pulsar B for one part of the orbit where B's emission is visible. The emission from pulsar A is also visible in the background, most prominently at orbital phase of $\sim$225$^{\circ}$. The drifting features are present in most of the data, but are most visible at orbital phases between $190^{\circ}-210^{\circ}$. The dots on the right hand panel denote the arrival of emission from A at the center of B. This image is from \citet{Maura_mod_2004}. Copyright AAS. Reproduced with permission.}
            \label{B_drifting_subpulses}
        \end{figure}
        
        Another consequence of this interaction is that pulsar A induces emission in pulsar B's magnetosphere, which was first discovered by \citet{Maura_mod_2004}. This manifests itself as drifting sub-pulses in the observed emission from B, as shown in Fig.~\ref{B_drifting_subpulses}. As we can see, the emission in the drifting sub-pulses is correlated with the arrival of emission from pulsar A at the center of pulsar B. \citet{Maura_mod_2004} also found that this observed modulation has a period equal to the rotational period of pulsar A, implying that this induced emission is a result of the magnetic dipole radiation of A interacting with B's magnetosphere instead of A's beamed emission or its intensity or pressure, both of which have a period of twice the rotation period of A since we observe emission from both magnetic poles of A \citep{Ferdman_snevidence_2013}. This is the first time that we have observed an external stimulation of emission in a pulsar. We exploit these drifting sub-pulses to infer the sense of rotation of pulsar A in Sec.~\ref{chap:sense_rotn_A}.
        
    \section{Multi-messenger astrophysics with DNS systems}
        
        As described above, DNS systems have rich potential for science with observations in the electromagnetic band. However, with the recent advent of gravitational-wave (GW) astronomy, DNS systems are one of the most promising sources for GW observatories. The observations with GWs can be combined with observations using electromagnetic light to perform multi-messenger studies of these sources that were impossible only $\sim$5 years ago.
        
        GW detectors operate by exploiting the space-time distortions produced by an incident GW. For example, terrestrial GW detectors like the LIGO--Virgo \citep{LIGO_detector_ref, VIRGO_detector_ref} network use Michelson interferometers to convert these space-time distortions into a constructive/destructive interference in the interferometer, which is then used to interpret the properties of the GW. Space-based detectors like the Laser Interferometer Space Antenna \citep[LISA,][]{LISA}, due to be launched in the 2030s, uses a similar concept on-board satellites orbiting the Earth and Sun, while pulsar timing arrays (PTAs) like the North American Nanohertz Observatory for Gravitational waves \citep[NANOGrav,][]{nanograv_foundation} uses the apparent Doppler shift induced by the passing GW in the times of arrival of the periodic pulsar emission.
        
        \begin{figure}
            \centering
            \includegraphics[width = \columnwidth]{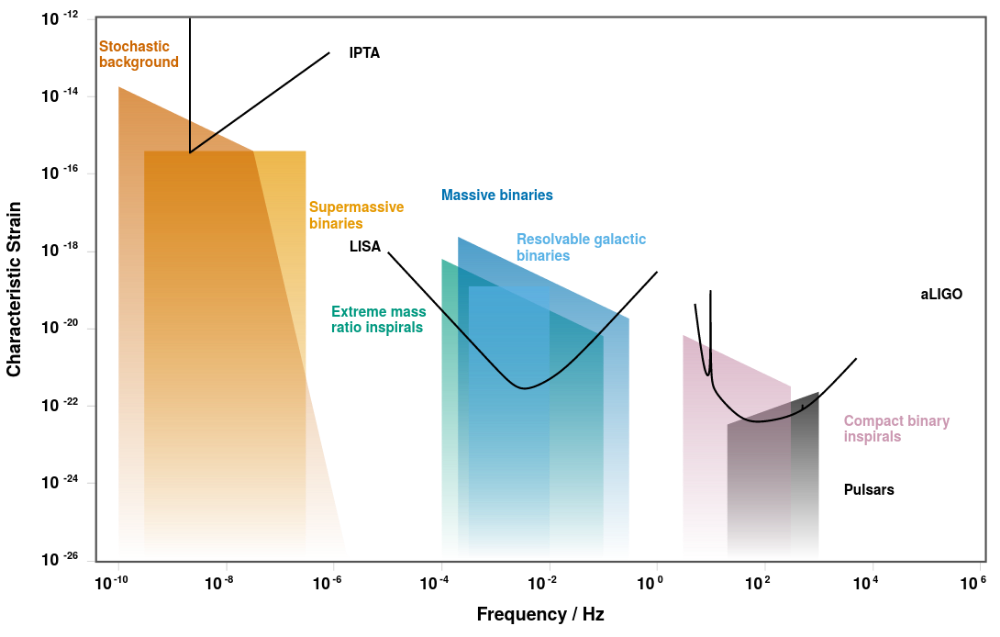}
            \caption{Illustration of the GW spectrum along with the characteristic strain sensitivity curves for different GW detectors. The figures also shows the sources at different frequency bands. This image was generated using the software provided by \citet{gw_spectrum_plotter}.}
            \label{gw_spectrum}
        \end{figure}
        
        However, despite the similarity in their underlying principle of operation, the GW frequencies (and thus, sources) that these GW detectors are sensitive to are drastically different. As shown in Fig.~\ref{gw_spectrum}, LIGO--Virgo detectors operate at frequencies of few tens to hudreds of hertz, LISA operates at frequencies ranging from a few tenths of a millihertz to a few hertz, and NANOGrav operates at $\sim$nanohertz frequencies. Correspondingly, among other sources, LIGO--Virgo is mainly sensitive to merging compact binary systems, LISA is mainly sensitive to inspiraling (ultra)compact binaries and NANOGrav is sensitive to individual inspiraling supermassive black hole binaries. Thus, inspiraling BNS systems with compact orbits will be visible to LISA, while BNS systems that are merging will be visible to LIGO--Virgo.
        
        The era of multi-messenger astronomy was ushered in with the detection of the merger of a DNS system in 2017 \citep{THE_DNS_merger}. This merger was first detected using GWs by the LIGO--Virgo \citep{LIGO_detector_ref, VIRGO_detector_ref} detectors. On detection of this event, the LIGO--Virgo collaboration circulated an alert to other electromagnetic observatories around the world, which led to a detection of the same event across the electromagnetic band \citep{THE_DNS_merger_EM_assoc}. This discovery has led to important scientific results ranging from tests of General Relativity \citep{dns_merger_testofGR}, a new method of measuring the Hubble constant \citep{dns_merger_hubble_const}, placing constraints on the NS equation-of-state \citep{dns_merger_EoS} and provided the explanation for the existence of elements heavier than iron in the Universe \citep{dns_merger_rprocess}. A second detection of a merging DNS, while not detected in the electromagnetic band, provided a detection of a DNS system with a total mass significantly higher than that measured for the known DNS systems in the Galaxy \citep{dns_merger_2}.
        
        All of the above science comes from the detection of only two DNS mergers with ground-based GW observatories. We can predict the number of mergers that these observatories will detect each year by analyzing the observed population of DNS systems in the Galaxy which were discovered in radio pulsar surveys (see Sec.~\ref{search_BNS_systems}). However, these systems are $\sim$Myrs away from merging and thus being detectable by the aforementioned ground-based GW observatories. However, we can still use these DNS systems, especially those which will merge within the age of the Universe, to model the population of merging DNS systems in the Galaxy and thus derive the corresponding merger rate we expect for the LIGO--Virgo detectors. The process to calculate this merger rate is described in Chapter~\ref{chap:dns_merger_rate}.
        
        While the current ground-based GW observatories are only sensitive to the cataclysmic mergers of DNS systems, LISA \citep{LISA} will be sensitive to BNS systems with orbital periods $\sim$tens of minutes. However, no BNS system with such a small orbital period has been detected in the Galaxy to date. Thus, in preparation of the launch of LISA, in Chapter~\ref{chap:lisa_bns_systems} we place an upper limit on the population of these ``ultra-compact" binary systems in the Galaxy. We also derive an optimum survey integration time that will maximize the probability of detecting these systems for the current radio pulsar surveys.
\newpage
\renewcommand{\thechapter}{2}

\chapter[Calculating the Galactic Double Neutron Star Merger Rate]{Future prospects for ground-based gravitational wave detectors --- The Galactic double neutron star merger rate revisited}
\label{chap:dns_merger_rate}
\blfootnote{This chapter combines the results published in Pol et al., 2019, ApJ, 870, 71, and Pol et al., 2020, RNAAS, 4, 22. The text from the former has been updated to reflect the results derived in the latter. \\
\textbf{Contributing authors:} Nihan Pol, Maura McLaughlin, Duncan Lorimer
}

\section{Abstract} \label{abs}
    We present the Galactic merger rate for double neutron star (DNS) binaries using the observed sample of eight DNS systems merging within a Hubble time. This sample includes the recently discovered, highly relativistic DNS systems J1757$-$1854 and J1946+2052, and is three times the sample size used in previous estimates of the Galactic merger rate by Kim et al. Using this sample, we calculate the vertical scale height for DNS systems in the Galaxy to be $z_0 = 0.4 \pm 0.1$~kpc. We calculate a Galactic DNS merger rate of $\mathcal{R}_{\rm MW} = 37^{+24}_{-11}$~Myr$^{-1}$ at the 90\% confidence level. The corresponding DNS merger detection rate for Advanced LIGO is $\mathcal{R}_{\rm LIGO} = 1.9^{+1.2}_{-0.6} \times \left( D_{\rm r}/100 \ \rm Mpc \right)^3 \rm yr^{-1}$, where $D_{\rm r}$ is the range distance. Using this merger detection rate and the predicted range distance of 120--170~Mpc for the third observing run of LIGO (Laser Interferometer Gravitational-wave Observatory, Abbott et al., 2018), we predict, accounting for 90\% confidence intervals, that LIGO--Virgo will detect anywhere between two and fifteen DNS mergers. We explore the effects of the underlying pulsar population properties on the merger rate and compare our merger detection rate with those estimated using different formation and evolutionary scenario of DNS systems. As we demonstrate, reconciling the rates are sensitive to assumptions about the DNS population, including its radio pulsar luminosity function. Future constraints from further gravitational wave DNS detections and pulsar surveys anticipated in the near future should permit tighter constraints on these assumptions. 

\section{Introduction}      \label{intro}
    
    The first close binary with two neutron stars (NSs) discovered was PSR~B1913+16 \citep{1913_ht_discovery}. This double neutron star (DNS) system, known as the Hulse-Taylor binary, provided the first evidence for the existence of gravitational waves through measurement of orbital period decay in the system \citep{1913_gw_emission}. This discovery resulted in a Nobel Prize being awarded to Hulse and Taylor in 1993. The discovery of the Hulse-Taylor binary opened up exciting possibilities of studying relativistic astrophysical phenomena and testing the general theory of relativity and alternative theories of gravity in similar DNS systems \citep{stairs_dns_review}.
    
    Despite the scientific bounty on offer, relatively few DNS systems have been discovered since the Hulse-Taylor binary, with only 15 more systems discovered since. DNS systems are intrinsically rare since they require the binary system to remain intact with both components of the system undergoing supernova explosions to reach the final neutron star stage of their evolution. In addition, DNS systems are very hard to detect because of the large accelerations experienced by the two neutron stars in the system, which results in large Doppler shifts in their observed rotational periods \citep{Bagchi_odf}.
    
    As demonstrated in the Hulse-Taylor binary, the orbit of these DNS systems decays through the emission of gravitational waves which eventually leads to the merger of the two neutron stars in the system \citep{1913_gw_emission}. DNS mergers are sources of gravitational waves that can be detected by ground-based detectors such as the Laser Interferometer Gravitational-Wave Observatory \citep[LIGO,][]{LIGO_detector_ref} in the USA and the Virgo detector \citep[][]{VIRGO_detector_ref} in Europe. Very recently, one such double neutron star (DNS) merger was observed by the LIGO-Virgo network
    \citep[][]{THE_DNS_merger} which was also detected across the electromagnetic spectrum \citep[][]{THE_DNS_merger_EM_assoc}, heralding a new age of multi-messenger gravitational wave astrophysics.
    
    We can predict the number of such DNS mergers that the LIGO-Virgo network will be able to observe by determining the merger rate in the Milky Way, and then extrapolate it to the observable volume of the LIGO-Virgo network. The first such estimates were provided by \citet[][]{Phinney_blue_lum_scaling} and \citet[][]{Narayan_1991} based on the DNS systems B1913+16 \citep{1913_ht_discovery} and B1534+12 \citep{1534_disc}. A more robust approach for calculating the merger rate was developed by \citet[][hereafter KKL03]{dunc_merger_1}, on the basis of which \citet{0737A_disc} and \citet[][]{A_effect_on_merger_rate, Kim_B_merger} were able to update the merger rate by including the Double Pulsar J0737--3039 system \citep{Lyne_Bdiscovery_2004, 0737A_disc}. 
    
    In the method described in KKL03, which we adopt in this work, we simulate the population of DNS systems like the ones we have detected by modeling the selection effects introduced by the different pulsar surveys in which these DNS systems are discovered or re-detected. This population of the DNS systems is then suitably scaled to account for the lifetime of the DNS systems and the number of such systems in which the pulsar beam does not cross our line of sight. We are only interested in those DNS systems that will merge within a Hubble time. Using this methodology, \citet[][]{Kim_B_merger} estimated the Galactic merger rate to be $\mathcal{R}_{\rm g} = 21^{+28}_{-14}$~Myr$^{-1}$ and the total merger detection rate for LIGO to be $\mathcal{R}_{\rm LIGO} = 8^{+10}_{-5}$~yr$^{-1}$, with errors quoted at the 95\% confidence interval, assuming a horizon distance of $D_{\rm h} = 445$~Mpc, and with the B1913+16 and J0737--3039 systems being the largest contributors to the rates.
    
    
    In this work, we include six new DNS systems into the estimation of the merger rate. Out of these six systems, two, J1757--1854 \citep{1757_disc}, with a time to merger of 76~Myr, and 1946+2052 \citep{1946_disc}, with a time to merger of 46~Myr, are highly relativistic systems that will merge on a timescale shorter than that of the Double Pulsar, which had the previous shortest time to merger of 85~Myr. The other DNS systems that we include in our analysis, J1906+0746 \citep{1906_disc}, J1756--2251 \citep{1756_disc}, J1913+1102 \citep{1102_disc}, and J0509+3801 \citep{0509_disc_paper} are not as relativistic, but are important to accurately modeling the complete Milky Way merger rate. These systems were not included in the previous studies due to insufficient evidence for them being DNS systems. However, \citet[][for J1906+0746]{1906_dns_evidence}, \citet[][for J1756--2251]{1756_dns_evidence}, and \citet[][for J1913+1102]{1102_dns_evidence} have established through timing observations that these are DNS systems.
    
    We tabulate the properties of all the known DNS binaries in the Milky Way, sorted by their time to merger, in Table~\ref{psr_table}. With the inclusion of the six additional systems, our sample size for calculating the merger rate is three times the one used in \citet[][]{Kim_B_merger}. In Section~\ref{survey_sims}, we describe the pulsar population characteristics and survey selection effects that are implemented in this study. In Section~\ref{stat_analysis}, we briefly describe the statistical analysis methodology presented in KKL03 and present our results on the individual and total merger rates. In Section~\ref{discuss}, we discuss the implications of our merger rates and compare our total merger rate with that predicted by the LIGO-Virgo group and that estimated through studying the different formation and evolutionary scenarios for DNS systems.
    
\begin{landscape}
\begin{table*}
\centering
\begin{tabular}{rrrrrrrrrrr}
\toprule
& & & & & & & & & \\
{} &          \multicolumn{1}{c}{$l$} &         \multicolumn{1}{c}{$b$} &        \multicolumn{1}{c}{$P$} &        \multicolumn{1}{c}{$\dot{P}$} &       \multicolumn{1}{c}{DM} &       \multicolumn{1}{c}{$P_b$} &       \multicolumn{1}{c}{$a$} &        \multicolumn{1}{c}{$e$} &        \multicolumn{1}{c}{$z$} & \multicolumn{1}{c}{Merger time} \\
PSR        &    (deg)         &     (deg)       &     (ms)      &     ($10^{-18}$ s/s)      &     (pc cm$^{-3}$)     &    (days)      &   (lt-s)       &            &     (kpc)      &       (Gyr)        \\
& & & & & & & & & \\
\midrule
& & & & Non-merging systems & & & & & \\
\midrule
& & & & & & & & & \\
J1518+4904  &   80.8 &  54.3 &  40.9 &  0.027 &   12 &   8.63 &  20.0 &  0.25 &  0.78 &  2400 \\
J0453+1559  &  184.1 & $-$17.1 &  45.8 &  0.19 &   30 &   4.07 &  14.5 &  0.11 & $-$0.15 &  1430 \\
J1811$-$1736  &   12.8 &   0.4 & 104.2 &  0.90 &  476 &  18.78 &  34.8 &  0.83 &  0.03 &  1000 \\
J1411+2551  &   33.4 &  72.1 &  62.4 &  0.096 &   12 &   2.62 &   9.2 &  0.17 &  1.08 &  460 \\
J1829+2456  &   53.3 &  15.6 &  41.0 &  0.052 &   14 &   1.18 &   7.2 &  0.14 &  0.24 &  60 \\
J1753$-$2240  &    6.3 &   1.7 &  95.1 &  0.97 &  159 &  13.64 &  18.1 &  0.30 &  0.09 & - \\
J1930$-$1852  &   20.0 & $-$16.9 & 185.5 &  18.0 &   43 &  45.06 &  86.9 &  0.40 & $-$0.58 &  - \\
& & & & & & & & & & \\
\midrule
& & & & Merging systems & & & & & \\
\midrule
& & & & & & & & & \\
B1534+12    &   19.8 &  48.3 &  37.9 &  2.4 &   12 &   0.42 &   3.7 &  0.27 &  0.79 &  2.70 \\
J1756$-$2251  &    6.5 &   0.9 &  28.5 &  1.0 &  121 &   0.32 &   2.8 & 0.18 &  0.01 &  1.69 \\
J0509+3801  &    168.3 &   $-$1.2 &  76.5 &  7.9 &  69 &   0.38 &   2.0 & 0.58 &  $-$0.04 &  0.59 \\
J1913+1102  &   45.2 &   0.2 &  27.3 &  0.16 &  339 &   0.21 &   1.7 &  0.09 &  0.02 &  0.50 \\
J1906+0746  &   41.6 &   0.1 &  144.0 &  20000 &  218 &   0.17 &   1.4 &  0.08 &  0.02 &  0.30 \\
B1913+16    &   50.0 &   2.1 &  59.0 &  8.6 &  169 &   0.32 &   2.3 &  0.62 &  0.19 &  0.30 \\
J0737$-$3039A &  245.2 &  $-$4.5 &  22.7 &  1.8 &   49 &   0.10 &   1.4 & 0.09 & $-$0.09 &  0.085 \\
J0737$-$3039B &  245.2 &  $-$4.5 &  2773.5 &  890 &   49 &   0.10 &   1.5 &  0.09 & $-$0.09 &  0.085 \\
J1757$-$1854  &   10.0 &   2.9 &  21.5 &  2.6 &  378 &   0.18 &   2.2 & 0.60 & 0.37 &  0.076 \\
J1946+2052  &   57.7 &  $-$2.0 &  16.9 &     0.90 &   94 &   0.08 &   1.1 &  0.06 &  $-$0.14 &  0.046 \\
\bottomrule
\end{tabular}
\caption[Properties of the observed DNS systems in the Galaxy.]{The current sample of DNS systems ranked in decreasing order of time to merger, along with the relevant pulsar and orbital properties. We only consider those systems that will merge within a Hubble time for the merger rate analysis.}
\label{psr_table}
\end{table*}
\end{landscape}

\section{Pulsar survey simulations}     \label{survey_sims}
    
    To model the pulsar population and survey selection effects, we make use of the freely available PsrPopPy\footnote{\url{https://github.com/devanshkv/PsrPopPy2}} software \citep[][]{psrpoppy, psrpoppy_dag} for generating the population models and writing our own Python code\footnote{\url{https://github.com/NihanPol/2018-DNS-merger-rate}} \citep{code_for_paper, deg_fac_code} to handle all the statistical computation. Here, we describe some of the important selection effects that we model using PsrPopPy.
    
    \subsection{Physical, luminosity and spectral index distribution} \label{physical_dist}
        
        Since we want to calculate the total number of DNS systems like the ones that have been observed, we fix the physical parameters of the pulsars generated in our simulation to represent the DNS systems in which we are interested. These physical parameters include the pulse period, pulse width, and orbital parameters like eccentricity, orbital period, and semi-major axis.
        
        However, even if the physical parameters of the pulsars are the same, their luminosity will not be the same. Thus, to model the luminosity distribution of these pulsars, we use a log-normal distribution with a mean of $\left<\text{log}_{10}L\right> = -1.1$ ($L = 0.07$~mJy~kpc$^2$) and standard deviation, $\sigma_{\text{log}_{10} L} = 0.9$ \citep[][]{FG_kaspi_lum}.
        
        We also vary the spectral index of the simulated pulsar population. We assume the spectral indices have a normal distribution, with mean, $\alpha = -1.4$, and standard deviation, $\beta = 1$ \citep[][]{Bates_si_2013}.
        
    \subsection{Surveys chosen for simulation} \label{surveys}
        
        All of the DNS systems that merge within a Hubble time have either been detected or discovered in the following surveys: the Pulsar Arecibo L-band Feed Array survey \citep[PALFA,][]{PALFA_survey_1}, the High Time-Resolution Universe pulsar survey \citep[HTRU,][]{htru_low_mid}, the Parkes High-latitude pulsar survey \citep[][]{PHSURV_1}, the Parkes Multibeam Survey \citep[][]{PMSURV}, the Green Bank North Celestial Cap survey \citep[GBNCC,][]{gbncc} and the survey carried out by \citet[][]{1534_disc} in which B1534+12 was discovered. All of these surveys together cover more area on the sky than that covered by the 18 surveys simulated in KKL03 and by \citet[][]{Kim_B_merger}, who included the Parkes Multibeam Pulsar survey in addition to the 18 surveys simulated in KKL03.
        
        We implement these surveys in our simulations with PsrPopPy. We generate a survey file \citep[see Sec.~4.1 in ][]{psrpoppy} for each of these surveys using the published survey parameters. These parameters are then used to estimate the radiometer noise in each survey, which, along with a fiducial signal-to-noise cut-off, will determine whether a pulsar from the simulated population can be detected with a given survey. For example, one important difference in these surveys is their integration time, which ranges from 34~s for Arecibo drift-scan surveys to 2100~s in the Parkes Multibeam survey. Other selection effects can be introduced through differences in the sensitivity of the different surveys, the portion and area of the sky covered and minimum signal-to-noise ratio cut-offs.
        
        PSR J1757--1854 was discovered in the HTRU low-latitude survey using a novel search technique \citep{1757_disc}. As described in \citet[][]{ng_accel_search_tech}, the original integration time of 4300~s of the HTRU low-latitude survey was successively segmented by a factor of two into smaller time intervals until a pulsar was detected. This has the effect of reducing Doppler smearing due to extreme orbital motion in tight binary systems (see Sec.~\ref{doppler_smearing} for more on Doppler smearing). The shortest segmented integration time used in their analysis is 537~s (one-eighth segment), which implies that the data are sensitive to binary systems with orbital periods $P_{\rm b} \geq 1.5$~hr \citep{ng_accel_search_tech}. All of these segments are searched for pulsars in parallel. We use the integration time of 537~s in our analysis to ensure that the HTRU survey is sensitive to all the DNS systems included in this analysis. We demonstrate the effect of this choice in Sec.~\ref{lum_dist_eff}.
        
        The survey files are available in the GitHub repository associated with this paper and the properties of these surveys are listed in Table~\ref{survey_table}.
    
    \subsection{Spatial distribution} \label{spatial_dist}
        
        For the radial distribution of the DNS systems in the Galaxy, like \citet[][]{Kim_B_merger}, we use the model proposed in \citet[][]{lorimer_rad_dist}. For the distribution of pulsars in terms of their height, $z$, with respect to the Galactic plane, we use the standard two-sided exponential function \citep[][]{Lyne_stats_1998, lorimer_rad_dist},
        \begin{equation}
            \displaystyle f(z) \propto {\rm exp} \left( \frac{-|z|}{z_0} \right)
            \label{z_scale_ht}
        \end{equation}
        where $z_0$ is the vertical scale height. To constrain $z_0$, we simulate DNS populations with a uniform period distribution ranging from 15~ms to 70~ms, consistent with the periods of the recycled pulsars in the DNS systems listed in Table~\ref{psr_table}, and the aforementioned luminosity and spectral index distribution. We generate these populations with vertical scale heights ranging from $z_0 = 0.1$~kpc to $z_0 = 2$~kpc. We run the surveys described in Section~\ref{surveys} on these populations to determine the median vertical scale height of the pulsars that are detected in these surveys. We also calculate the median DM$\times {\rm sin}(|b|)$, which is more robust against errors in converting from dispersion measure to a distance using the NE2001 Galactic electron density model \citep[][]{ne2001}.
        
        We compare these values at different input vertical scale heights with the corresponding median values for the real DNS systems. We show the median DM$\times {\rm sin}(|b|)$ value and the median vertical $z$-height of the pulsars detected in the simulations as a function of the input $z_0$ in Figs.~1~and~2 respectively. In both of these plots, the median values of the real DNS population are plotted as the red dashed line, with the error on the median shown by the shaded cyan region. As can be seen, the analysis using DM$\times {\rm sin}(|b|)$ predicts a vertical scale height of $z_0 = 0.4 \pm 0.1$~kpc, while the analysis using the $z$-height estimated using the NE2001 model \citep{ne2001} returns a vertical scale height of $z_0 = 0.4^{+0.3}_{-0.2}$~kpc. While both these values are consistent with each other, the vertical scale height returned by the DM$\times {\rm sin}(|b|)$ analysis yields a better constraint on the scale height which is more in line with vertical scale heights for ordinary pulsars \citep[$0.33 \pm 0.029$~kpc,][]{z_dist, lorimer_rad_dist} and millisecond pulsars \citep[0.5~kpc,][]{Levin_zheight_2013}. We expect the neutron stars that exist in DNS systems, and particularly those DNS systems that merge within a Hubble time, to be born with small natal kicks so as not to disrupt the orbital system. Consequently, we would expect these systems to be closer to the Galactic plane than the general millisecond pulsar population. As a result, we adopt a vertical scale height of $z = 0.33$~kpc as a conservative estimate on the vertical scale height of the DNS population distribution. This difference in the vertical scale height does not result in a significant change in the merger rates.
        
        \begin{figure}
            \centering
            \includegraphics[width = \columnwidth]{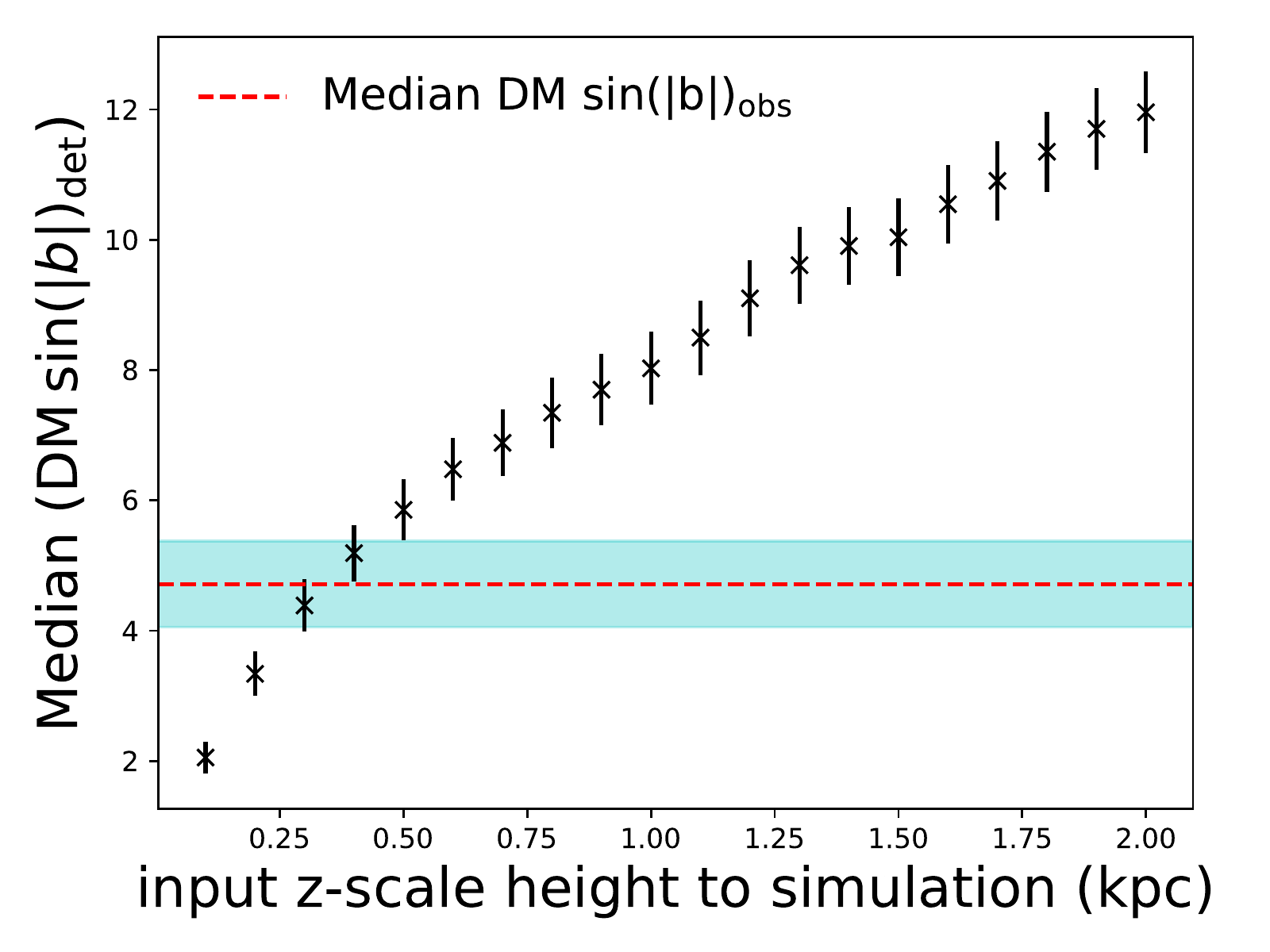}
            \caption[Median observed DM$\times {\rm sin}(|b|)$ plotted versus the input scale height $z_0$ used for the simulated population.]{Median observed DM$\times {\rm sin}(|b|)$ plotted versus the input scale height $z_0$ used for the simulated population. The median DM$\times {\rm sin}(|b|)$ value of the real observed population is shown as the horizontal dashed line, with the shaded cyan region depicting 1$\sigma$ error. The populations generated with vertical scale heights ranging from 0.3~kpc to 0.5~kpc agrees with the observed sample.}
            \label{dm_sinb}
        \end{figure}
        
        \begin{figure}
            \centering
            \includegraphics[width = \columnwidth]{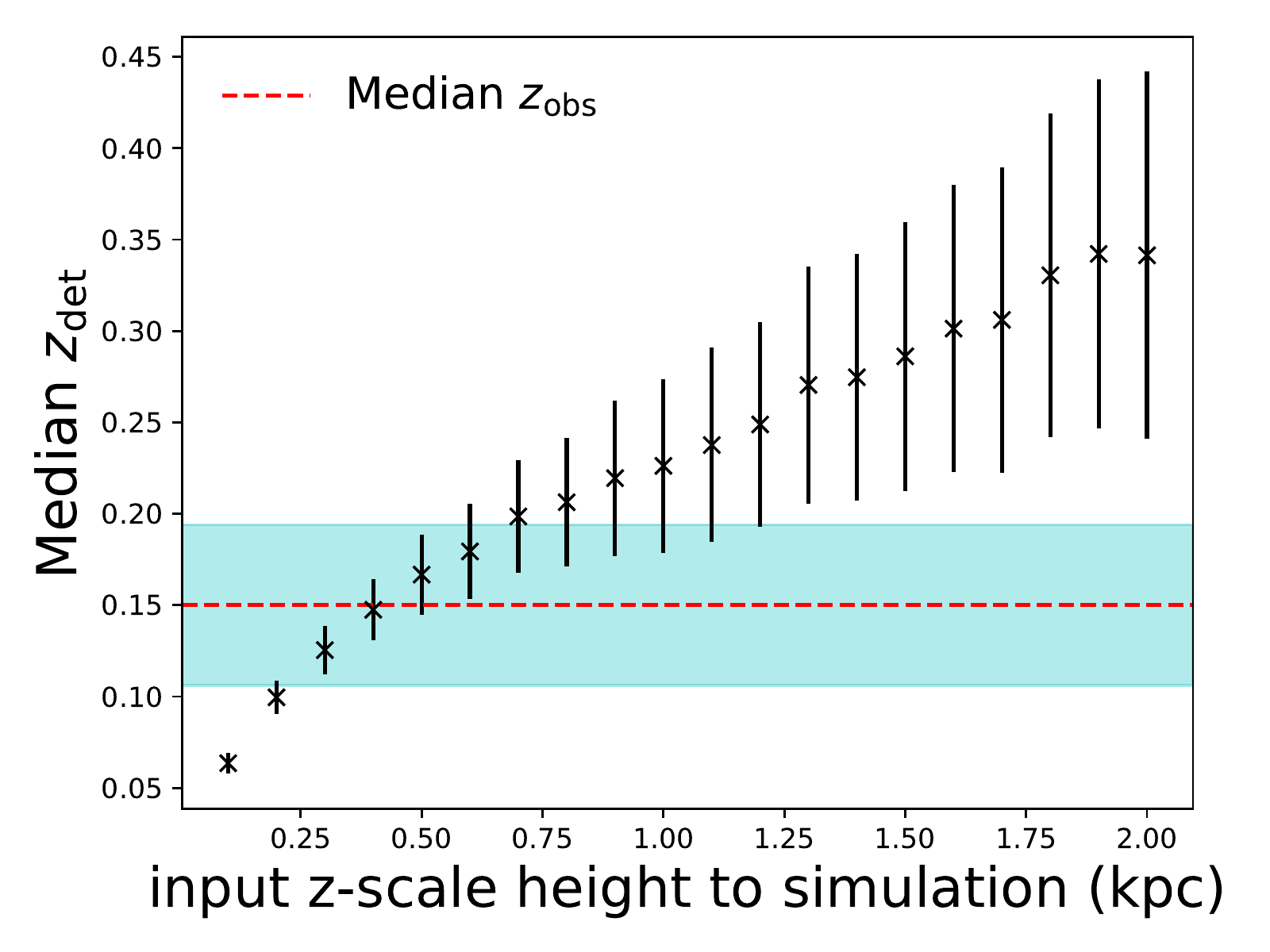}
            \caption[Median vertical scale height calculate using the NE2001 model plotted versus the input scale height $z_0$ used for the simulated population.]{Median vertical scale height calculate using the NE2001 model plotted versus the input scale height $z_0$ used for the simulated population. The median vertical scale height of the real observed population is shown as the horizontal dashed line, with the shaded cyan region depicting 1$\sigma$ error. The populations generated with vertical scale heights ranging from 0.2~kpc to 0.7~kpc agree with the observed sample.}
            \label{z_ht}
        \end{figure}
            
    \subsection{Doppler Smearing} \label{doppler_smearing}
        
        The motion of the pulsar in the orbit of the DNS system introduces a Doppler shifting of the observed pulse period. The extent of the Doppler shift depends on the velocity and acceleration of the pulsar in different parts of its orbit. This Doppler shift results in a reduction in the signal-to-noise ratio for the observation of the pulsar \citep{Bagchi_odf}.
        
        To account for this effect, we use the algorithm developed by \citet[][]{Bagchi_odf}, which quantifies the reduction in the signal-to-noise ratio as a degradation factor, $0 < \gamma < 1$, averaged over the entire orbit. This degradation factor depends on the orbital parameters of the DNS system (such as eccentricity and orbital period), the mass of the two neutron stars, the integration time for the observation, and the search technique used in the survey \citep[for example, HTRU and PALFA surveys use acceleration-searches;][]{Bagchi_odf}. A degradation factor $\gamma \sim 1$ implies very little Doppler smearing, while a degradation factor $\gamma \sim 0$ implies heavy Doppler smearing in the pulsar's radio emission.
        
        The implementation of the algorithm as a Fortran program was kindly provided to us by the authors of \citet[][]{Bagchi_odf}, which we make available\footnote{\url{https://github.com/NihanPol/SNR\_degradation\_factor\_for\_BNS\_systems}} with their permission. Since PsrPopPy does not include functionality to handle this degradation factor, we had to manually introduce the degradation factor into the source code of PsrPopPy. The modified PsrPopPy source files are also available on the GitHub repository.
            
    \subsection{Beaming correction factor} \label{beaming_fraction}
        
        The beaming correction fraction, $f_b$, is defined as the inverse of the pulsar's beaming fraction, i.e. the solid angle swept out by the pulsar's radio beam divided by $4 \pi$. PSRs B1913+16, B1534+12, and J0737--3039A/B have detailed polarimetric observation data, from which precise measurement of their beaming fractions, and thus, their beaming corrections factors has been possible. These beaming corrections are collected in Table~2 of \citet[][]{Kim_B_merger}.
        
        However, the other merging DNS systems are relatively new discoveries and do not have measured values for their beaming fractions. Thus, we assume that the beaming correction factor for these new pulsars is the average of the measured beaming correction factor for the three aforementioned pulsars, i.e. 4.6. We list these beaming fractions in Table~\ref{result_table}, and defer discussion of their effect on the merger rate for Section~\ref{discuss}.
        
    \subsection{Effective lifetime} \label{eff_life}
        
        The effective lifetime of a DNS binary, $\tau_{\rm life}$, is defined as the time interval during which the DNS system is detectable. Thus, it is the sum of the time since the formation of the DNS system and the remaining lifetime of the DNS system,
        \begin{multline}
            \displaystyle \tau_{\rm life} = \tau_{\rm age} + \tau_{\rm obs} 
             = {\rm min} \left( \tau_{\rm c}, \tau_{\rm c}\left[1 - \frac{P_{\rm birth}}{P_{\rm s}} \right]^{n - 1} \right) + {\rm min}(\tau_{\rm merger}, \tau_{\rm d}).
            \label{lifetime}
        \end{multline}
        Here $\tau_{\rm c} = P_{\rm s} / (n - 1) \dot{P_{\rm s}}$ is the characteristic age of the pulsar, $n$ is the braking index, assumed to be 3, $P_{\rm birth}$ is the period of the millisecond pulsar at birth, i.e.~when it begins to move away from the fiducial spin-up line on the $P-\dot{P}$ diagram, $P_{\rm s}$ is the current spin period of the pulsar, $\tau_{\rm merger}$ is the time for the DNS system to merge, and $\tau_{\rm d}$ is the time in which the pulsar crosses the ``death line'' beyond which pulsars should not radiate significantly \citep[][]{death_line_1}.
        
        Unlike normal pulsars, the characteristic age $\tau_{\rm c}$, for millisecond and recycled pulsars may not be a very good indicator of the true age of the pulsar. This is due to the fact that the period of the pulsar at birth is much smaller than the current period of the pulsar, which is not true for recycled millisecond pulsars found in DNS systems. A better estimate for the age of a recycled millisecond pulsar can be calculated by measuring the distance of the pulsar from a fiducial spin-up line on the $P-\dot{P}$ diagram \citep[][]{true_age}, represented by the second part of the first term in Eq.~\ref{lifetime}.
        
        Finally, the time for which a given DNS system is detectable after birth depends on whether we are observing the non-recycled companion pulsar (J0737--3039B, J1906+0746) or the recycled pulsar in the DNS system (e.g.~B1913+16, J1757--1854, J1946+2052, etc.). In the latter case, the combination of a small spin-down rate and millisecond period ensures that the DNS system remains detectable until the epoch of the merger. However, for the former case, both the period and spin-down rate are at least an order of magnitude larger than their recycled counterparts. As such, the time for which these systems are detectable depends on whether they cross the pulsar ``death line" before their epoch of merger \citep[][]{death_line_1}. The radio lifetime of any pulsar is defined as the time it takes the pulsar to cross this fiducial ``death line'' on the $P-\dot{P}$ diagram \citep{death_line_1}.
        
        We estimate the radio lifetime for J1906+0746 using two different techniques. One estimate is described by \citet[][]{death_line_1} and assumes a simple dipolar rotator to find the time to cross the deathline. Using Eq.~6 in their paper we calculate a radio lifetime of $\tau_{\rm d} \sim 3$~Myr for J1906+0746. However, as discussed in \citet[][]{death_line_1}, the death line for a pure dipolar rotator might not be an accurate turn-off point for pulsars, with many observed pulsars lying past this line on the $P-\dot{P}$ diagram. A better estimate of the radio lifetime might be given by Eq.~9 in \citet[][]{death_line_1}, which assumes a twisted field configuration for pulsars. Using this, we find $\tau_{\rm d} \sim 30$~Myr. Another estimate for the radio lifetime can be made from spin-down energy loss considerations. Adopting the formalism given in \citet[][]{death_line_2}, we find, for a simple dipolar spin-down model, that a pulsar with a current spin-down energy loss rate $\dot{E}_{\rm now}$ and characteristic age $\tau$ will reach a cut-off $\dot{E}$ value of $10^{30}$~ergs/s below which radio emission through pair production is suppressed on a timescale
        \begin{equation}
            \displaystyle \tau_{\rm d} = \tau_{\rm c} \left( \sqrt{\frac{\dot{E}_{\rm now}}{\dot{E} = 10^{30}}} - 1 \right).
            \label{death_line_1}
        \end{equation}
        Using this formalism, we calculate a radio lifetime of $\tau_{\rm d} \sim 60$~Myr. This method of estimation has been used in previous estimates \citep{comp_merger_rate_analysis, dunc_merger_1, Kim_B_merger} of the merger rates, and represents a conservative estimate on the radio lifetime of J1906+0746. We adopt it here as the fiducial radio lifetime of J1906+0746 for consistency, and defer the discussion of the implications of variation in the calculated radio lifetime to Sec.~\ref{discuss}.
        
        A similar analysis could be done for pulsar B in the J0737--3039 system. However, unlike \citet[][]{Kim_B_merger}, we do not include B in our merger rate calculations. The uncertainties in the radio lifetime are very large, as for PSR J1906+0746, and therefore pulsar A provides a much more reliable estimate of the numbers of such systems. In addition, unlike J1906+0746, pulsar B also shows large variations in its equivalent pulse width \citep{Kim_B_merger}, and thus, its duty cycle, due to pulse profile evolution through geodetic precession \citep{B_pulse_ev_perera}. This also leads to an uncertainty in its beaming correction factor \citep[see Fig.~4 in ][]{Kim_B_merger}. There are additional uncertainties introduced by pulsar B exhibiting strong flux density variations over a single orbit around A. All these factors introduce a large uncertainty in the merger rate contribution from B, and do not provide better constraints on the merger rate compared to when only pulsar A is included \citep{Kim_B_merger}. Finally, the Double Pulsar system was discovered through pulsar A and will remain detectable through pulsar A long after B crosses the death line. Due to these reasons, we do not include pulsar B in our analysis.
    
\section{Statistical Analysis and Results}      \label{stat_analysis}
    
    Our analysis is based on the procedure laid out in \citet[][]{dunc_merger_1} (hereafter KKL03). For completeness, we briefly outline the process below.
    
    We generate populations of different sizes $N_{\rm tot}$, for each of the known, merging DNS systems which are beaming towards us in physical and radio luminosity space using the observed pulse periods and pulse widths. The choice of the physical and luminosity distribution is discussed in Sec.~\ref{physical_dist}. On each population, we run the surveys described in Sec.~\ref{surveys} to determine the total number of pulsars that will be detected $N_{\rm obs}$, in those surveys. The population size $N_{\rm tot}$, that returns a detection of one pulsar, i.e. $N_{\rm obs} = 1$, will represent the true size of the population of that DNS system.
    
    For a given $N_{\rm tot}$ pulsars of some type in the Galaxy, and the corresponding $N_{\rm obs}$ pulsars that are detected, we expect the number of observed pulsars to follow a Poisson distribution:
    \begin{equation}
        \displaystyle P(N_{\rm obs}; \lambda) = \frac{\lambda^{N_{\rm obs}} e^{-\lambda}}{N_{\rm obs}!}
        \label{poisson}
    \end{equation}
    where, by definition, $\lambda \equiv \left< N_{\rm obs} \right>$. Following arguments presented in KKL03, we know that the linear relation
    \begin{equation}
        \displaystyle \lambda = \alpha N_{\rm tot}
        \label{alpha_eq}
    \end{equation}
    holds. Here $\alpha$ is a constant that depends on the properties of each of the DNS system populations and the pulsar surveys under consideration.
    
    The likelihood function, $P(D|HX)$, where $D = 1$ is the real observed sample, $H$ is our model hypothesis, i.e. $\lambda$ which is proportional to $N_{\rm tot}$, and $X$ is the population model, is defined as:
    \begin{equation}
        \displaystyle P(D|HX) = P(1|\lambda(N_{\rm tot}), X) = \lambda(N_{\rm tot}) e^{-\lambda(N_{\rm tot})}
        \label{likelihood_fn}
    \end{equation}
    Using Bayes' theorem and following the derivation given in KKL03, the posterior probability distribution, $P(\lambda| DX)$, is equal to the likelihood function. Thus,
    \begin{equation}
        \displaystyle P(\lambda| DX) \equiv P(\lambda) = P(1|\lambda(N_{\rm tot}), X) = \lambda(N_{\rm tot}) e^{-\lambda(N_{\rm tot})}.
        \label{posterior_fn}
    \end{equation}
    Using the above posterior distribution function, we can calculate the probability distribution for $N_{\rm tot}$,
    \begin{equation}
        \displaystyle P(N_{\rm tot}) = P(\lambda) \left| \frac{d\lambda}{dN_{\rm tot}} \right| = \alpha^2 N_{\rm tot} e^{-\alpha N_{\rm tot}}
        \label{ntot_prob}
    \end{equation}
    For a given total number of pulsars in the Galaxy, we can calculate the corresponding Galactic merger rate $\cal{R}$, using the beaming fraction $f_{\rm b}$, of that pulsar and its lifetime $\tau_{\rm life}$, as follows:
    \begin{equation}
        \displaystyle \mathcal{R} = \frac{N_{\rm tot}}{ \tau_{\rm life}} f_{\rm b}.
        \label{rate_eq}
    \end{equation}
    Finally, we calculate the Galactic merger rate probability distribution
    \begin{equation}
        \displaystyle P(\mathcal{R}) = P(N_{\rm tot}) \frac{dN_{\rm tot}}{d\mathcal{R}} = \left( \frac{\alpha\, \tau_{\rm life}}{f_b} \right)^2                                 \mathcal{R} \, e^{-(\alpha \tau_{\rm life} / f_b)\mathcal{R}}.
        \label{rate_pdf}
    \end{equation}
    
    \begin{figure*}[htb]
        \centering
        \includegraphics[width = \textwidth]{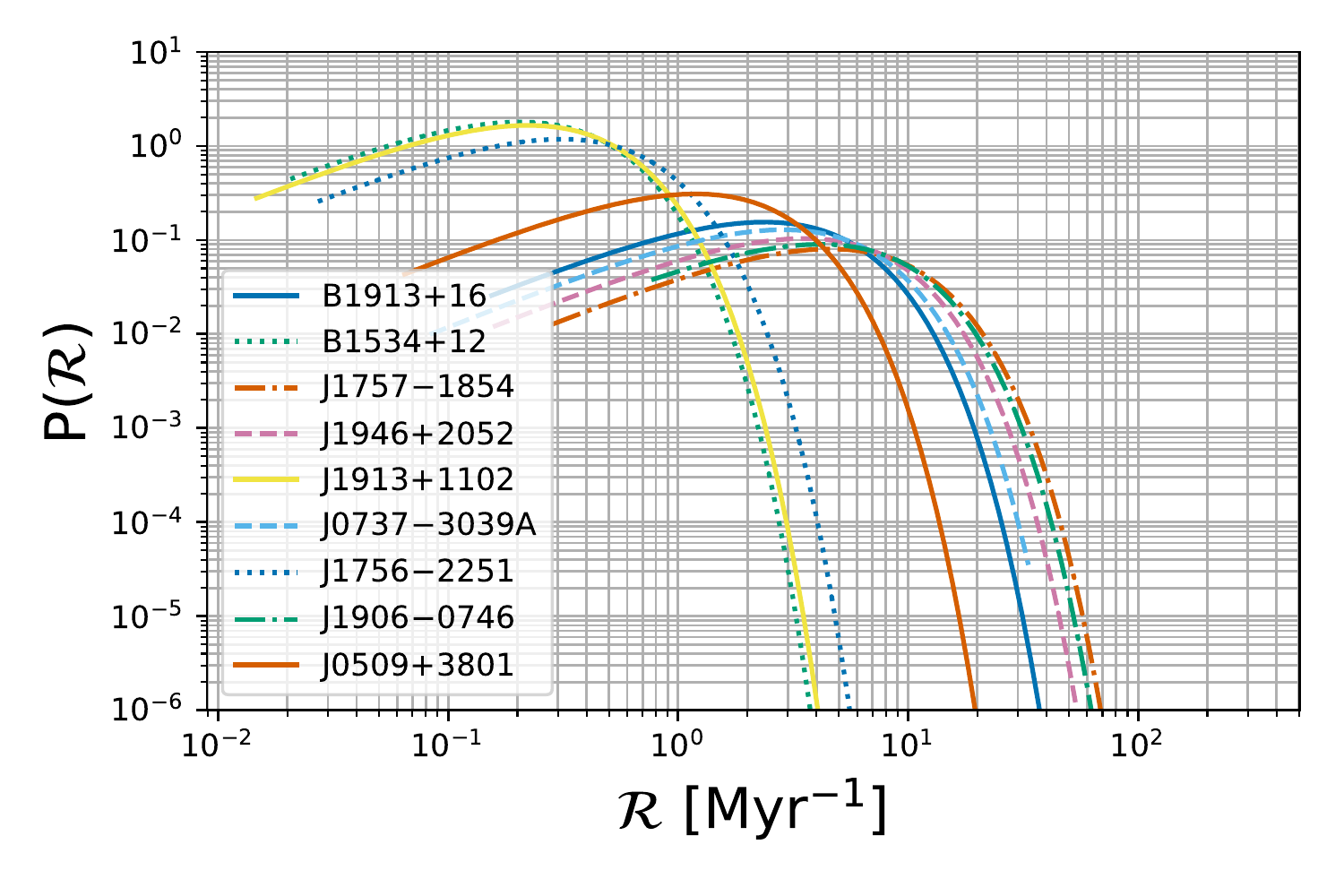}
        \caption[Probability distribution function of the Galactic merger rate of each DNS system in the Galaxy.]{Probability distribution function of the Galactic merger rate of each individual DNS system. J1757--1854, J1946+2052, J1906+0746, J0737--3039 and B1913+16 have the largest individual merger rates, followed by J0509+3801 and then B1534+12, J1756--2251 and J1913+1102.}
        \label{individual_rates}
    \end{figure*}
    
    Following the above procedure for all the merging DNS systems, we obtain the individual Galactic merger rates for each system, which are shown in Fig.~\ref{individual_rates}.
    
    \subsection{Calculating the total Galactic merger rate}
                
        After calculating individual merger rates from each DNS system, we need to combine these merger rate probability distributions to find the combined Galactic probability distribution. We can do this by treating the merger rate for the individual DNS systems as independent continuous random variables. In that case, the total merger rate for the Galaxy will be the arithmetic sum of the individual merger rates
        \begin{equation}
            \displaystyle \mathcal{R}_{\rm MW} = \sum_{i = 1}^{9} \mathcal{R}_{i}
            \label{gal_rate}
        \end{equation}
        with the total Galactic merger rate probability distribution given by a convolution of the individual merger rate probability distributions,
        \begin{equation}
            \displaystyle P(\mathcal{R}_{\rm MW}) = \prod_{i = 1}^{9}  P(\mathcal{R}_{i})
            \label{total_gal_rate}
        \end{equation}
        where $\prod$ denotes convolution. As the number of known DNS systems increases over time, the method of convolution of individual merger rate PDFs is more efficient than computing an explicit analytic expression as in KKL03 and \citet[][]{Kim_B_merger}.
        
        \begin{figure}[htb]
            \centering
            \includegraphics[width = \columnwidth]{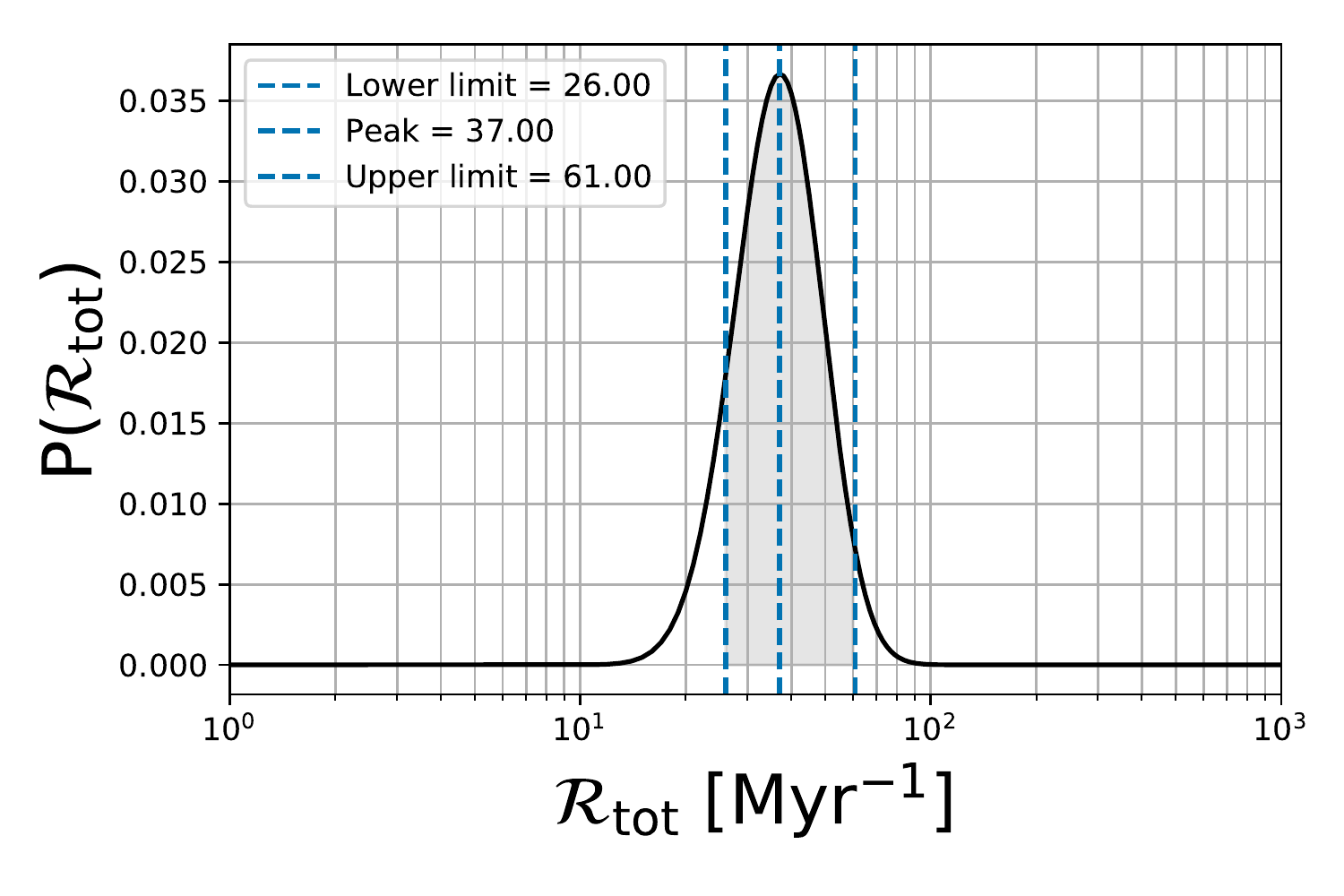}
            \caption[The total Milky Way DNS merger rate probability distribution function.]{The total Milky Way DNS merger rate probability distribution function. This distribution is obtained by convolution of the individual merger rate probability distributions as described in Section~\ref{stat_analysis}. The shaded region denotes the 90\% confidence interval, which was calculated starting from the peak of the distribution and collecting 45\% probability on both sides of the peak independently, while the vertical dashed lines represent the limits on the Milky Way DNS merger rate at the 90\% confidence interval.}
            \label{total_mw_rate}
        \end{figure}
        
        Combining all the individual Galactic merger rates, we obtain a total Galactic merger rate of $\mathcal{R}_{\rm MW} = 37^{+24}_{-11}$~Myr$^{-1}$, which is shown in Fig.~\ref{total_mw_rate}.
        
    \subsection{The merger detection rate for advanced LIGO}
        
        The Galactic merger rate calculated above can be extrapolated to calculate the number of DNS merger events that LIGO will be able to detect. Assuming that the DNS formation rate is proportional to the formation rate of massive stars, which is in turn proportional to the $B$-band luminosity of a given galaxy \citep[][]{Phinney_blue_lum_scaling, comp_merger_rate_analysis}, the DNS merger rate within a sphere of radius $D$ is given by \citep[][]{extrapolate_to_get_LIGO_rate}
        \begin{equation}
            \displaystyle \mathcal{R}_{\rm LIGO} = \mathcal{R}_{\rm MW} \left( \frac{L_{\rm total} (D)}{L_{\rm MW}} \right)
            \label{b_band_luminosity}
        \end{equation}
        where $L_{\rm total} (D)$ is the total blue luminosity within a distance $D$, and $L_{\rm MW} = 1.7 \times 10^{10} L_{B, \odot}$, where $L_{B, \odot} = 2.16 \times 10^{33}$~ergs/s, is the $B$-band luminosity of the Milky Way \citep[][]{extrapolate_to_get_LIGO_rate}.
        
        Using a reference LIGO range distance of $D_{\rm r} = 100$~Mpc \citep[][]{LIGO_horizon_dist}, and following the arguments laid out in \citet[][]{extrapolate_to_get_LIGO_rate}, we can calculate the rate of DNS merger events visible to LIGO \citep[equation 19 in ][]{extrapolate_to_get_LIGO_rate}
        \begin{multline}
            \displaystyle \mathcal{R}_{\rm LIGO} = \frac{N}{T} 
            = 7.4 \times 10^{-3} \left( \frac{\mathcal{R}}{(10^{10} L_{B, \odot})^{-1}\,{\rm Myr}^{-1}} \right) \left( \frac{D_{\rm h}}{100\,{\rm Mpc}} \right)^3 \rm yr^{-1}
            \label{extrapol}
        \end{multline}
        where $N$ is the number of mergers in $T$ years, and $\mathcal{R} = \mathcal{R_{\rm MW}} / L_{\rm MW}$ is the Milky Way merger rate weighted by the Milky Way B-band luminosity and $D_{\rm h} = 2.26 \times D_{\rm r}$ is the horizon distance.
        
        Using the above equation, we calculate the DNS merger detection rate for LIGO,
        \begin{equation}
            \displaystyle \mathcal{R}_{\rm LIGO} \equiv \mathcal{R}_{\rm PML18} = 1.9^{+1.2}_{-0.6} \times \left( \frac{D_{\rm r}}{100 \ \rm Mpc} \right)^3 \rm yr^{-1},
            \label{ligo_rate}
        \end{equation}
        where we use $\mathcal{R}_{\rm PML18}$ to distinguish our merger detection rate estimate from the others that will be referred to later in the paper.
    
\section{Discussion}        \label{discuss}
    
    \begin{table*}[]
        \centering
        \begin{tabular}{ccccccc}
            \toprule
            PSR & $f_b$ & $\delta$ & $\tau_{\rm age}$ & $N_{\rm obs}$ & $N_{\rm pop}$ & $\mathcal{R}$ \\
             & & & (Myr) & & & (Myr$^{-1}$) \\
            \midrule
            & & & & & & \\
            B1534+12 & 6.0 & 0.04 & 208 & $98^{+455}_{-64}$ & $591^{+2750}_{-386}$ & $0.2^{+0.9}_{-0.1}$ \\
            & & & & & & \\
            J1756$-$2251 & 4.6 & 0.03 & 396 & $114^{+523}_{-80}$ & $523^{+2403}_{-367}$ & $0.3^{+1.4}_{-0.2}$ \\
            & & & & & & \\
            J1913+1102 & 4.6 & 0.06 & 2625 & $150^{+691}_{-104}$ & $688^{+3171}_{-477}$ & $0.2^{+1.0}_{-0.1}$ \\
            & & & & & & \\
            J0509+3801 & 4.6 & 0.06 & 710 & $186^{+838}_{-136}$ & $853^{+3849}_{-624}$ & $1.2^{+5.4}_{-0.9}$ \\
            & & & & & & \\
            J1906+0746 & 4.6 & 0.01 & 0.11 & $54^{+248}_{-32}$ & $248^{+1136}_{-147}$ & $4.1^{+19.1}_{-2.4}$ \\
            & & & & & & \\
            B1913+16 & 5.7 & 0.169 & 77 & $154^{+700}_{-108}$ & $880^{+4000}_{-617}$ & $2.4^{+10.8}_{-1.7}$ \\
            & & & & & & \\
            J0737$-$3039A & 2.0 & 0.27 & 159 & $342^{+1565}_{-252}$ & $683^{+3131}_{-503}$ & $2.9^{+13.0}_{-2.1}$ \\
            & & & & & & \\
            J1757$-$1854 & 4.6 & 0.06 & 87 & $162^{+739}_{-116}$ & $743^{+3391}_{-532}$ & $4.6^{+21}_{-3.3}$ \\
            & & & & & & \\
            J1946+2052 & 4.6 & 0.06 & 247 & $226^{+1034}_{-164}$ & $1036^{+4748}_{-751}$ & $3.5^{+16.2}_{-1.0}$ \\
            & & & & & & \\
            \bottomrule
        \end{tabular}
        \caption[Size and merger rate of the observed DNS systems in the Galaxy.]{Parameters and results from the DNS merger rate analysis described in Section~\ref{stat_analysis}. Here $f_b$ is the beaming correction factor, $\delta$ is the pulse duty cycle, $\tau_{\rm age}$ is the effective age described in Section~\ref{lifetime}, $N_{\rm obs}$ is the number of each DNS system that are beaming towards the Earth, $N_{\rm pop}$ is the total number of each DNS system in the Milky Way, and $\mathcal{R}$ is the merger rate of each individual DNS system population, the probability distribution function for which is shown in Fig.~\ref{individual_rates}. The errors on the quantities represent 95\% confidence interval.}
        \label{result_table}
    \end{table*}
    
    In this paper, we consider nine DNS systems that merge within a Hubble time, and using the procedure described in KKL03 estimate the Galactic DNS merger rate to be $\mathcal{R}_{\rm MW} = 37^{+24}_{-11}$~Myr$^{-1}$ at 90\% confidence. This is a modest increase from the most recent rate calculated by \citet[][$\mathcal{R}_{\rm MW} = 21^{+28}_{-14}$~Myr$^{-1}$ at the 95\% confidence level]{Kim_B_merger}, despite the addition of six new DNS systems in our analysis. This is due to the addition of three large scale surveys (the PALFA, HTRU and GBNCC surveys) to our analysis, as a result of which we are sampling a significantly larger area on the sky than \citet[][]{Kim_B_merger}. This larger fraction of the sky surveyed, coupled with only a few new DNS discoveries, contributes to the overall reduction in the population of the individual DNS systems. For example, \citet[][]{Kim_B_merger} predict that there should be $\sim$907 J0737--3039A-like systems in the galaxy, while our analysis predicts a lower value of $\sim$683 such systems. This reduced population of individual DNS systems leads to a reduction in their respective contribution to the merger rate. 
    
    Irrespective of the reduction in the individual DNS system population, the six new DNS systems added in this analysis cause an overall increase in the Galactic merger rate. As shown in Fig.~\ref{individual_rates}, J1757--1854, J1946+2052 and J1906+0746 have the highest contributions to the merger rate along with J0737--3039, B1534+12 and B1913+16, while the other three DNS systems of J0509+3801, J1913+1102 and J1756--2251 round out the Galactic merger rate with relatively smaller contributions. We do not consider pulsar B from the J0737--3039 system in our analysis. The inclusion of pulsar A is sufficient to model the contribution of the Double Pulsar to the merger rates \citep[][]{Kim_B_merger} and the inclusion of B does not lead to a better constraint on the merger rate.
    
    \subsection{Comparison with the LIGO DNS merger detection rate}
        
        The recent detections of DNS mergers by LIGO \citep[][]{THE_DNS_merger} enabled a calculation of the rate of DNS mergers visible to LIGO \citep[][]{THE_DNS_merger, dns_merger_2}. The rate that was calculated using the two observed DNS mergers in \citet[][]{dns_merger_2}, converted to the units used in our calculations, is
        \begin{equation}
            \displaystyle \mathcal{R}_{\rm LIGO} \equiv \mathcal{R}_{\rm A20} = 4.6^{+7.1}_{-3.4} \times \left( \frac{D_{\rm r}}{100 \ \rm Mpc} \right)^3 \rm yr^{-1}
            \label{LIGO_rate}
        \end{equation}
        where $\mathcal{R}_{\rm A17}$ is the merger detection rate and the errors quoted are 90\% confidence intervals.
        
\begin{figure*}[htb]
            \centering
            \includegraphics[width = \textwidth]{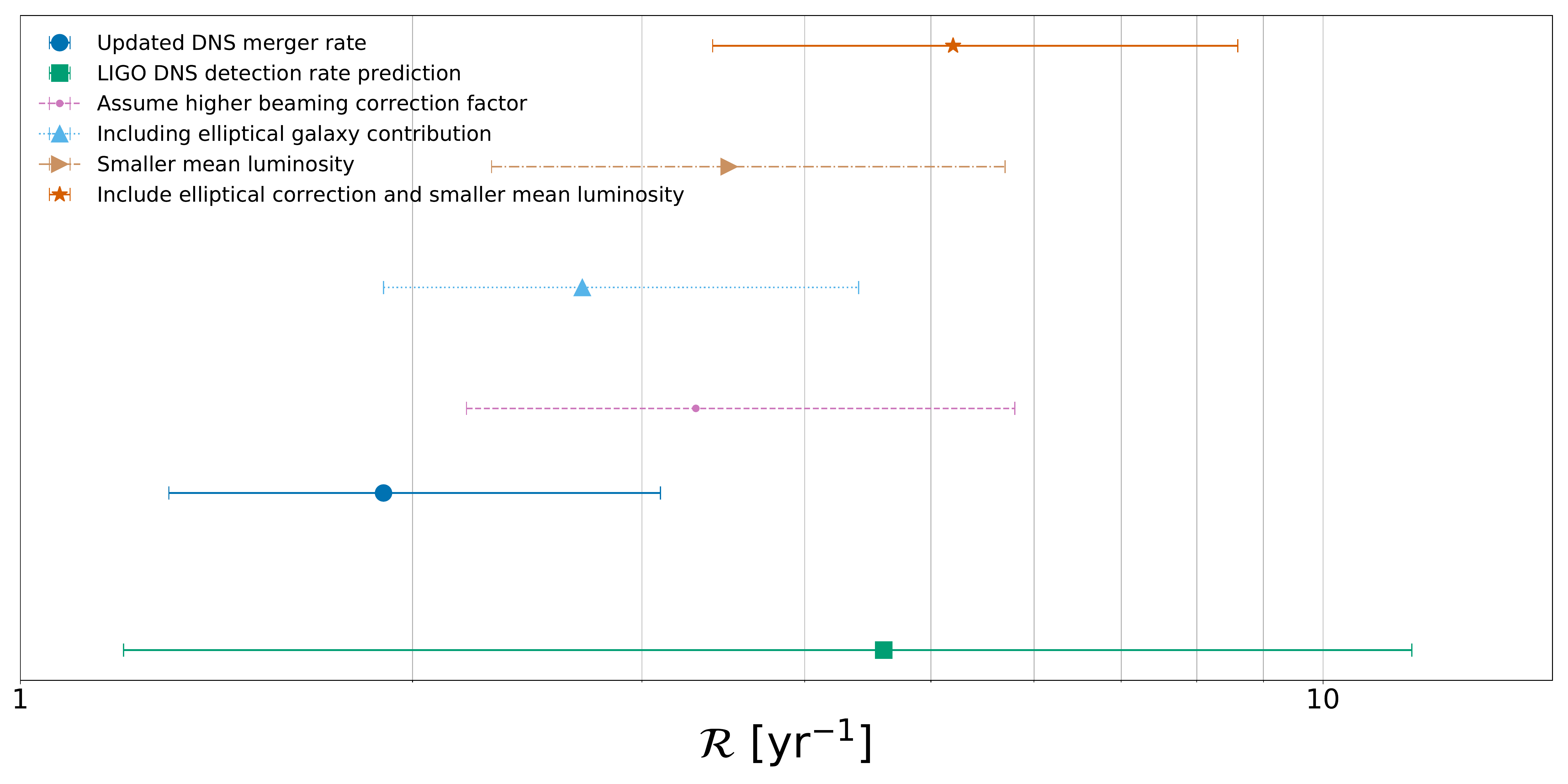}
            \caption[Comparison of the variation in the merger detection rate with change in the different underlying assumptions used in the derivation of the rate.]{We compare the variation in the merger detection rate calculated in this work with change in the different underlying assumptions used in the derivation of the rate. We show the effects of variation in the luminosity distribution of the DNS population, assuming a large beaming correction factor, and including the contribution of elliptical galaxies in LIGO's observable volume (see text for details). We also plot the modified merger detection rate that includes both the correction for elliptical galaxies and a fainter DNS population. We do not include in this the overestimated beaming correction factor effect as we do not think the beaming correction factors will differ significantly from those assumed in this work.}
            \label{var_sim}
        \end{figure*}

        We plot both the rate estimates in Figure~\ref{var_sim}. This rate estimated by LIGO is in agreement with the DNS merger detection rate that we calculate using the Milky Way DNS binary population, $\mathcal{R}_{\rm PML18}$, at the lower end of LIGO's 90\% confidence level range.
        
    \subsection{Caveats on our merger and detection rates}
        
        \subsubsection{Luminosity distribution} \label{lum_dist_eff}
            
            In generating the populations of each type of DNS system in the Galaxy, we assumed a log-normal distribution with a mean of $\left<\text{log}_{10}L\right> = -1.1$ and standard deviation, $\sigma_{\text{log}_{10} L} = 0.9$ \citep[][]{FG_kaspi_lum}. This distribution was found to adequately represent ordinary pulsars by \citet[][]{FG_kaspi_lum}. However, the DNS system population might not be well represented by this distribution. The dearth of known DNS systems prevents an accurate measurement of the mean and standard deviation of the log-normal distribution for the DNS population.
            
            The sample of DNS systems in the Galaxy might be well represented by the sample of recycled pulsars in the Galaxy. \citet[][]{Bagchi_gc_lum_func} analyzed the luminosity distribution of the recycled pulsars found in globular clusters, and concluded that both powerlaw and log-normal distributions accurately model the observed luminosity distribution, though there was a wide spread in the best-fit parameters for both distributions. They found that the luminosity distribution derived by \citet[][]{FG_kaspi_lum} is consistent with the observed luminosity distribution of recycled pulsars.
            
            We also assumed an integration time for the HTRU low-latitude survey (537~s) that is one-eighth of the integration time of the survey (4300~s) \citep[see Sec.~\ref{surveys} and][]{ng_accel_search_tech}. Based on the radiometer equation, this implies a reduction in sensitivity by a factor of $\sim 2.8$ \citep{lorimer_kramer} in searching for a given pulsar.
            
            To test the effect of the above on $\mathcal{R}_{\rm PML18}$, we used the results from \citet[][]{Bagchi_gc_lum_func} to pick a set of parameters for the log-normal distribution that represents a fainter population of DNS systems in the Galaxy. We pick a mean of $\left<\text{log}_{10}L\right> = -1.5$ (consistent with the lower flux sensitivity of the HTRU low-latitude survey) and standard deviation, $\sigma_{\text{log}_{10} L} = 0.94$ \citep[][]{Bagchi_gc_lum_func}. This increases our merger detection rate to $3.5^{+2.2}_{-1.2}$~yr$^{-1}$, which is a factor of 1.8 larger than our calculated merger detection rate. This demonstrates that if the DNS population is fainter than the ordinary pulsar population, we would see a marked increase in the merger detection rate.
            
        \subsubsection{Beaming correction factors}
            
            In our analysis, we use the average of the beaming correction factors measured for B1913+16, B1534+12, and J0737--3039A (see Table~\ref{result_table}) as the beaming correction factors for the newly added DNS systems. However, the Milky Way merger rate that we calculate is sensitive to changes in the beaming correction factors for the newly added DNS systems. To demonstrate this, we changed the beaming correction factors of all the new DNS systems to 10, i.e. slightly more than twice the values that we use. The resulting merger detection rate then increases to $3.3^{+2.5}_{-1.1}$~yr$^{-1}$, which is a 77.77\% increase from the original merger detection rate $\mathcal{R}_{\rm PML18}$. 
            
            Even though this is a significant increase in the merger detection rate, it is highly unlikely to see beaming correction factors as large as 10. The study by \citet[][]{beaming_fraction_review_1} demonstrates that pulsars with periods between 10~ms $< P <$ 100~ms are likely to have beaming correction factors $\sim 6$, with predictions not exceeding 8 in the most extreme cases \citep[see Figs.~3~and~4 in][]{beaming_fraction_review_1}. As a result, we do not expect a huge change in the merger detection rate due to variations in the beaming correction factors for the new DNS systems added in this analysis.
            
        \subsubsection{The effective lifetime of J1906+0746}
            
            PSR~J1906+0746 is an interesting DNS system which highlights the significance of the effective lifetime in the Galactic merger rate and the merger detection rate calculations. The properties of J1906+0746 suggest that it is similar to pulsar B in the Double Pulsar system. However, all searches for a companion pulsar in the J1906+0746 system have been negative \citep{1906_dns_evidence}. Just like J0737--3039B, the combination of a long period and high period derivative implies that the radio lifetime of J1906+0746 might be shorter than the coalescence timescale of the system through emission of gravitational waves. 
            
            As shown in Sec.~\ref{eff_life}, there is more than an order of magnitude variation in the estimated radio lifetime of J1906+0746. Including the gravitational wave coalescence timescale, the range of possible radio lifetimes, and hence the effective lifetimes (the characteristic age of J1906+0746 is a tender 110~kyr), for J1906+0746 ranges from $3~{\rm Myr} < \tau_{\rm eff} < 300~{\rm Myr}$. This has a significant impact on the contribution of J1906+0746 towards the merger detection rate through Eq.~\ref{rate_eq}, and thus the complete merger detection rate. For example, if $\tau_{\rm d} = 3$~Myr is an accurate estimate of the effective lifetime of J1906+0746, our merger detection rate would increase to $5.9^{+15.6}_{-2.5}$. In this scenario, J1906+0746 would contribute as much as $\sim 95$~Myr$^{-1}$ to the Galactic merger rate, compared to its contribution of $\sim 5$~Myr$^{-1}$ in the fiducial scenario. However, as pointed out earlier, it is unlikely that the effective lifetime of J1906+0746 will be as short as $\tau_{\rm d} = 3$~Myr. At the other extreme, an effective lifetime of $\tau_{\rm d} = 300$~Myr would reduce our merger detection rate to $2.2^{+1.7}_{-0.7}$. This effective lifetime is almost certainly longer than the true effective lifetime of J1906+0746 by about an order of magnitude as shown by the different calculations in Sec.~\ref{eff_life}.
            
            Thus, the effective lifetime of a DNS system is a significant source of uncertainty in the merger rate contribution of each DNS system. Fortunately, the effect of the variation in the radio lifetime is seen only in pulsars of the type of J0737--3039B and J1906+0746, i.e. the second-born, non-recycled younger constituents of the DNS systems. The recycled pulsars in DNS systems have radio lifetimes longer than the coalescence time by emission of gravitational waves. In the Double Pulsar system, since both NSs have been detected as pulsars, we can ignore pulsar B in that system. However, the companion neutron star in the J1906+0746 system has not yet been detected as a pulsar, and we have to account for the uncertainty in the radio lifetime of the detected pulsar.

        \subsubsection{Extrapolation to LIGO's observable volume}
            
            In extrapolating from the Milky Way merger rate to the merger detection rate, we assumed that the DNS merger rate is accurately traced by the massive star formation rate in galaxies, which in turn can be traced by the B-band luminosity of the galaxies. This assumption might lead to an underestimation of the contribution of elliptical and dwarf galaxies to the merger detection rate for LIGO. As an example, the lack of current star formation in elliptical galaxies implies that binaries of the J1757--1854, J1946+2052 and J0737--3039 type might have already merged. However, there might be a population of DNS systems like B1534+12 and J1756--2251 in those galaxies which are due for mergers around the current epoch. However, as we see in this analysis, systems such as B1534+12 and J1756--2251 are not large contributors to the Galactic merger rate, and should not drastically affect the merger detection rate. 
            
            The GW170817 DNS merger event was localized to an early type host galaxy \citep[][]{THE_DNS_merger_EM_assoc}, NGC 4993. \citet[][]{THE_DNS_host_gal_prop} concluded that NGC 4993 is a normal elliptical galaxy, with an SB profile consistent with a bulge-dominated galaxy. However, this galaxy shows evidence for having undergone a recent merger event \citep[][]{THE_DNS_host_gal_prop}, which might have triggered star formation in the galaxy. Thus, the GW170817 merger cannot conclusively establish the presence of a significant number of DNS mergers in elliptical galaxies. NGC 4993 is also included in the catalog published by \citet[][]{extrapolate_to_get_LIGO_rate}, with a $B$-band luminosity of $L_{B} = 1.69 \times 10^{10} L_{B, \odot}$ and contributes in the derivation of Eq.~\ref{extrapol} \citep{extrapolate_to_get_LIGO_rate}.
            
            \citet[][]{extrapolate_to_get_LIGO_rate} estimate that the correction to the merger detection rate from the inclusion of elliptical galaxies should not be more than a factor of 1.5. Folding this constant factor into our calculation, our merger detection rate for LIGO increases to $2.7^{+1.7}_{-0.8}$~yr$^{-1}$.
            
        \subsubsection{Unobserved underlying DNS population in the Milky Way}
            
            In this analysis, we assume that the population of the DNS systems that has been detected accurately represents the ``true" distribution of the DNS systems in the Milky Way. It is possible that there exists a population of DNS systems which has been impossible to detect due to a combination of small fluxes from the pulsar in the system, extreme Doppler smearing of the orbit (for relativistic systems such as J0737--3039) and extremely large beaming correction factors (i.e. very narrow beams). Addition of more DNS systems, particularly highly relativistic systems with large beaming correction factors, would lead to an increase in the Milky Way merger rate, which would consequently lead to an increase in the merger detection rate for LIGO.
            
    \subsection{Comparison with other DNS merger rate estimates}
        
        \begin{figure*}[htb]
            \centering
            \includegraphics[width = \textwidth]{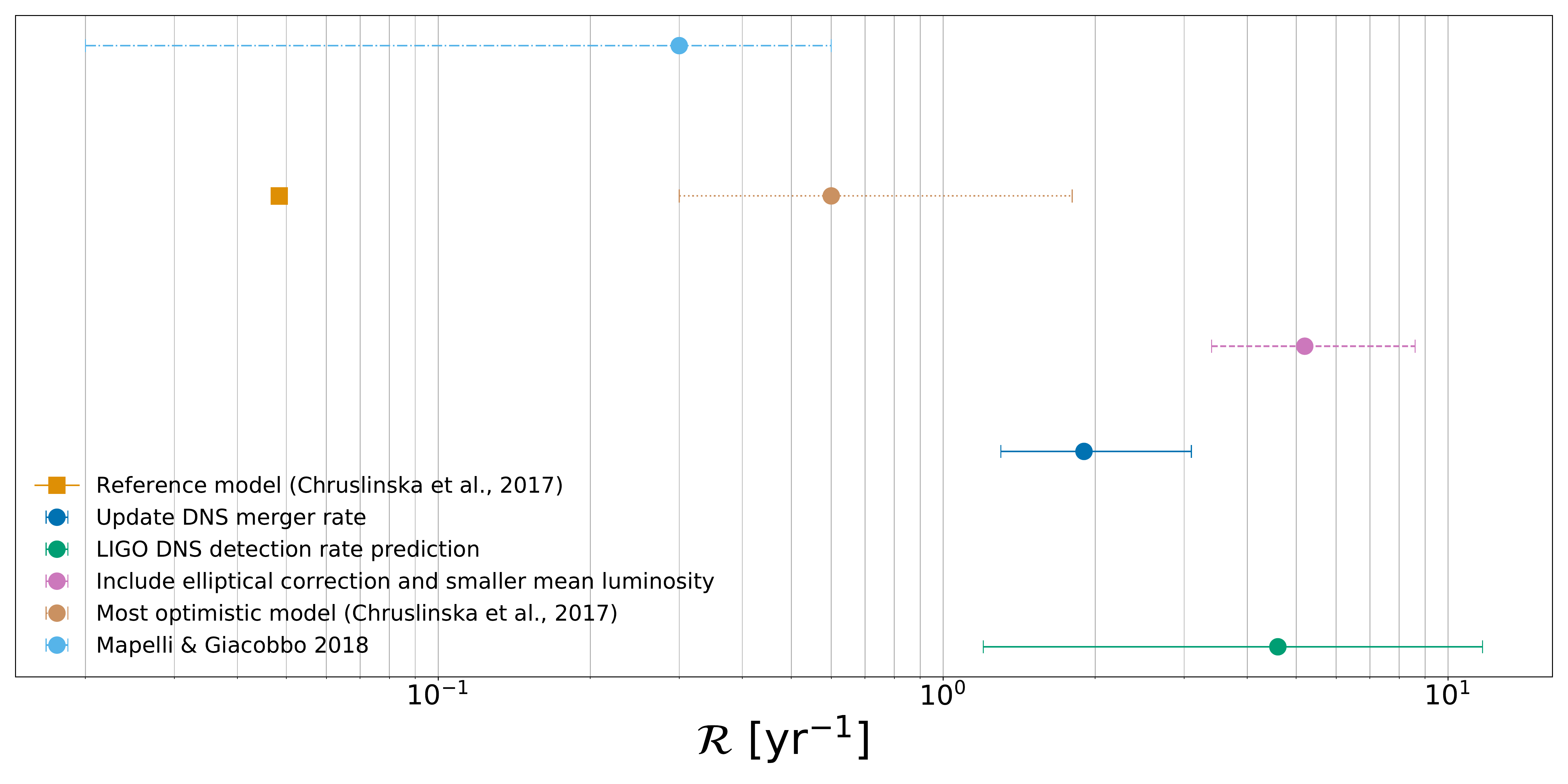}
            \caption[Comparison of the DNS merger detection rate with other contemporary predictions.]{We compare the merger detection rate derived in this work with those derived in works by \citet[][]{chruslinska_rate} and \citet[][]{mapelli_rate}. We also plot our estimate of the merger detection rate including the correction for elliptical galaxies and a lower luminosity population of DNS systems in the Galaxy. We can see that the rate derived in \citet[][]{chruslinska_rate} with their reference model is significantly lower than that predicted in this work, while their most optimistic model is consistent with our results at 90\% confidence. The merger detection rates predicted by \citet[][]{mapelli_rate} are inconsistent with those derived in this work.}
            \label{compare_rates}
        \end{figure*}
        
        We can also compare our merger detection rate to that predicted through theoretical studies and simulations of the formation and evolution of DNS binary systems. This approach to calculating the merger detection rate factors in the different evolutionary scenarios leading to the formation of the DNS system, including modeling stellar wind in progenitor massive star binaries, core-collapse and electron-capture supernovae explosions, natal kicks to the NSs and the common-envelope phase \citep[][]{abadie_dns_merger_rate, dominik_dns_merger_rate_1, dominik_dns_merger_rate_2}. We compare our merger detection rate to the predictions made using the above methodology following the DNS merger detected by LIGO \citep{THE_DNS_merger}, i.e. the studies by \citet[][]{chruslinska_rate} and \citet[][]{mapelli_rate}. We plot their estimates along with those calculated in this work in Fig.~\ref{compare_rates}.
        
        \citet[][]{chruslinska_rate}, using their reference model, calculated a merger detection rate density for LIGO of 48.4~Gpc$^{-3}$~yr$^{-1}$, which scaled to a range distance of 100~Mpc is equivalent to a merger detection rate of 0.0484~yr$^{-1}$. This value is significantly lower than our range of predicted merger detection rates. In addition to the reference model, they also calculate the merger detection rate densities for a variety of different models, with the most optimistic model predicting a merger detection rate density of $600^{+600}_{-300}$~Gpc$^{-3}$~yr$^{-1}$. Scaling this to our reference range distance of 100~Mpc, we obtain a merger detection rate of $0.6^{+0.6}_{-0.3}$~yr$^{-1}$, which is consistent with the LIGO calculated merger detection rate ($\mathcal{R}_{\rm A17} = 1.54^{+3.20}_{-1.22}$~yr$^{-1}$). However, this optimistic model assumes that Hertzsprung gap (HG) donors avoid merging with their binary companions during the common-envelope phase. Applying the same evolutionary scenario to black hole binaries (BHBs) overestimates their merger detection rate from that derived using the BHB mergers observed by LIGO \citep{chruslinska_rate}. Thus, for the optimistic model to be correct would need the common-envelope process to work differently for BHB systems as compared to DNS systems, or that BHB systems would endure a bigger natal kick in the same formation scenario than DNS systems would \citep{chruslinska_rate}.
        
        \citet[][]{mapelli_rate} showed that the above problem could be avoided and a rate consistent with the LIGO prediction of the merger detection rate ($\mathcal{R}_{\rm A17} = 1.54^{+3.20}_{-1.22}$~yr$^{-1}$) could be obtained if there is high efficiency in the energy transfer during the common-envelope phase coupled with low kicks for both electron capture and core-collapse supernovae. Based on their population synthesis, they calculate a merger detection rate density of $\sim 600$~Gpc$^{-3}$~yr$^{-1}$ \citep[for $\alpha = 5$, low $\sigma$ in Fig.~1][]{mapelli_rate}. The full range of merger detection rate densities predicted by \citet[][]{mapelli_rate} ranges from $\sim 20$~Gpc$^{-3}$~yr$^{-1}$ to $\sim 600$~Gpc$^{-3}$~yr$^{-1}$, which at a range distance of 100~Mpc corresponds to a merger detection rate ranging from $0.02$~yr$^{-1}$ to $0.6$~yr$^{-1}$. This merger detection rate is inconsistent with that derived in this work, and thus the hypotheses of high energy transfer efficiency in the common-envelope phase and low natal kicks in DNS systems made by \citet[][]{mapelli_rate} is not sufficient by itself to produce the higher merger detection rate derived in this work and by LIGO.
        
    \subsection{Future prospects}
        
        In the short term, the difference between our merger rate and that calculated by LIGO can be clarified from the results of the third operating run (O3), which is scheduled to run sometime in early 2019. Based on the fiducial model in our analysis and the predicted range distance of 120--170~Mpc for O3 \citep{LIGO_horizon_dist}, we predict, accounting for 90\% confidence intervals, that LIGO--Virgo will detect anywhere between two and fifteen DNS mergers. Further detections or non-detections by LIGO will be able to shed light on the detection rate within LIGO's observable volume. In addition, the localization of these mergers to their host galaxies as demonstrated by GW170817 \citep[][]{THE_DNS_merger_EM_assoc} will determine the contribution of galaxies lacking in blue luminosity (such as ellipticals) to the total merger rate.
        
        In the long term, with the advent of new large scale telescope facilities such as the Square Kilometer Array \citep{SKA}, we should be able to survey our Galaxy with a much higher sensitivity. Such deep surveys might reveal more of the DNS population in our Galaxy, which would yield a better constraint on the Galactic merger rate.
        
        In addition to future radio surveys, a large number of LIGO detections of DNS mergers will allow us to probe the underlying DNS population directly. Assuming no large deviations from the DNS population parameters adopted in this study (see Sec.~\ref{survey_sims} and Sec.~\ref{discuss}), a significantly larger number of DNS merger detections by LIGO would imply a larger underlying DNS population. The localization of the DNS mergers to their host galaxies will allow us to test the variation in the DNS population with respect to host galaxy morphology. We might also be able to test if the DNS population in different galaxies is similar to the DNS population in the Milky Way. This will clarify the effect of the host galaxy morphology on the evolutionary scenario of DNS systems.

\newpage
\renewcommand{\thechapter}{3}

\chapter[Estimating the Galactic Population of Ultra-compact Binary Neutron Star systems and Optimizing the Chances for their Detection]{Fantastic binary neutron star systems and whether we can find them .I. Ultra-compact binary systems}
\label{chap:lisa_bns_systems}
\blfootnote{Submitted to ApJ. \\
\textbf{Contributing authors:} Nihan Pol, Maura McLaughlin, Duncan Lorimer, Nathan Garver-Daniels
}

\newcommand{\eg}{e.\,g.\ }
\newcommand{\pb}{$P_{\rm b}$}
\newcommand{\ma}{$m_{1}$}
\newcommand{\mb}{$m_{2}$}
\newcommand{\ps}{$P_{\rm s}$}
\newcommand{\ecc}{$e$}
\newcommand{\om}{$\omega_{\rm p}$}
\newcommand{\m}{$m$}
\newcommand{\tobs}{$t_{\rm obs}$}
\newcommand{\inc}{$i$}
\newcommand{\gam}{$\gamma_1$}
\newcommand{\gama}{$\gamma_2$}
\newcommand{\gamj}{$\gamma_3$}
\newcommand{\psrpoppy}{{\sc psrpoppy}}

\section{Abstract}
    Using neural networks, we integrate the ability to account for Doppler smearing due to a pulsar's orbital motion with the pulsar population synthesis package \psrpoppy\ to allow, for the first time, accurate modeling of the observed binary pulsar population. 
    As the first application, we show that binary neutron star systems which are asymmetric in mass are, on average, easier to detect than systems which are symmetric in mass. 
    We then investigate the population of ultra-compact ($1.5 \, {\rm min} \leq P_{\rm b} \leq 15\,\rm min$) neutron star--white dwarf (NS--WD) and double neutron star (DNS) systems which are promising sources for the Laser Interferometer Space Antenna (LISA) gravitational-wave detector. 
    Given the non-detection of these systems in radio surveys thus far, we estimate a 95\% confidence upper limit of $\sim$850 and $\sim$1100 ultra-compact NS--WD and DNS systems in the Milky Way that are beaming towards the Earth, respectively. This does not imply fewer ultra-compact NS--WD systems than DNS systems in the Galaxy, but merely that we can place better constraints on the size of the population of the former type of system. We also show that with their current setup, the radio pulsar surveys at the Arecibo radio telescope have $\sim$50\% chance of detecting at least one of these systems.
    We also show that using a survey integration time of $t_{\rm int} \sim 1$~min maximizes the signal-to-noise ratio as well as the probability of detection of these ultra-compact binary systems. 

\section{Introduction}
    
    The era of multi-messenger astronomy was ushered in with GW170817, a detection of the merger of two neutron stars using gravitational waves (GWs) by the Laser Interferometer Gravitational-wave Observatory \citep[LIGO,][]{LIGO_detector_ref}  and Virgo \citep[][]{VIRGO_detector_ref} detectors \citep{THE_DNS_merger} as well as across the electromagnetic spectrum by a range of ground and space-based telescopes \citep{THE_DNS_merger_EM_assoc}. While LIGO-Virgo has made another confirmed detection of a double neutron star (DNS) merger event \citep{gw190425} and released alerts for a few more potential DNS mergers, these are relatively rare cataclysmic events.
    On the other hand, the Laser Interferometer Space Antenna \citep[LISA,][]{LISA} is a space-based GW detector which is sensitive to compact objects in binary systems which are emitting GWs at frequencies, $f_{\rm GW}$, between $0.1 \, {\rm mHz} \lesssim f_{\rm GW} \lesssim 100 \, {\rm mHz}$. Given the abundance of binaries consisting of compact objects as well as their non-cataclysmic nature, these systems provide rich potential for long-term multi-messenger science. 
    
    The strongest sources for LISA are ultra-compact binary (UCB) systems, which are binary systems with stellar-mass components and orbital periods $P_{\rm b} < 15$~min. These UCB systems can consist of any combination of white dwarf, neutron star or black holes, with the most common source ($\sim 10^7$ in the Galaxy) being double white dwarf (DWD) binaries \citep{dwd_pop_synth_2, dwd_pop_synth_1}. However, population synthesis simulations have shown that LISA should also detect a few tens of ultra-compact double neutron star (DNS) and neutron star--white dwarf (NS--WD) systems \citep{lisa_dns_andrews, lau_detecting_dns_w_lisa}. UCB systems are ``verification binaries" for LISA, i.e. these systems should be detectable within weeks of LISA beginning operations. Verification binaries for DWD systems have already been identified in the electromagnetic (EM) band using optical surveys \citep{lisa_dwd_1, lisa_dwd_2, lisa_dwd_3}. However, no verification binary consisting of a neutron star has been detected yet.
    
    Joint, multi-messenger observations of these UCB systems can provide significantly more information than observations in the EM or GW bands alone. As shown by \citet{shah_lisa_mm_inc}, measuring the inclination of an UCB system through EM observations can improve the constraint on the GW amplitude of that system by a factor as large as six. In addition, knowing the sky position of an UCB can improve the GW parameter estimation by a factor of two \citep{shah_lisa_mm_skypos}. Additionally, for DNS systems, joint EM and GW observations can constrain the mass-radius relation to within $\approx$0.2\% \citep{thrane_lisa_dns}. Thus, it is important to find as many UCB systems as possible before LISA is launched in the 2030s to maximize the scientific potential of the mission.
    
    In the electromagnetic band, neutron star binaries are discovered by searching for pulsars, which are rapidly rotating neutron stars emitting beamed emission at radio wavelengths. So far, 185 pulsars have been discovered in binary systems with a white dwarf companion, while 20 pulsars have been discovered in binary systems with another neutron star \citep[ATNF pulsar catalog\footnote{https://www.atnf.csiro.au/research/pulsar/psrcat/},][]{psrcat}. The shortest orbital period for a pulsar--WD binary is $\sim$2~hours \citep[J1518+0204C,][]{shortest_pb_wd_1, shortest_pb_wd_2}, while for pulsar--NS systems the shortest orbital period is $\sim$1.8~hours \citep[J1946+2052,][]{1946_disc}. Henceforth in this dissertation, we assume that the binary system contains a pulsar whenever we refer to ultra-compact NS--WD or DNS systems.
    
    The limiting factor in detecting UCB systems in these radio-wavelength surveys is the Doppler smearing of the pulsar emission due to its orbital motion \citep{og_odf} and it causes a reduction in the signal-to-noise ratio (S/N) with which the pulsar is detected. This Doppler smearing is quantified using the orbital degradation factor \citep{og_odf}, which can take values between 0 and 1, and lower values of the orbital degradation factor signify higher Doppler smearing and thus a larger reduction in S/N for the pulsar. The orbital degradation factor depends on, among other things, the orbital period of the binary system and is smaller for systems with small orbital periods. Thus, UCB systems, with their extremely small orbital periods, are difficult to detect in normal radio pulsar surveys.
    
    To improve sensitivity to pulsars in binary systems, acceleration and jerk search techniques are employed in radio pulsar surveys \citep[see][for a review of implementation techniques]{lorimer_kramer}. Acceleration searches have now been widely implemented in the search pipelines for almost all large radio pulsar surveys \citep[for example,][]{accel_search_implementation_1}, while jerk searches are only recently being implemented \citep{jerk_search_implementation} due to the technique being significantly more computationally expensive than acceleration searches. The effect of the acceleration search technique on the S/N of the pulsar was quantified in \citet{og_odf} for circular binaries, while \citet{Bagchi_odf} expanded their work to include eccentric systems as well as the effect of jerk search techniques.
    
    While these techniques have been well known in the literature, they have never been fully incorporated into pulsar population synthesis simulations. While \citet{Bagchi_odf} did provide software to compute the orbital degradation factors, the calculations are time-intensive and thus not optimized for inclusion in large scale population synthesis analysis such as \psrpoppy\ \citep{psrpoppy}.
    As a result, there has not been any significant modeling of the observed binary pulsar populations.
    
    In this work, we develop a computationally efficient framework to calculate the orbital degradation factor using the software provided by \citet{Bagchi_odf} and integrate the orbital degradation factor into \psrpoppy\ (Sec.~\ref{sec_integration_w_poppy}), a pulsar population synthesis package designed to model the observed pulsar population discovered in multiple radio surveys at different radio frequencies \citep{psrpoppy}. We use this to place upper limits on the population of ultra-compact NS--WD and DNS systems in the Milky Way given that we have not yet detected any such system (Sec.~\ref{sec_ucb_pop}) and calculate the probability for any of the current large pulsar surveys to detect these UCB systems. Finally in  Sec.~\ref{sec_optimum_tint}, we calculate a range of optimum integration times that will maximize the S/N for UCB systems, thereby increasing the probability of detection of these systems in radio surveys.
    
\section{Integrating orbital degradation factor into {\sc psrpoppy}} \label{sec_integration_w_poppy}
    
    We use the framework developed in \citet{Bagchi_odf} to calculate the orbital degradation factor for a binary system. The orbital degradation factor $\gamma$ can take values between $0 \leq \gamma \leq 1$ and when calculated at the harmonic \m depends on the mass of the pulsar \ma and the companion \mb, the orbital period \pb, eccentricity \ecc, inclination \inc, and angle of periastron passage \om, as well as the spin period of the pulsar \ps and the integration time of the survey \tobs\ \citep{og_odf}. The orbital degradation factor can be calculated for the case of a normal pulsar search (\gam) as well as pulsar searches which apply acceleration (\gama) and jerk (\gamj) search techniques. The radiometer signal-to-noise ratio (S/N) for the pulsar in the binary system is reduced by a factor of $\gamma_{\rm i}^2$, where $i = 1,2,3$ depending on the type of search technique. A lower orbital degradation factor implies a lower recovered S/N for the pulsar in the binary system due to Doppler smearing of the pulsar signal from the pulsar's orbital motion. We assume all modern pulsar surveys use acceleration search techniques and present results based on \gama\ in this work.
    
    The software to calculate the orbital degradation factor that was provided by \citet{Bagchi_odf} is computationally inefficient for use in large-scale population synthesis simulations. To solve this problem, we used this software as a data generator to train a simple neural network to calculate the orbital degradation factor for a given binary system.
    
    \subsection{Data standardization} 
        The neural network takes the parameters described above as an input to calculate the orbital degradation factor. The ranges of training values for each of the input parameters are shown in Table~\ref{param_range}. Since some of the input parameters can span multiple orders of magnitude, it is necessary to normalize the data to ease the training of the neural network. Thus, we first take the logarithm of the parameters \tobs, \ma, \mb, \ps, and \pb\ so that they have a dynamic range similar to the other input parameters. Next we normalize all of the input parameters such that they fall in the range between $\pm 1$. 
        
        \begin{table}[]
            \centering
            \caption[Range of values over which input parameters of the orbital degradation factor neural network were trained.]{Range of values of the input parameters for which the neural network presented in this work is trained.}
            \begin{tabular}{cccc}
                \toprule
                Name of parameter & units & Minimum & Maximum  \\
                \midrule
                Harmonic \m & -- & 1 & 5 \\
                Survey integration time \tobs\ & seconds & 1 & $5 \times 10^3$\\
                Mass of pulsar \ma\ & $M_{\odot}$ & 1 & 2.4 \\
                Mass of companion \mb\ & $M_{\odot}$ & 0.2 & $10^9$ \\
                Spin period of pulsar \ps\ & seconds & $10^{-3}$ & 5 \\
                Inclination of binary system \inc\ & degrees & $0^{\circ}$ & $90^{\circ}$ \\
                Angle of periastron passage \om\ & degrees & $0^{\circ}$ & $360^{\circ}$ \\
                Eccentricity \ecc & -- & 0 & 0.9 \\
                Orbital period \pb\ & days & $10^{-3}$ & $10^{3}$ \\
                \bottomrule
            \end{tabular}
            \label{param_range}
        \end{table}
        
    \subsection{Network architecture}
        We use {\sc keras} \citep{keras_paper} with the {\sc TensorFlow} \citep{tf_paper} backend to develop our neural network model. 
        The neural network consists of five layers, with the input layer having 9 nodes (equal to number of inputs), three ``hidden" layers containing 32 nodes, and the final, output layer consisting of a single node. We use the ``swish" activation function \citep{swish_paper} for the hidden layers, while the output layer uses a linear activation function. Since we assume that all surveys use the acceleration search technique, we describe the training and performance of the neural network that models the acceleration search technique below (i.e. \gama). However, the results are similar for the neural networks modeling the other search technique.
        
    \subsection{Training the neural network}
        We generate $\sim 7 \times 10^4$ combinations of the parameters described in Table~\ref{param_range} and calculate the corresponding \gama\ values using the software provided with \citet{Bagchi_odf}. 
        We take care to ensure that the training dataset spans the entire range of parameters described in Table~\ref{param_range}. We extract 5\% of this dataset for use as a test dataset with which we can quantify the accuracy of the trained neural network. The remaining dataset has another 5\% of the data reserved to be used as the validation dataset.
        
        During training, the neural network uses the input parameters to predict the orbital degradation factor (referred to as the prediction) which is then compared to the orbital degradation factor calculated using the analytical calculation (referred to as the label) in \citet{Bagchi_odf}. We use the mean absolute percentage error MAPE,
        \begin{equation}
            \displaystyle {\rm MAPE} = \left< 100 \times \frac{\left| \rm prediction - label \right|}{\rm label} \right>
        \end{equation}
        as the loss function for our neural network. We use the Adaptive Moment \citep[``adam",][]{adam_paper} technique to optimize the learning for the neural networks and we stop training the neural network once the MAPE has stopped improving for the validation dataset. 
        
        \begin{figure}
            \centering
            \includegraphics[width = \columnwidth]{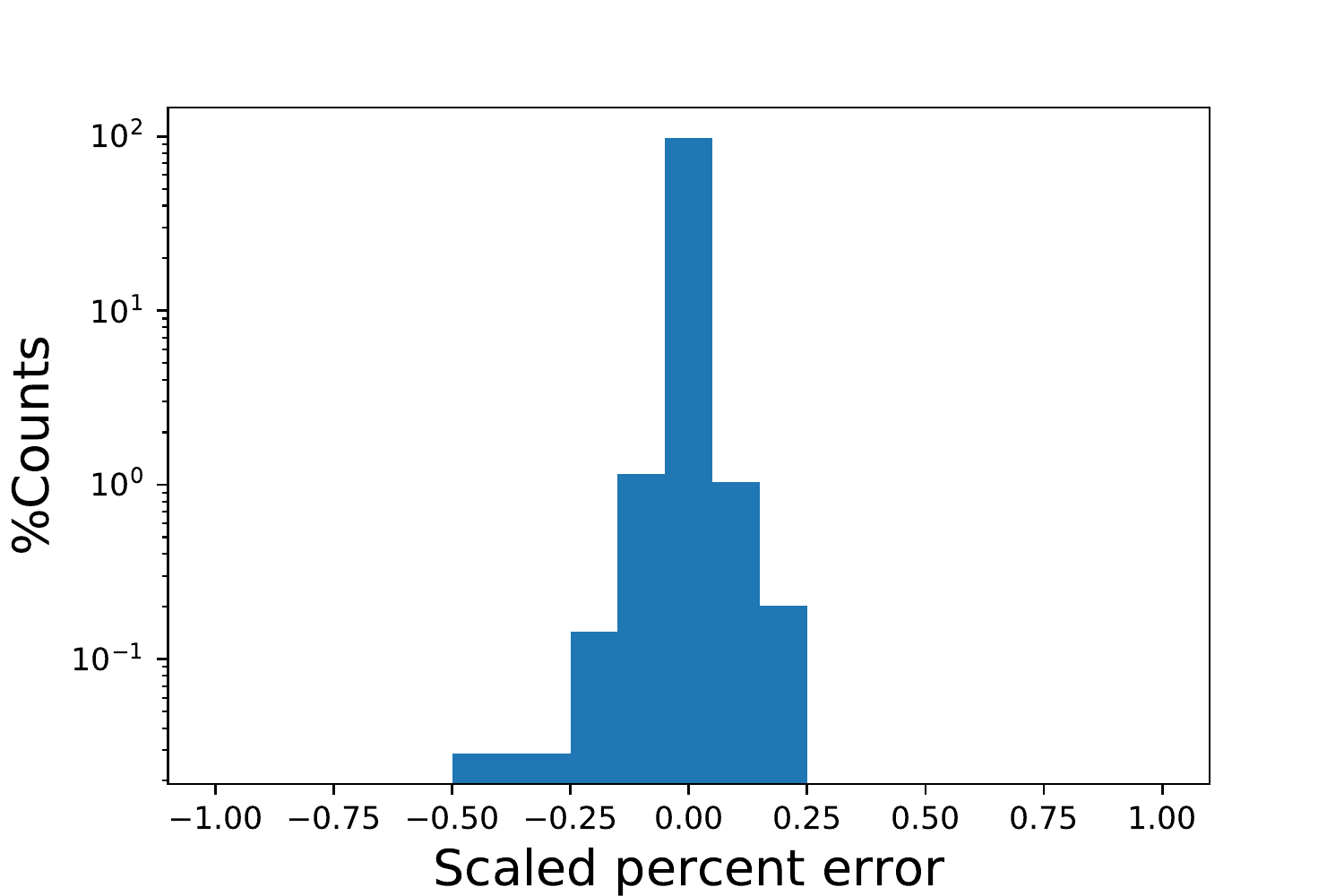}
            \caption[Distribution of the scaled percent error for the \gama\  neural network evaluated on the training dataset.]{The distribution of the scaled percent error (Eq.~\ref{spe}) for the \gama\ neural network evaluated on the training dataset. 97.44\% of the predictions made by the neural network have an error $\leq$5\%.}
            \label{accuracy}
        \end{figure}
        
        The accuracy of this trained neural network can be calculated by evaluating its performance on the test dataset. We quantify the accuracy of the neural network through the distribution of the scaled percent error SPE,
        \begin{equation}
            \displaystyle \rm SPE = 100 \times \frac{\rm prediction - label}{label},
            \label{spe}
        \end{equation}
        which is shown in Fig.~\ref{accuracy}. 
        As we can see, 97.44\% of the predictions made by the neural network have an error of $\leq$5\% compared to the values predicted using the analytic solution from \citet{Bagchi_odf}, while there are almost no values with an error $\gtrsim$25\%. This accuracy is sufficient for using the neural network model in large-scale population simulations. 
        
        The orbital degradation factor computation using the neural network framework is faster by a factor of $10^4$ compared to the same computation using the software provided by \citet{Bagchi_odf} demonstrating the suitability of the former for large-scale population synthesis simulations. In addition, the inherent parallelism of the neural network framework allows it to compute multiple orbital degradation factors in a single pass while the software provided with \citet{Bagchi_odf} was limited to a single computation. This provides an additional significant improvement in the computational efficiency of the neural network framework.
        
\begin{figure*}
            \centering
            \begin{subfigure}[b]{0.49\textwidth}
            \centering
            \includegraphics[width = \textwidth]{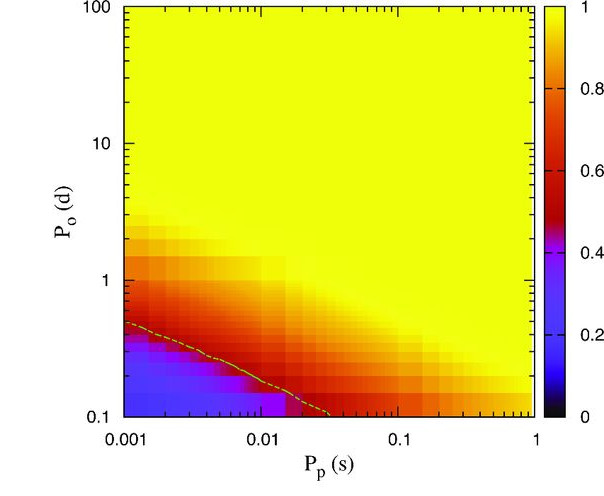}
            \caption{}
            \end{subfigure}
            \begin{subfigure}[b]{0.49\textwidth}
            \centering
            \includegraphics[width = \textwidth]{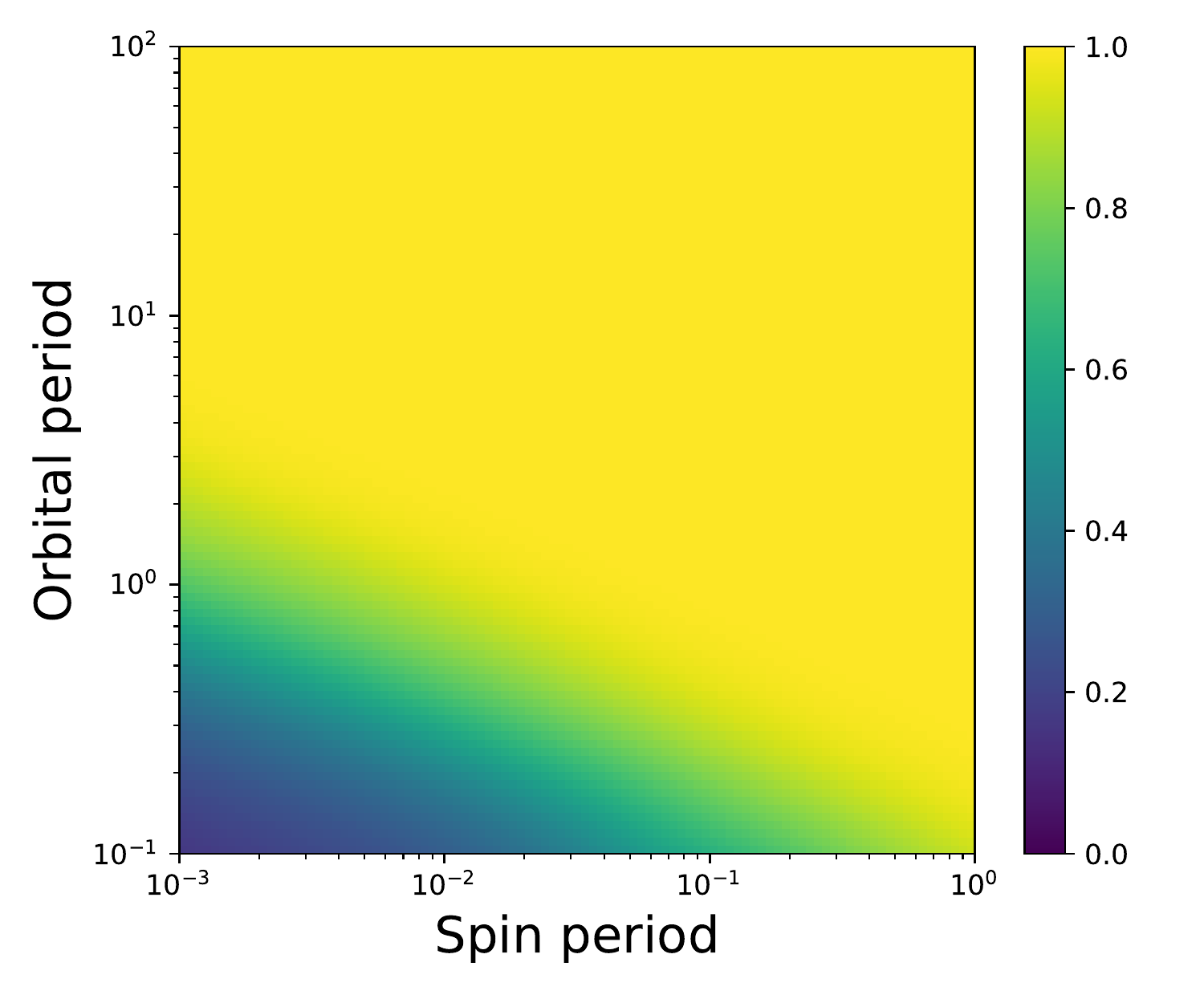}
            \caption{}
            \end{subfigure}
            \caption[Comparison of calculation of orbital degradation factor between the neural network implementation and the original implementation from \citet{Bagchi_odf}.]{Comparison of the results from \citet{Bagchi_odf} (panel (a)) to the \gama\ neural network trained in this work (panel (b)). In both plots, the horizontal axis shows the spin period of the pulsar, the vertical axis shows the orbital period of the BNS system and the color represents the orbital degradation factor. The two methods produce identical results for the orbital degradation factor \gama.}
            \label{fig_9b_comp}
        \end{figure*}

        We can directly compare the results produced by the trained neural network to those published in \citet{Bagchi_odf} by reproducing figures from that work. As an example, in Fig.~\ref{fig_9b_comp}, we compare the results presented in Figure 9(b) of \citet{Bagchi_odf} with those produced by our neural network. The results produced by the two methods are identical, which is another confirmation of the accuracy of our neural network.
        
    \subsection{Integration with \psrpoppy} \label{sec_integration}
        The orbital degradation factor calculated with the neural network can be directly integrated into \psrpoppy\ for modeling the different types of binary pulsar populations. 
        We add the ability for \psrpoppy\ to generate orbital parameters for a synthetic pulsar\footnote{\url{https://github.com/NihanPol/PsrPopPy2}} which are used to compute the orbital degradation factor.
        \psrpoppy\ calculates the S/N for a pulsar using the radiometer equation \citep{lorimer_kramer}, which can be directly scaled by $\gamma^2$ \citep{Bagchi_odf} to get the S/N for the same pulsar if it were in a binary. 
        
    \subsection{Selection bias against asymmetric mass DNS systems} \label{sec_asymm_bias}
        
        As an application of the orbital degradation factor, we investigate whether it is easier to detect a DNS system that is symmetric in mass as compared to a system asymmetric in mass.
        The question depends only on how the orbital degradation factor depends on the mass ratio of the binary system.  
        
        To investigate this, we perform a Monte Carlo simulation where we randomly draw samples from the distributions for all the input parameters to the orbital degradation factor. 
        The majority of the observed sample of NS--WD and DNS systems have spin periods less than $\sim$100~ms \citep{psrcat}.
        Similarly, the majority of the observed NS--WD systems have orbital periods less than $\sim$50~days, while the majority of the observed DNS systems have orbital periods less than $\sim$10~days \citep{psrcat}.
        To correspond to the observed sample, we restrict the spin and orbital periods for the pulsars to be in the range $1 \, {\rm ms} < P_{\rm s} < 100 \, {\rm ms}$ and $10^{-3} \, {\rm days} < P_{\rm b} < 50 \, {\rm days}$, respectively. 
        Other parameters are allowed to vary across their full range as listed in Table~\ref{param_range}.
        However, in place of using the companion mass directly, we instead define a new parameter, the mass ratio, $q = m_1 / m_2$. We define symmetric systems as those having $0.9 \leq q \leq 1.0$ and asymmetric systems as $0.1 \leq q < 0.9$.
        
\begin{figure}
            \centering
            \includegraphics[width = \columnwidth]{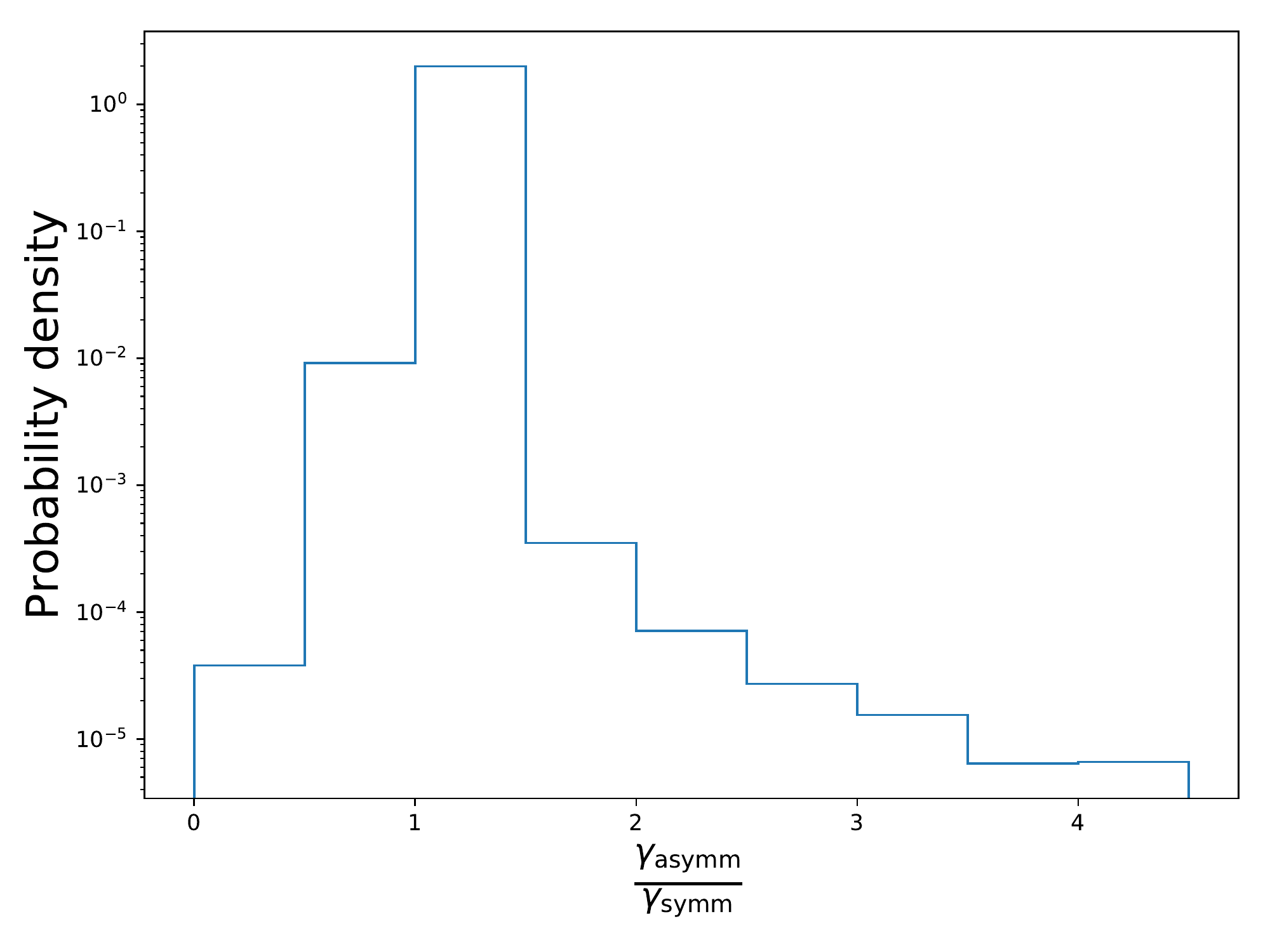}
            \caption[The ratio of the degradation factor for asymmetric mass systems to that of symmetric mass systems.]{The ratio of the degradation factor for asymmetric mass systems to that of symmetric mass systems. The mass of the pulsar and all other orbital parameters, except the mass of the companion, are the same between the symmetric and asymmetric systems. The mass of the companion is calculated using the mass ratio $q$, where $q \geq 0.9$ for symmetric mass systems and $q < 0.9$ for asymmetric mass systems. The histogram shows that it is easier to detect asymmetric mass systems than symmetric mass systems.}
            \label{asymm_pref}
        \end{figure}

        We randomly draw a value for the mass ratio for symmetric and asymmetric systems as defined above. Both of these mass ratio values are then assigned the same set of remaining input parameters required to calculate the orbital degradation factor. We then calculate and plot the distribution of the ratio of the orbital degradation factor for asymmetric systems to that for symmetric systems. The distribution obtained after $10^7$ sample draws is shown in Fig.~\ref{asymm_pref}. 
        
        We find there are fewer systems in which the ratio has a value less than one for asymmetric systems, implying that the orbital degradation factor for asymmetric DNS systems is on average greater than that for symmetric DNS systems. Consequently, asymmetric DNS systems are easier to detect in surveys than symmetric DNS systems.
        
        However, despite the preference for the detection of asymmetric mass DNS systems, only two such systems have been detected, J0453+1559 \citep[$q = 0.75$][]{most_asymm_bns_0453} and J1913+1102 \citep[$q = 0.78$,][]{1913_nature_paper}, compared to eighteen DNS systems with mass ratios $q \gtrsim 0.9$. This result suggests that this discrepancy in the number of detected asymmetric systems might not be due to selection effects, but rather due to differences in the evolutionary scenarios between the two types of systems.
        
\section{Ultra-compact binary population statistics} \label{sec_ucb_pop}
    
    \subsection{Size of population} \label{subsec_pop_num}
        
        Given the non-detection of UCB systems by current radio pulsar surveys, we can place an upper limit on the number of these systems in the Galaxy. To do so, we use the version of \psrpoppy\ \citep{psrpoppy} that is integrated with the orbital degradation factor described in Sec.~\ref{sec_integration}. We compute separately the upper limit of the population of ultra-compact NS--WD and DNS systems.
        
        We follow a procedure that is based on the framework described in \citet{kkl}. For any given type of binary system, if $N_{\rm obs}$ is the number of observed systems, we expect the probability distribution of the number of observed systems to follow a Poisson distribution:
        \begin{equation}
            \displaystyle P(N_{\rm obs}; \lambda) = \frac{\lambda^{N_{\rm obs}} e^{-\lambda}}{N_{\rm obs}!}
            \label{poisson}
        \end{equation}
        where, by definition, $\lambda = \left< N_{\rm obs} \right>$. As described in \citet{kkl}, we know that the relation
        \begin{equation}
            \displaystyle \lambda = \alpha N_{\rm tot}
            \label{hypo}
        \end{equation}
        is true, where $N_{\rm tot}$ is the number of UCB pulsars that are beaming towards Earth and $\alpha$ is a constant that depends on the properties of the UCB system and the pulsar surveys under consideration. Since no UCB systems have been detected, we can set $N_{\rm obs} = 0$, which reduces Eq.~\ref{poisson} to $P(0; \lambda) = e^{-\lambda}$.
        
        As described in \citet{kkl}, the likelihood function $P(D|HX)$, where $D = 0$ is the real observed sample, $H$ is our model hypothesis (i.e. Eq.~\ref{hypo}), and X is the population model, is defined as   
        \begin{equation}
            \displaystyle P(D|HX) = P(0|\lambda(N_{\rm tot}), X) = e^{-\lambda(N_{\rm tot})}.
            \label{likely}
        \end{equation}
        Using Bayes' theorem and the justification given in \citet{kkl}, the posterior $P(\lambda|DX)$ is equal to the likelihood function, i.e.,
        \begin{equation}
            \displaystyle P(\lambda|DX) \equiv P(\lambda) = P(0| \lambda, X) = e^{-\lambda(N_{\rm tot})}.
            \label{posterior}
        \end{equation}
        Using this posterior, we can calculate the probability distribution for $N_{\rm tot}$,
        \begin{equation}
            \displaystyle P(N_{\rm tot}) = P(\lambda) \left| \frac{d\lambda}{dN_{\rm tot}} \right| = \alpha e^{- \alpha N_{\rm tot}}.
            \label{pop_prob}
        \end{equation}
        
        With {\psrpoppy}, we generate populations of different sizes and calculate $\lambda$ for each population using Eq.~\ref{poisson}, which in combination with Eq.~\ref{hypo} gives us the value of $\alpha$. Using Eq.~\ref{pop_prob} with the value of $\alpha$ gives us the probability density for the population of the UCB systems that are beaming towards Earth.
        
        For all UCB systems, we allow the mass of the pulsar \ma, inclination of the system \inc, the angle of periastron passage \om, and the eccentricity \ecc, to have the range listed in Table~\ref{param_range}. For ultra-compact NS--WD systems, we restrict the companion mass to the range $0.2 \, M_{\odot} < m_2 < 1.4 \, M_{\odot}$, while for ultra-compact DNS systems, we restrict the companion mass to the range $1.0 \, M_{\odot} < m_2 < 2.4 \, M_{\odot}$. We also assume the pulsar is an orthogonal rotator and thus, most of the power from the pulsar emission is constrained in the second harmonic and set $m = 2$. In the case that the pulsar is not an orthogonal rotator, a choice of $m = 2$ results in a more conservative upper limit. For both ultra-compact NS--WD and DNS systems, we constrain the orbital period to the range $1.5 \ {\rm minutes} < P_{\rm b} < 15 \ {\rm minutes}$. All of these parameters have uniform distributions.
        
        In addition, for both of these types of systems, we also constrain the spin period for the pulsar to the range $1 \, {\rm ms} < P_{\rm s} < 100 \, {\rm ms}$ to correspond to the spin periods of the observed DNS systems as described in Sec.~\ref{sec_asymm_bias}. We model the pulsar luminosity distribution using a log-normal distribution with a mean $\left< {\rm log}_{10}L \right> = -1.1$ ($L = 0.07 \, {\rm mJy \, kpc^2}$) and standard deviation $\sigma_{\rm log_{10}L} = 0.9$ \citep{fk06}. Since we consider surveys at different radio frequencies, we also model the pulsar spectral index as having a normal distribution with mean $\alpha = -1.4$ and standard deviation $\beta = 1$ \citep{bates_si_dist}. We assume the radial distribution for the UCB systems as described in \citet{lorimer_rad_dist} and the two-sided exponential function for the $z$-height distribution, with a scale height of $z_0 = 0.33$~kpc.

\begin{figure*}
            \centering
            \begin{subfigure}[b]{0.49\textwidth}
                \centering
                \includegraphics[width = \textwidth]{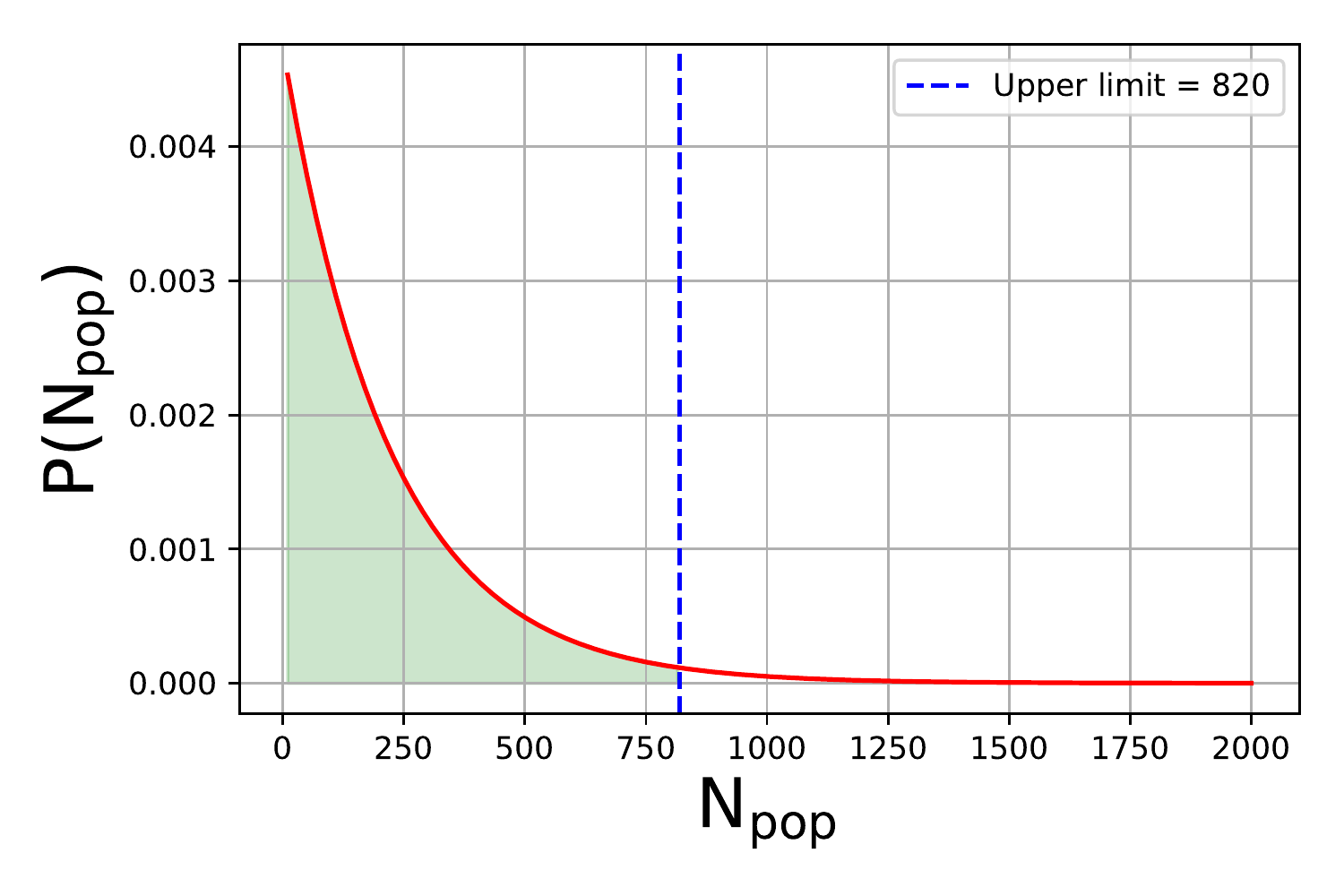}
                \caption{Ultra-compact NS--WD systems}
            \end{subfigure}
            \begin{subfigure}[b]{0.49\textwidth}
                \centering
                \includegraphics[width = \textwidth]{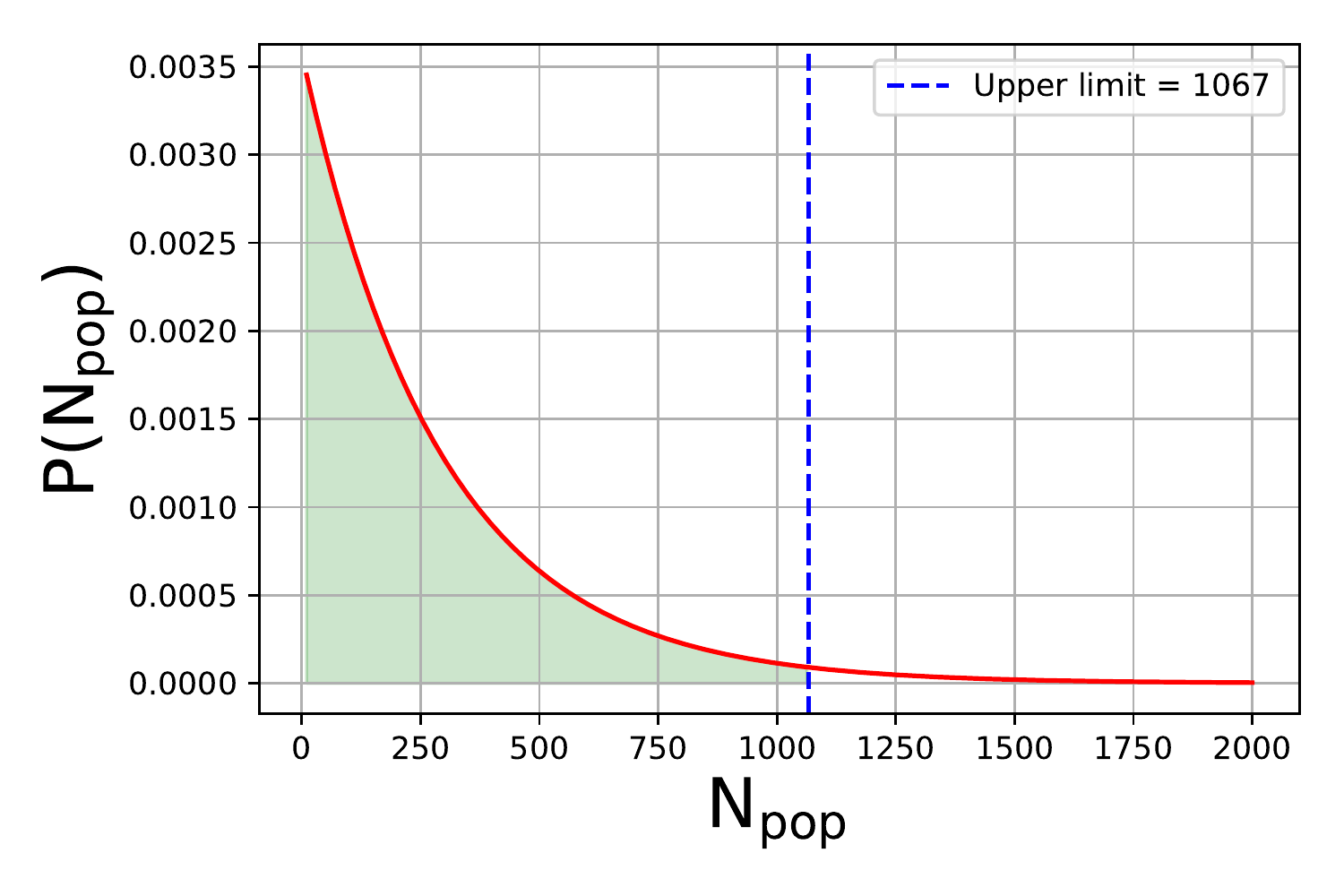}
                \caption{Ultra-compact DNS systems}
            \end{subfigure}
            \caption[Probability distribution function for the number of UCB systems that are beaming towards the Earth.]{Probability distribution function for the number of UCB systems that are beaming towards Earth. The 95\% upper limit on the number of ultra-compact NS--WD systems (panel (a)) is $\sim$850 systems, while that for ultra-compact DNS systems (panel (b)) is $\sim$1100 systems.}
            \label{pops}
        \end{figure*}

        The surveys that we consider are listed in Table~\ref{survey_table}. These are the largest radio pulsar surveys conducted to date. The survey integration times from the individual surveys are used in the calculation of the orbital degradation factor. Using these parameters, the probability distribution for the size of the population of the UCB systems that are beaming towards Earth is shown in Fig.~\ref{pops}. The 95\% upper limit on the number of ultra-compact NS--WD systems in the Galaxy is $\sim$850 systems, while that for ultra-compact DNS systems in the Galaxy is $\sim$1100 systems.
        
        \begin{landscape}
        \begin{table*}
            \centering
            \caption[Telescope and survey parameters for the largest pulsar surveys.]{The telescope and survey parameters for the large pulsar surveys that are considered in this work.}
            \begin{tabular}{ccccccc}
                \toprule
                Survey & Gain $G$ & Center Frequency $f_{\rm c}$ & Bandwidth $B$ & System temperature $T_{\rm sys}$ & Integration time $t_{\rm int}$ \\
                -- & (K/Jy) & (MHz) & (MHz) & (K) & (s) \\
                \midrule
                PALFA\footnote{Pulsar Arecibo L-band Feed Array, \citet{PALFA}} & 8.5 & 1374 & 300 & 25 & 268 \\
                PMSURV\footnote{Parkes Multibeam SURvey, \citet{PMSURV}} & 0.6 & 1374 & 288 & 25 & 2100 \\
                AODRIFT\footnote{ArecibO DRIFT scan survey, \citet{aodrift_1}} & 10 & 327 & 25 & 100 & 50 \\
                GBNCC\footnote{Green Bank North Celestial Cap Survey, \citet{gbncc}} & 2 & 350 & 100 & 46 & 120 \\
                HTRU--low\footnote{High Time Resolution Universe low-latitude survey \citet{htru_low_mid}} & 0.6 & 1352 & 340 & 25 & 340 \\
                HTRU--mid\footnote{High Time Resolution Universe mid-latitude survey \citet{htru_low_mid}} & 0.6 & 1352 & 340 & 25 & 540 \\
                \bottomrule
            \end{tabular}
            \label{survey_table}
        \end{table*}
        \end{landscape}
        
        The upper limit on the number of ultra-compact NS--WD systems is slightly smaller than that of ultra-compact DNS systems due to the fact that the orbital degradation factor for the former is on average larger than that for the latter. Consequently, ultra-compact NS--WD systems are easier to detect than ultra-compact DNS system and thus their non-detection so far places a more stringent constraint on their population as compared to ultra-compact DNS systems. Note that this does not imply that there are fewer ultra-compact NS--WD binaries than DNS binaries, but merely states that we can constrain the population of the former type of system better than that of the latter.
        
        As stated earlier, the upper limits above are the number of these UCB systems that are beaming towards Earth. The total number of such systems in the Galaxy can be calculated by scaling $N_{\rm tot}$ by the beaming correction factor $f_{\rm b}$ \citep{kkl, my_merger_rate}. Given the large uncertainty in the beaming correction factors, if we use the average beaming correction factor measured for the merging DNS systems, $f_{\rm b} = 4.6$ \citep{my_merger_rate}, the upper limit on the total number of ultra-compact NS--WD and DNS systems comes out to $\sim$4000 and $\sim$5000 systems, respectively. Since we used uninformative priors in our Monte Carlo simulations, we do not convert these numbers into an estimate of the merger rate for this type of system.
        
        The number of ultra-compact DNS systems derived here is less than the total number of merging DNS systems derived in \citet{my_merger_rate} and \citet{1913_nature_paper}. This difference can be explained by the fact that the UCB systems that we consider in this work have lifetimes $\sim$ few Myr, significantly smaller than that for the merging DNS systems studied in the aforementioned studies. As a result, these systems are closer to merger and spend a relatively short amount of time in this subclass of DNS systems compared to the larger orbital period merging DNS systems from \citet{my_merger_rate}, which results in an overall smaller population size of ultra-compact DNS systems. The upper limit on the number of ultra-compact DNS systems is also consistent with recent estimates of the size of this population made by \citet{lau_detecting_dns_w_lisa} and \citet{lisa_dns_andrews}.
        
    \subsection{Probability of pulsar surveys detecting an UCB system} \label{subsec_survey_eff}
        
        Knowing the upper limit on the number of UCB systems that are beaming towards Earth, we can calculate the probability of the radio pulsar surveys listed in Table~\ref{survey_table} in detecting these systems. To do so, we assume that the number of the UCB systems (both NS--WD and DNS) in the Galaxy is equal to their upper limits, i.e. we calculate the probability for these surveys in the most optimistic scenario.
        
\begin{figure*}
            \centering
            \begin{subfigure}[b]{0.49\textwidth}
                \centering
                \includegraphics[width = \textwidth]{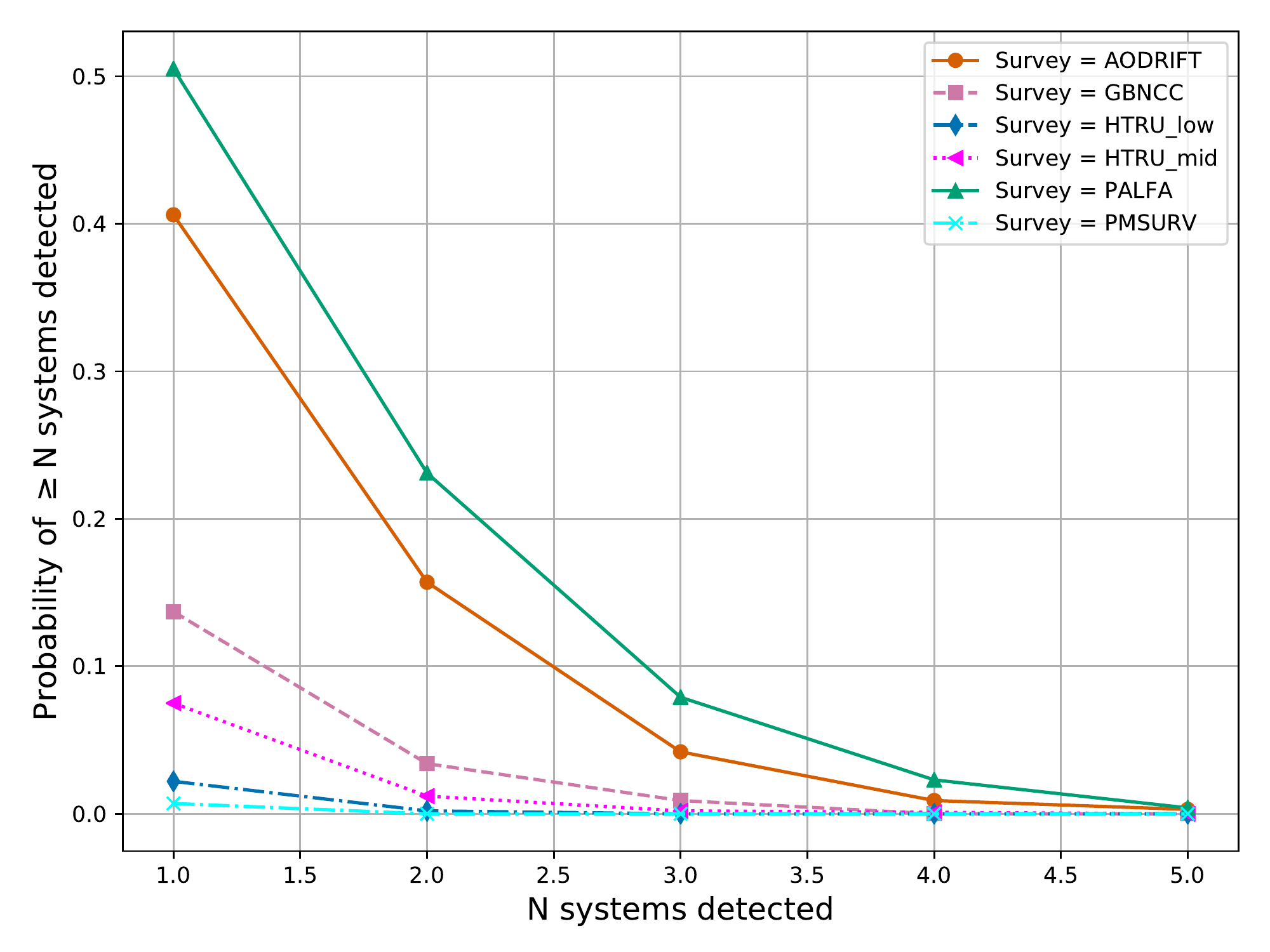}
                \caption{Ultra-compact NS--WD systems}
            \end{subfigure}
            \begin{subfigure}[b]{0.49\textwidth}
                \centering
                \includegraphics[width = \textwidth]{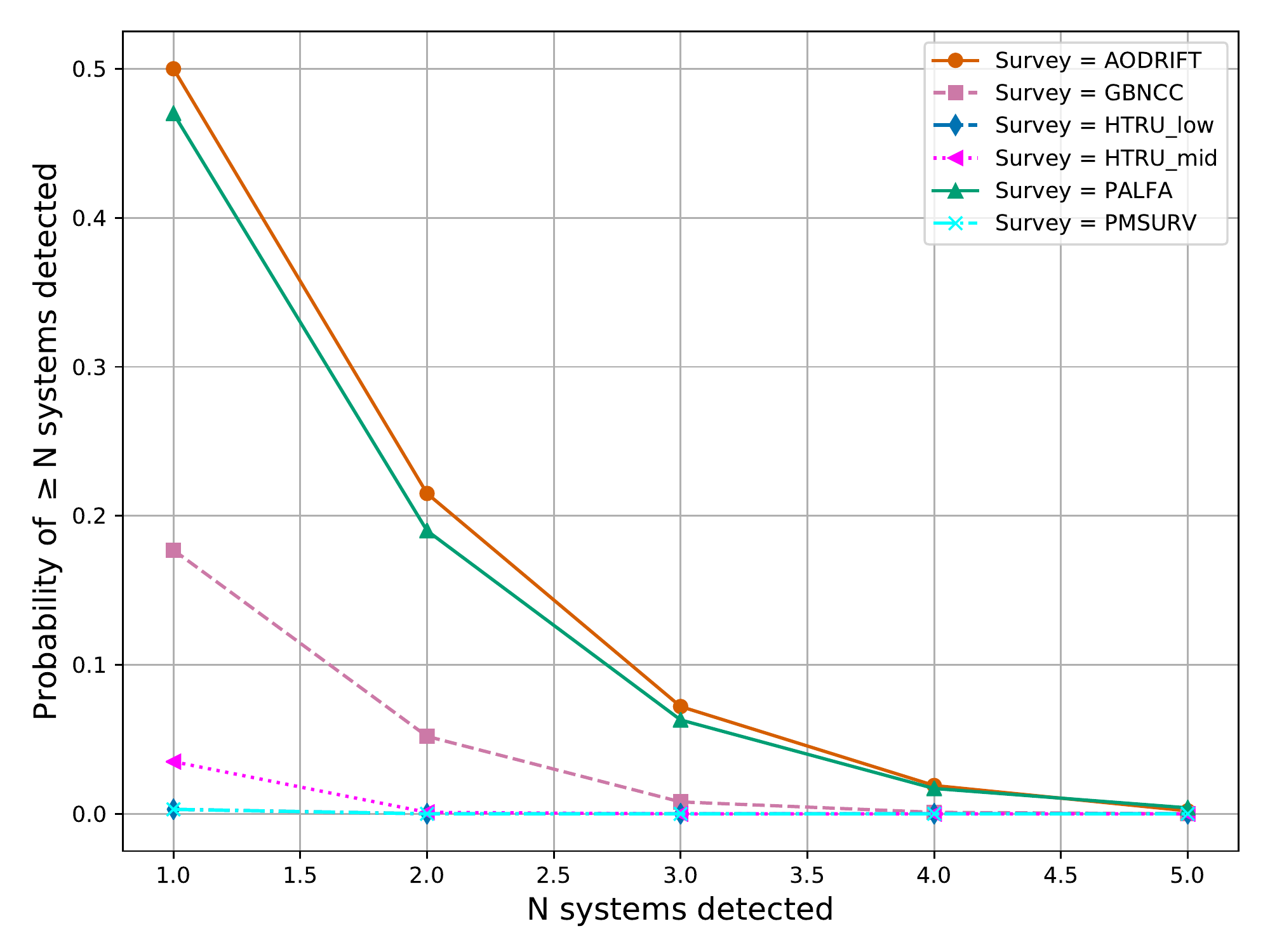}
                \caption{Ultra-compact DNS systems}
            \end{subfigure}
            \caption[Complementary cumulative distribution function (CDF) of the number of UCB systems that are detectable by the radio pulsar surveys listed in Table 3.2.]{Complementary cumulative distribution function (CDF) of the number of UCB systems that are detectable by the radio pulsar surveys listed in Table~\ref{survey_table}. The horizontal axis shows the number of detectable systems $N$, while the vertical axis shows the probability that $\geq N$ systems will be detected in the given survey. The surveys conducted with the Arecibo telescope, i.e. PALFA and AODRIFT, are most likely to detect at least one of these UCB systems.}
            \label{success_survey}
        \end{figure*}

        Next, we use \psrpoppy\ to generate this number of pulsars in the Galaxy, with the orbital, spin, luminosity, spatial and spectral index distributions being the same as described in Sec.~\ref{subsec_pop_num}. We generate $10^3$ different versions of these populations to ensure that we are efficiently sampling all of the prior distributions. Accounting for the orbital degradation factor, we then ``run" each of the surveys listed in Table~\ref{survey_table} on each of these populations and count the number of systems that are detected by each survey. We then calculate the complementary cumulative distribution function for the number of detections by each survey, which is shown in Fig.~\ref{success_survey}.
        
        The surveys with the Arecibo radio telescope, i.e. PALFA and AODRIFT, have the highest probability to detect $\geq$1 of these UCB systems. This is followed by the GBNCC and HTRU mid-latitude survey, while the HTRU low-latitude and PMSURV have the lowest probability of detecting any of these UCB systems.
        
        The difference in the efficiency of these surveys at detecting the UCB systems is due to the integration times used for processing these surveys. As can be seen in Table~\ref{survey_table}, AODRIFT has the shortest integration time of all the surveys, followed by GBNCC, PALFA, and the HTRU mid-latitude survey. A shorter integration time always produces a larger degradation factor, thereby providing a larger S/N detection for these systems.
        
        However, it is possible to use a longer integration time and still maintain sensitivity to UCB systems, as demonstrated by the PALFA survey. While the PALFA survey has the third-shortest integration time in Table~\ref{survey_table}, it is able to offset the relative loss in S/N due to the orbital degradation by increasing the overall sensitivity of the radio telescope. We discuss the balancing of the survey integration time with the orbital degradation factor further in Sec.~\ref{sec_optimum_tint}.
        
    \subsection{Multi-messenger prospects for detectable ultra-compact binaries}
        
        In this optimistic scenario where the number of UCB systems beaming towards Earth corresponds to the 95\% upper limit derived above, we can expect to detect as many as four of these UCB systems with the radio pulsar surveys at Arecibo alone. Given their short orbital periods, these UCB systems could be promising sources for LISA, the space-based gravitational wave observatory scheduled to launch in the 2030s \citep{LISA}. To see if these systems will be detectable with LISA, we need to compute the S/N for these systems with respect to LISA's sensitivity curve \citep{lisa_sensitivity_curve}.
        
        For a binary system with eccentricity \ecc, the GW emission from the system is spread over multiple harmonics of the orbital frequency, $f_{\rm n} = n / P_{\rm b}$, where $n$ represents the n'th harmonic. The total S/N for these systems as observed by LISA can be calculated as the quadrature sum of the S/N at each of these harmonics \citep{lisa_ecc_strain_1, lisa_ecc_strain_2},
        \begin{equation}
            \displaystyle {\rm S/N^2} \approx \sum_{n = 1}^{\infty} \frac{h_{\rm n}^2(f_{\rm n}) T_{\rm LISA}}{S_{\rm LISA}(f_{\rm n})}
            \label{lisa_snr}
        \end{equation}
        where $S_{\rm LISA}(f_{\rm n})$ is the LISA sensitivity curve as defined by \citet{lisa_sensitivity_curve}, $T_{\rm LISA} = 4$~yrs is the timespan of the LISA mission, and $h_{\rm n}(f_{\rm n})$ is the strain amplitude
        \begin{equation}
            \displaystyle h_{\rm n}(f_{\rm n}) = \frac{8}{\sqrt{5}} \left( \frac{2}{n} \right)^{5/3} \frac{(\pi f_{\rm n})^{2/3} (\mathcal{G} \mathcal{M})^{5/3}}{c^4 d} \sqrt{g(n, e)},
            \label{lisa_strain}
        \end{equation}
        where $\mathcal{G}$ is the Gravitational constant, $c$ is the speed of light, $d$ is the distance to the binary system, $\mathcal{M} = m_1^{3/5} m_2^{3/5} (m_1 + m_2)^{-1/5}$ is the chirp mass of the binary, and $g(n, e)$ provides the relative amplitude between the different harmonics \citep[see Eq.~20 in][]{peters_mathews_1963}.
        
        Using these relations, we can calculate the S/N with which LISA would observe the UCB systems that are detected with the radio pulsar surveys described above. For the UCB binaries that were detected in the simulations described in Sec.~\ref{subsec_survey_eff} (both NS--WD and DNS), we extract the masses \ma\ and \mb\ of the components of the UCB, the orbital period \pb, eccentricity \ecc, and radial distance to the UCB $d$. We remind the reader that the radial distribution of the pulsars in the Galaxy was assumed to be the one described in \citet{lorimer_rad_dist}, while the $z$-height distribution was the one described in \citet{lyne_z_dist}. Given these parameters, we calculate the strain using Eq.~\ref{lisa_strain} at harmonics $2 \leq n \leq 30$ and then calculate the S/N for each system by summing over these harmonics as described in Eq.~\ref{lisa_snr}.
        
        All of the UCB systems that were detected in the radio pulsar surveys have an S/N that is comfortably above the LISA threshold $\rm S/N = 7$ assuming a four year LISA mission. 
        Thus, as shown in Sec.~\ref{subsec_survey_eff}, even if radio pulsar surveys are able to detect only a couple of these UCB systems, these should be strong detections for LISA and will allow for multi-messenger studies of neutron stars \citep[for example, see][]{thrane_lisa_dns}. 
        
\section{Optimum integration time for detecting ultra-compact BNS systems} \label{sec_optimum_tint}
    
    The impact of the orbital motion of the pulsar on the S/N is most acutely felt when the pulsar is part of an ultra-compact binary system (UCB).
    The modified radiometer equation for pulsars, including the orbital degradation factor $\gamma$, can be written as \citep{lorimer_kramer}
    \begin{multline}
        \displaystyle {\rm S/N} = \left[ \frac{G \sqrt{B N_{\rm p}}}{T_{\rm sys}} \right] \, S \, \left[ \sqrt{\frac{t_{\rm int} (P_{\rm s} - w)}{w}} \right] \, \gamma(t_{\rm int}, P_{\rm s}, ...)^2 \\
        = \xi \times S \times f_1(t_{\rm int}, P_{\rm s}, w) \times \gamma(t_{\rm int}, P_{\rm s},...)^2 ,
        \label{radiometer_eq}
    \end{multline}
    where $\xi$ is a constant that depends on the telescope and survey setup, $S$ is the pulsar flux, $G$ is the receiver gain, $B$ is the receiver bandwidth, $N_{\rm p} = 2$ is the number of polarizations, $T_{\rm sys}$ is the receiver system temperature (which includes the sky temperature $T_{\rm sky}$), $P_{\rm s}$ is the spin period of the pulsar and $w$ is the effective pulse width (which includes effects of dispersion smearing and scattering). We list the telescope and survey parameters, as well as the integration times for the large pulsar surveys that we analyze in this work, in Table~\ref{survey_table}.
    
    \begin{figure}
        \centering
        \includegraphics[width = \columnwidth]{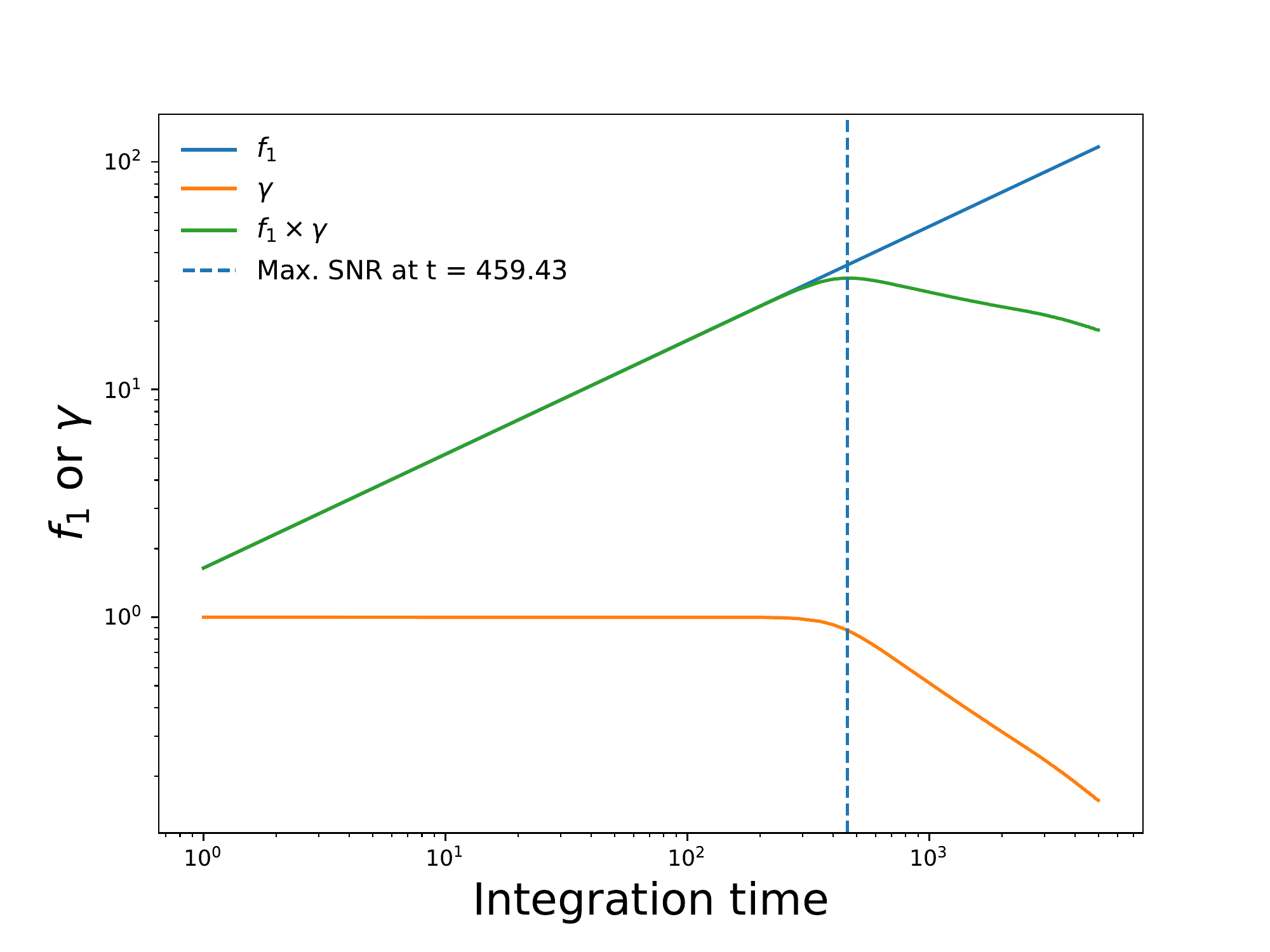}
        \caption[An example of maximizing the S/N by optimizing the selection of the integration time for the pulsar survey.]{An example of maximizing the S/N by optimizing the selection of the integration time for the pulsar survey. This example uses the spin and orbital parameters of J0737--3039A (Double Pulsar, \citep{0737A_disc, Lyne_Bdiscovery_2004}). The horizontal axis shows the survey integration time, while the vertical axis represents the magnitude of functions $f_1$ and $\gamma$ from Eq.~\ref{radiometer_eq}. The function $f_1$ is directly proportional to the integration time, while there is a ``knee" in the orbital degradation factor, $\gamma$.
        As a result, the maximum S/N for this pulsar would be at an integration time of $t_{\rm int} = 459$~s.}
        \label{motivate_opt_tint}
    \end{figure}
    
    For a pulsar with flux $S$, we need to select an integration time $t_{\rm int}$ that maximizes the observed S/N for the pulsar in the BNS system. To motivate the discussion, we show in Fig.~\ref{motivate_opt_tint} an example for a system that has the orbital and spin properties of the millisecond pulsar in the Double Pulsar system
    \citep{Lyne_Bdiscovery_2004, 0737A_disc}. As we can see, when the orbital degradation factor is not considered in the S/N calculation, the S/N for the system grows as a function of the integration time, ${\rm S/N} \propto \sqrt{t_{\rm int}}$. However, the orbital degradation factor function for this system has a ``knee" at an integration time of $\sim$460~s, after which the orbital degradation factor decreases with increasing integration time. As a result of this ``knee" feature, the total S/N for the system peaks at the position of the ``knee" introduced by the orbital degradation factor. Thus, the integration time corresponding to this peak would be the optimum integration time to detect systems like the Double Pulsar. Similarly, each binary pulsar system will have its own unique optimum integration time.
    
    Note that $\xi$ from Eq.~\ref{radiometer_eq} does not affect the optimum integration time, but does affect the final S/N of the system. This factor encodes the instrumental sensitivity of the telescope that is used for a given pulsar survey and, thus, this factor is larger for a more sensitive telescope. For example, for the PALFA survey at Arecibo, $\xi = 8.33$, while for the GBNCC survey at the Green Bank Telescope, $\xi = 0.61$. In fact, PALFA has the largest $\xi$ value for any survey listed in Table~\ref{survey_table}, which explains why it has a high probability of detecting an UCB system despite having the third-shortest integration time (see Sec.~\ref{subsec_survey_eff}). Despite this, it is still important to derive and use an optimum integration time to maximize the probability of detecting an UCB system with all the surveys.
    
    To generalize the example described above, we use a Monte Carlo simulation similar to the one described in Sec.~\ref{sec_asymm_bias}. Since we are interested in UCB systems, we constrain the mass of the companion and the orbital period to the range $0.2 {\rm M}_{\odot} < m_2 < 2.4 {\rm M}_{\odot}$ and $1.5 \, {\rm min} < P_{\rm b} < 15 \,$min. We also constrain the spin periods to the range $1 \, {\rm ms} < P_{\rm s} < 100 \, \rm ms$ to correspond to the periods seen for the BNS systems in the Galaxy (see Sec.~\ref{sec_asymm_bias}), assume a fixed duty cycle of $\delta = 0.06$ \citep{lorimer_kramer} and fix the harmonic to $m = 2$ (see Sec.~\ref{subsec_pop_num}). The other parameters have the same range as listed in Table~\ref{param_range}. We draw $10^7$ random samples from these distributions and calculate the optimum integration time as described above for each UCB system.
    
    The above analysis yields an optimum integration time of $t_{\rm opt} = 42^{+153}_{-22}$~s, where the errors represent the 95\% confidence intervals on the peak of the distribution. Comparing this time to the integration times used for the large pulsar surveys in Table~\ref{survey_table}, we can see that the AODRIFT and GBNCC surveys are ideally placed towards detecting UCB systems, while PALFA is able to compensate for the loss in S/N by having a high $\xi$ value as described above. This is also seen in Fig.~\ref{success_survey}, where these three surveys have the highest probability of detecting at least one UCB system.
    
    \begin{figure*}
            \centering
            \begin{subfigure}[b]{0.49\textwidth}
                \centering
                \includegraphics[width = \textwidth]{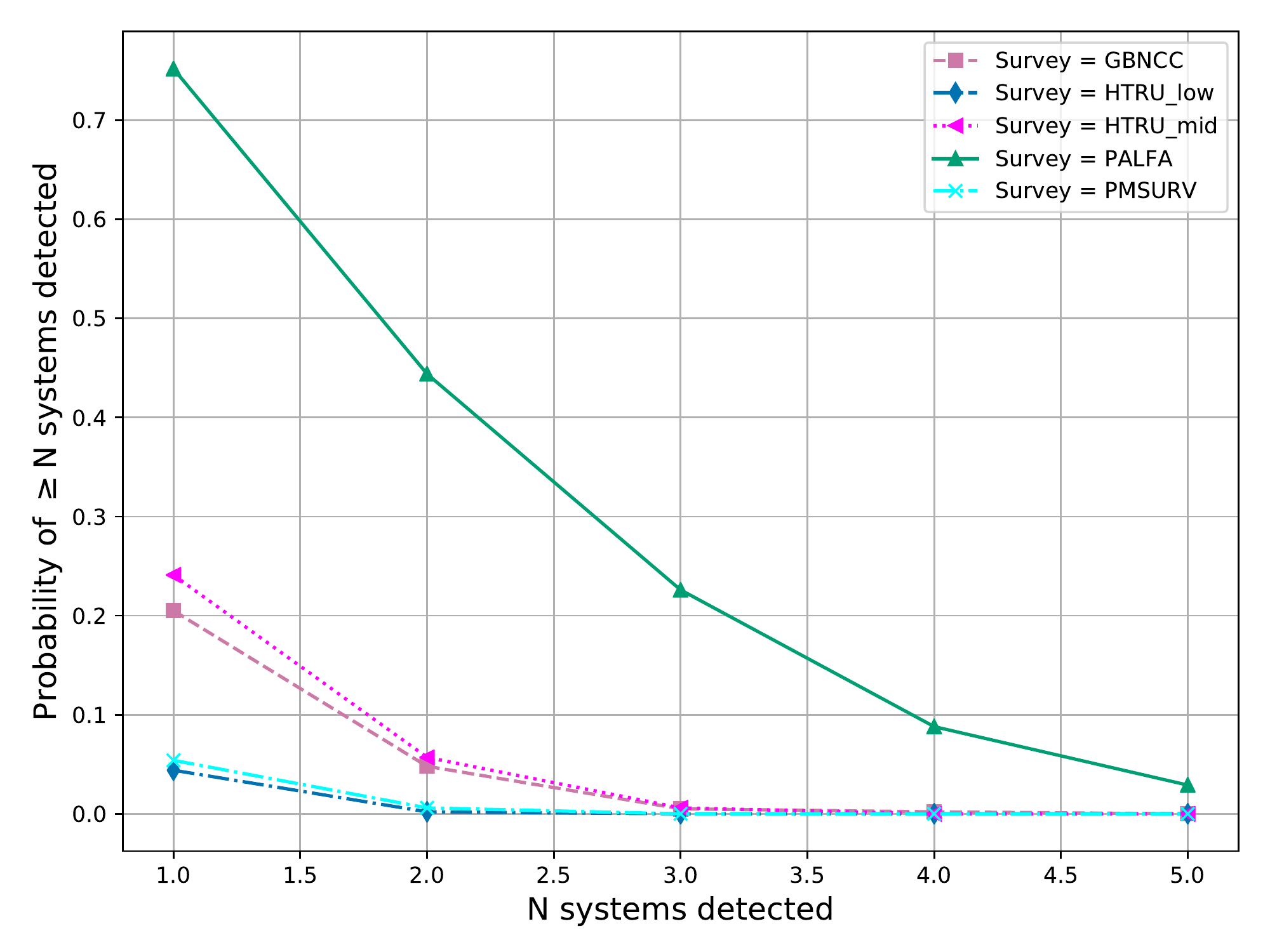}
                \caption{Ultra-compact NS--WD systems}
            \end{subfigure}
            \begin{subfigure}[b]{0.49\textwidth}
                \centering
                \includegraphics[width = \textwidth]{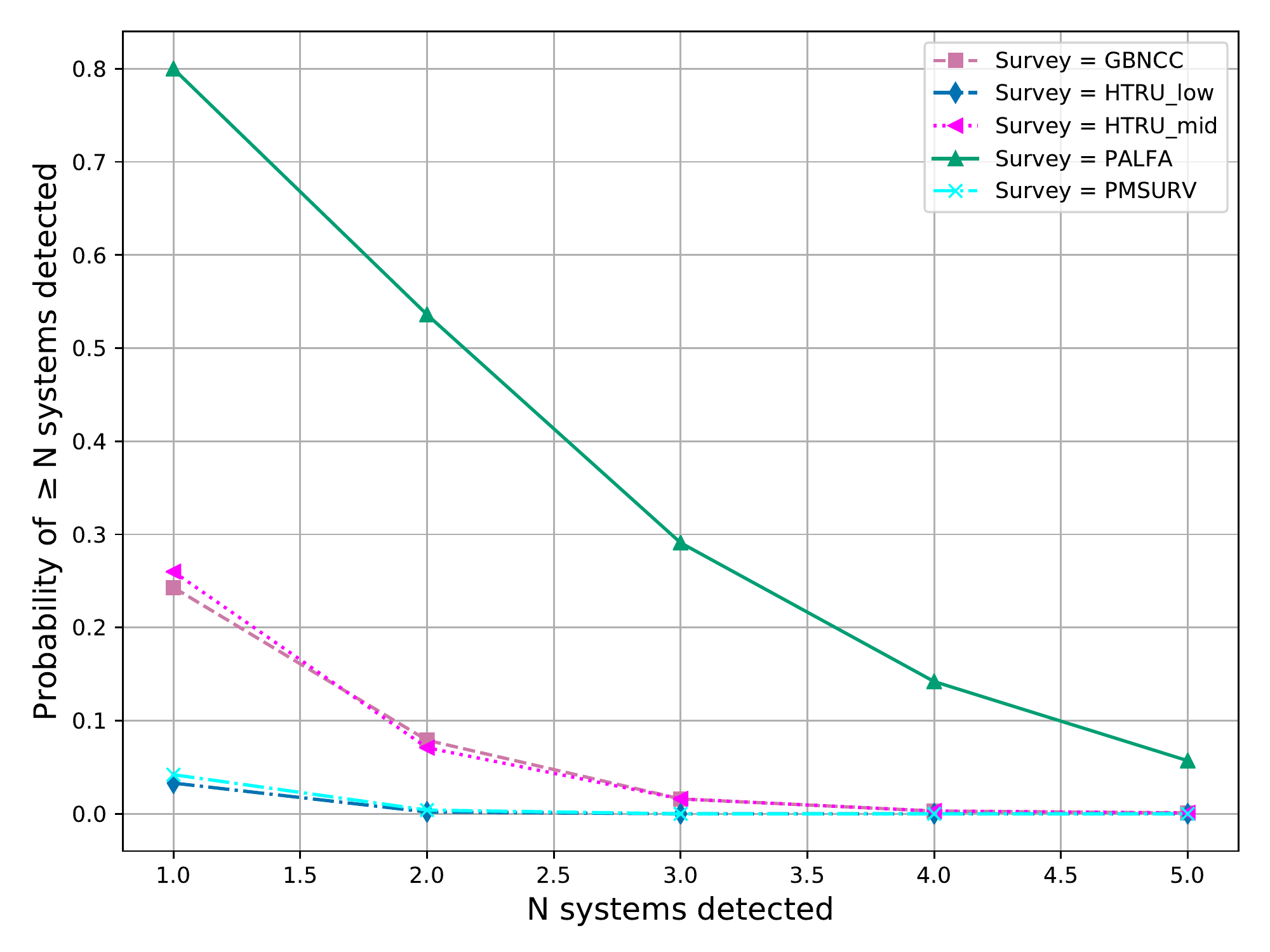}
                \caption{Ultra-compact DNS systems}
            \end{subfigure}
            \caption[Complementary cumulative distribution function (CDF) of the number of UCB systems that are detectable by the radio pulsar surveys listed in Table 3.2, but with integration time set to the optimum value of 50 s derived in Sec. 3.5.]{Complementary cumulative distribution function (CDF) of the number of UCB systems that are detectable by the radio pulsar surveys listed in Table~\ref{survey_table}, but with integration time set to the optimum value of 50~s derived in Sec.~\ref{sec_optimum_tint}. The horizontal axis shows the number of detectable systems $N$ while the vertical axis shows the probability that $\geq N$ systems will be detected in the given survey. Compared to Fig.~\ref{success_survey}, there is a significant improvement in the detection probability for the HTRU and PMSURV surveys.}
            \label{ideal_success_survey}
    \end{figure*}
    
    Choosing an integration time from the range described above also leads to an average increase in the radiometer S/N, with the biggest effect seen for surveys whose integration times are much higher than the range derived above.  For example, for the PALFA survey, reducing the integration time from 268~s to 42~s increases the S/N of the UCB systems by an average factor of 2.3. On the other hand, for PMSURV, which has the largest integration time of 2100~s, the S/N increases by an average factor of 4.5. The effect of this reduction in the integration time and increase in the S/N on the probability of detecting UCBs is shown in Fig.~\ref{ideal_success_survey}, where we can see a significant increase in the detection probability for the HTRU and PMSURV surveys.
    
    However, given the relatively large range of the optimum survey integration times and the fact that each binary system will have its own optimum integration time, rather than picking a single integration time, we recommend implementing the ``time domain resampling" technique \citep{og_odf}. In this method, the integration time for a given survey is progressively reduced by a factor of 2 and each chunk of data is searched individually for binary systems. 
    Using this method and starting with their design integration times, the survey will be most sensitive to UCB systems when the integration times are between 256~s and 32~s, which correspond approximately to the 95\% confidence limits on the optimum integration time derived above.
    The ``time domain resampling" method was most recently used in the HTRU survey \citep{ng_accel_search_tech}, but only up to a minimum integration time of $\sim$537~s, which optimized their search to binaries with orbital periods $P_{\rm b} \geq 1.5$~hours. This led to the discovery of J1757--1854, which has an eccentricity $e = 0.61$, orbital period of $P_{\rm b} = 4.4$~hours and is the most eccentric DNS system detected \citep{1757_accel_search_tech}. 
    Implementing the same time domain resampling technique on all surveys (except AODRIFT, due to its already low integration time) should increase the sensitivity of all the surveys to these UCB systems.
    
\section{Conclusion}
    
    Using the framework developed by \citet{Bagchi_odf}, we develop a neural network to calculate the orbital degradation factor for any given binary system.
    We combine this neural network with \psrpoppy, opening the possibility for modeling the observed binary pulsar population. We show that, on average, it is easier to detect binary systems which are asymmetric in mass as compared to systems which are symmetric in mass. 
    
    We also investigate the population of UCB systems in the Milky Way as these systems are promising targets for the future space-based gravitational wave observatory LISA. We place upper limits of $\sim$850 and $\sim$1100 ultra-compact NS--WD and DNS systems beaming towards Earth, respectively. Note that this does not imply that there are fewer ultra-compact NS--WD binaries than DNS binaries, but merely states that we can constrain the population of the former type of system better than that of the latter. We also show that the radio pulsar surveys with the Arecibo radio telescope have the highest probability of detecting at least one UCB system. We also show that a survey integration time of $t_{\rm opt} \sim 1$~min will maximize the S/N of the UCB systems.
    
\newpage
\renewcommand{\thechapter}{4}

\chapter[A Direct Measurement of Sense of Rotation of PSR J0737--3039A]{A direct measurement of sense of rotation of PSR J0737--3039A}
\label{chap:sense_rotn_A}
\blfootnote{Published as Pol et al., 2018, ApJ, 853, 73. \\
\textbf{Contributing authors:} Nihan Pol, Maura McLaughlin, Michael Kramer, Ingrid Stairs, Benetge B. P. Perera, Andrea Possenti
}

\section{Abstract}
	We apply the algorithm published by \citet{Liang2014} to describe the Double Pulsar system J0737--3039 and extract the sense of rotation of first-born recycled pulsar PSR J0737--3039A. We find that this pulsar is rotating prograde in its orbit. This is the first direct measurement of the sense of rotation of a pulsar with respect to its orbit and a direct confirmation of the rotating lighthouse model for pulsars. This result confirms that the spin angular momentum vector is closely aligned with the orbital angular momentum, suggesting that the kick of the supernova producing the second-born pulsar J0737-3039B was small.

\section{Introduction}
	
	The Double Pulsar PSR J0737--3039 \citep{0737A_disc, Lyne_Bdiscovery_2004} is the first and only neutron star binary system that has had two detectable radio pulsars. The recycled PSR J0737--3039A (hereafter `A') has a period of 22.7 ms and the younger, non-recycled PSR J0737--3039B (hereafter `B') has a spin period of 2.8~s. The 2.45-hour orbit makes this system the most relativistic binary known, providing a unique laboratory to conduct the most stringent tests of Einstein's theory of general relativity in the strong-field regime \citep{Kramer_GRtest_2006}.
	
	In addition to strong-field tests of gravity, the Double Pulsar also offers a unique laboratory to test plasma physics and magnetospheric emission from pulsars \citep{Breton_eclipse_I, emission_height, Lyutikov_eclipses}. We originally were able to detect bright single pulses from B in two regions of its orbit, $190^{\circ} \sim 230^{\circ}$, referred to as bright phase I (BP I), and $260^{\circ} \sim 300^{\circ}$, referred to as bright phase II (BP II) \citep{Lyne_Bdiscovery_2004, B_pulse_ev_perera}.  \citet{Maura_mod_2004} discovered drifting features in the sub-pulse structure from pulsar B. They showed that this phenomenon was due to the direct influence of the magnetic-dipole radiation from A on B. These drifting features (henceforth referred to as the `modulation signal') are only visible in BP I when the electromagnetic radiation from A meets the beam of B from the side \citep{david_lyutikov_modelling}. These modulation features were observed to have a frequency of $\approx 44$ Hz which suggests that this emission is not from the beamed emission of A, which has a frequency of $\approx 88$ Hz due to the visibility of emission from both the magnetic poles of A.
	
	\citet{Freire_model_2009} proposed a technique to measure, among other things, the sense of rotation of A with respect to its orbit using the time of arrival of pulsed radio emission from A and the modulation feature from B. A complementary technique was proposed by \citet{Liang2014} (henceforth LLW2014) to uniquely determine the sense of rotation of A using an approach based on the frequency of the modulation signal. LLW2014 argued that we should be able to observe an effect similar to the difference between solar and sidereal periods observed in the Solar System in the Double Pulsar. Thus, if pulsar A is rotating prograde with respect to its orbit, the modulation signal should have a period slightly greater than that if it were not rotating, and it would have a slightly smaller period for the case of retrograde motion. LLW2014 provide an algorithm to apply this concept to the observations of the Double Pulsar, and we refer the reader to that paper for more details on the calculations and details of the algorithm.
	
	In this dissertation, we implement this algorithm on the Double Pulsar data and determine the sense of rotation of pulsar A. In Sec.~\ref{Procedure}, we briefly describe the data used and the implementation of the algorithm from LLW2014. We present our results in Sec.~\ref{Results} and we discuss the implications of these results in Sec.~\ref{Discussion}.

\section{Procedure} \label{Procedure}
	
	\subsection{Observations and Data Preparation} \label{obs}
		
		We have carried out regular observations of the Double Pulsar since December 23, 2004 (MJD 52997) with the Green Bank Telescope (GBT). The radio emission of B has shown a significant reduction in flux density ($0.177 \ \textrm{mJy} \ \textrm{yr}^{-1}$) and evolution from a single peaked profile to a double peaked profile due to relativistic spin precession with B's radio emission disappearing in March 2008 \citep{B_pulse_ev_perera}. As a result, we choose the data where B's emission is brightest, which also corresponds to the modulation signal being the brightest, for this analysis, i.e. the data collected on MJD 52997.
		
\begin{figure}
			\includegraphics[width = \linewidth, clip]{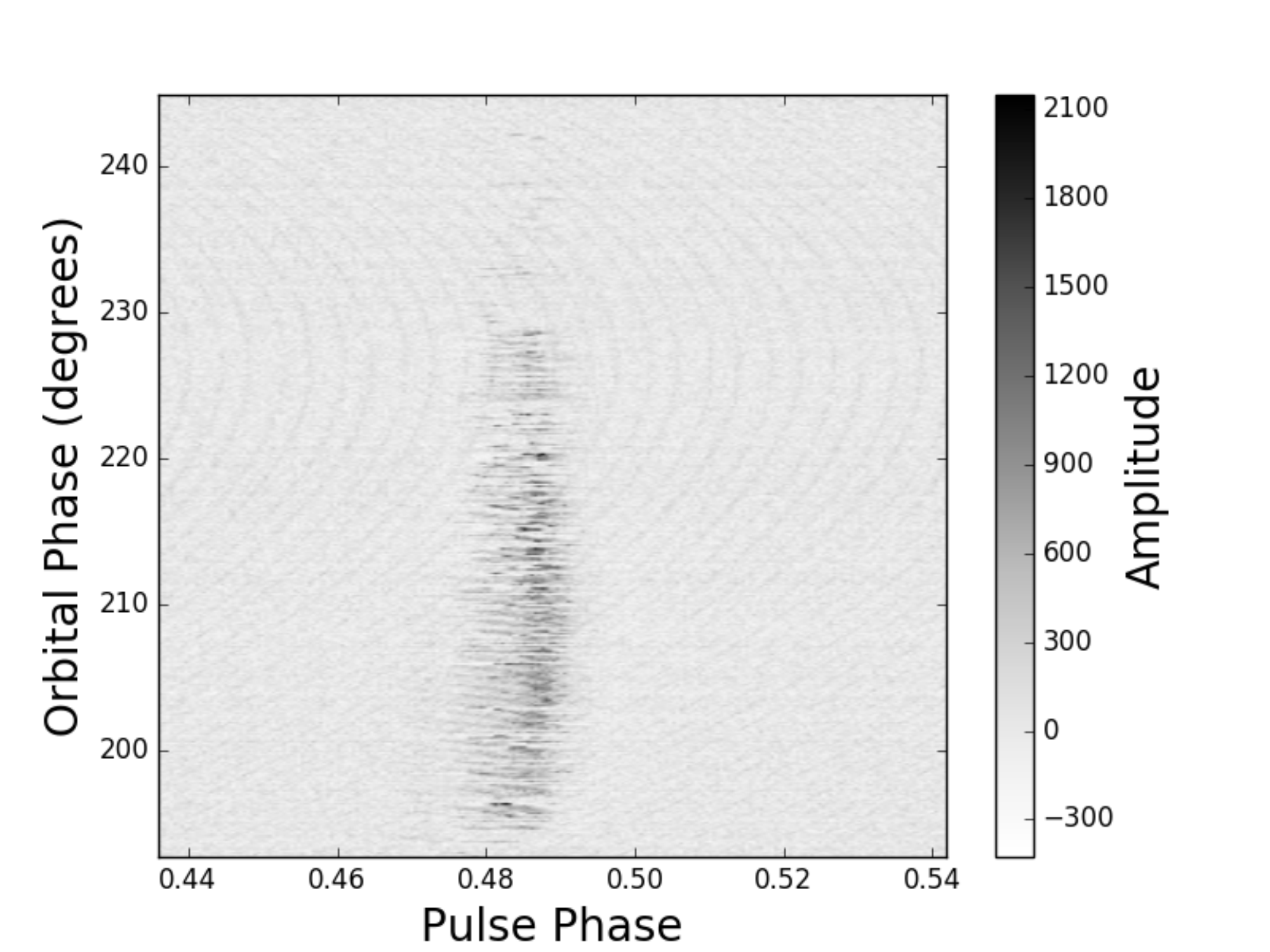}
			\caption[Single pulses of B for MJD 52997 for orbital phase $190^{\circ} - 240^{\circ}$ in the first BP I.]{Single pulses of B for MJD 52997 for orbital phase $190^{\circ} - 240^{\circ}$ in the first BP I. The drifting features are most prominent in the orbital phase range of $195^{\circ} \sim 210^{\circ}$. A's pulses are also visible in the background, and are most visible at $\sim 225^{\circ}$. Note that only a fraction of the pulse phase is shown for clarity and the units on the amplitude are arbitrary. This figure is adapted from \citet[][]{Maura_mod_2004}.}
			\label{original_signal}
		\end{figure}

		These data on MJD 52997 were taken at a center frequency of $800$ MHz with the GBT spectrometer SPIGOT card \citep[][]{Spigot_ref}. This observation had a sampling time of $40.96 \ \mu$s and the observation length was 5 hours, covering more than two complete orbits. We barycenter these data using the barycenter program from the SIGPROC\footnote{This software can be found at: \url{http://sigproc.sourceforge.net/}} software package. We decimate the data from its native resolution of $24.41$ kHz to $2048 \times f_{B,0} \approx 738.43$ Hz where $f_{B,0}$ is the rest-frame frequency of B. This is equivalent to splitting up a single rotational period of B into 2048 bins. We do this to increase the signal-to-noise ratio S/N of the modulation signal. Since B's drifting pulses are observed only in BP I, we focus only on these orbital phases. Since this data set covers more than two orbits, we obtain two such BP I time series. The first of these BP I time series is shown in Fig.~\ref{original_signal}.
		
		Since the drifting pulses are seen only at the beginning of this phase range, we analyze the orbital phase range (defined as the longitude from ascending node, i.e. the sum of longitude of periastron and true anomaly) $195^{\circ}$ to $210^{\circ}$. This final data set is approximately 344 seconds long. With the data set prepared as described above, and noting that the time series length $T = N \Delta t$ where $N$ is total number of samples and $\Delta t$ is the sampling time, our Fourier spectra have a frequency resolution
		\begin{equation}
			\displaystyle f_s = \frac{1}{T} = \frac{1}{N \Delta t} = 2.93 \textrm{ mHz}
		\end{equation}
		
	\subsection{Transformation}
		
		We apply the algorithm from LLW2014 (see Sec.~3.4 therein) to the time series. We found a typographic error in the algorithm from LLW2014. 
		All programming is done in Python. We write a function that returns values for the longitude of periastron of B's orbit $\omega_B$, and the true anomaly for B $\theta$ as a function of time \citep[see Chapter 8 in][]{lorimer_kramer}. All orbital parameters such as eccentricity $e$, orbital inclination angle $i$ and semi-major axis $a_B$ are obtained from the timing solution of B \citep{Kramer_GRtest_2006}. With all these parameters in place, the implementation of the algorithm was straight-forward. For completeness, we briefly list the transformation described in LLW2014, and refer the reader to that paper for more details.
		
		The basic idea of the transformations is to remove the Doppler smearing produced by eccentric orbits in the Double Pulsar by suitably resampling the data and obtaining the time at which the modulation signal left A. This can be done by first computing the resampled time series $t_B[k]$ which represents the time of the $k^{th}$ sample measured at B, by correcting for B's orbital motion (Eq. 10 in LLW2014):
		
		\begin{equation}
		\displaystyle t_B[k] = t[k] - \frac{L}{c} - \frac{a_B \ \textrm{sin } i \ (1 - e^2) \ \textrm{sin}(\omega_B + \theta)}{c \ (1 + e \ \textrm{cos } \theta)},
		\label{tb}
		\end{equation}
		where $t[k]$ is the time corresponding to the $k^{th}$ sample measured at the solar system barycenter (ssb), $L$ is the distance to the Double Pulsar and $a_B$ is the semi-major axis of B's orbit. The $L/c$ term is a constant offset which can be neglected without loss of information. Now, we can calculate the time $t_A[k]$, when the signal causing the modulation features left A by correcting for an additional propagation time delay along the path length from A to B,
		 \begin{equation}
		 \displaystyle t_A[k] = t_B[k] - (a_A + a_B) \ \frac{1 - e^2}{1 + e \ \textrm{cos} (\theta)},
		 \label{tA}
		 \end{equation}
		where $a_A$ is the semi-major axis of A. Eq.~\ref{tA} can be used to transform $I[k]$, the intensity data sampled at the ssb at time $t[k]$, into a frame where the time-variable Doppler shifts have been removed from the data.
		
		Using the resampled time series, we compute the Fourier power spectrum (Eq. 19 in LLW2014),
		\begin{equation}
		\displaystyle P_{n} (z f_{A,0})_s = \left| \sum_{k} I[k] \ \textrm{exp} ( -i \ n \ \left| \Phi_{\textrm{\textbf{m}}} [k,z] \right|_s) \right| ^ {2}
		\label{power spectrum}
		\end{equation}
		where $z$ is a frequency scaling factor, $ P_{n} (z f_{A,0})_s$ is the power in the $n^{th}$ harmonic of the modulation corresponding to the trial spin frequency $zf_{A,0}$, and
		\begin{equation}
		\displaystyle \left| \Phi_{\textrm{\textbf{m}}} [k,z] \right|_s = 2\pi \ (z \ f_{A,0}) \ t_A[k] - s \ \theta(t_B[k])
		\label{phi}
		\end{equation}
		is the ``modulation phase'' which is simply the rotational (or pulsational) phase of A corrected for the sense of its rotation, with $s = 1, -1, 0$ corresponding to prograde, retrograde and no rotation (pulsation), respectively. Here, $f_{A,0}$ is the sidereal frequency of A's rotation or pulsation, which we know from timing measurements to be $44.05406$ Hz \citep{Kramer_GRtest_2006} on MJD 52997.
		
		If we compute the power spectrum in Eq.~\ref{power spectrum} for each value of $s$, then, based on the arguments in LLW2014, we should observe a peak at a frequency corresponding to $z = 1$, i.e. at $f = f_{A,0}$ and the value of $s$ with the highest power in this peak will determine the sense of rotation of A.
	
\section{Results} \label{Results}
	
	\begin{figure*}
		\includegraphics[width = \textwidth]{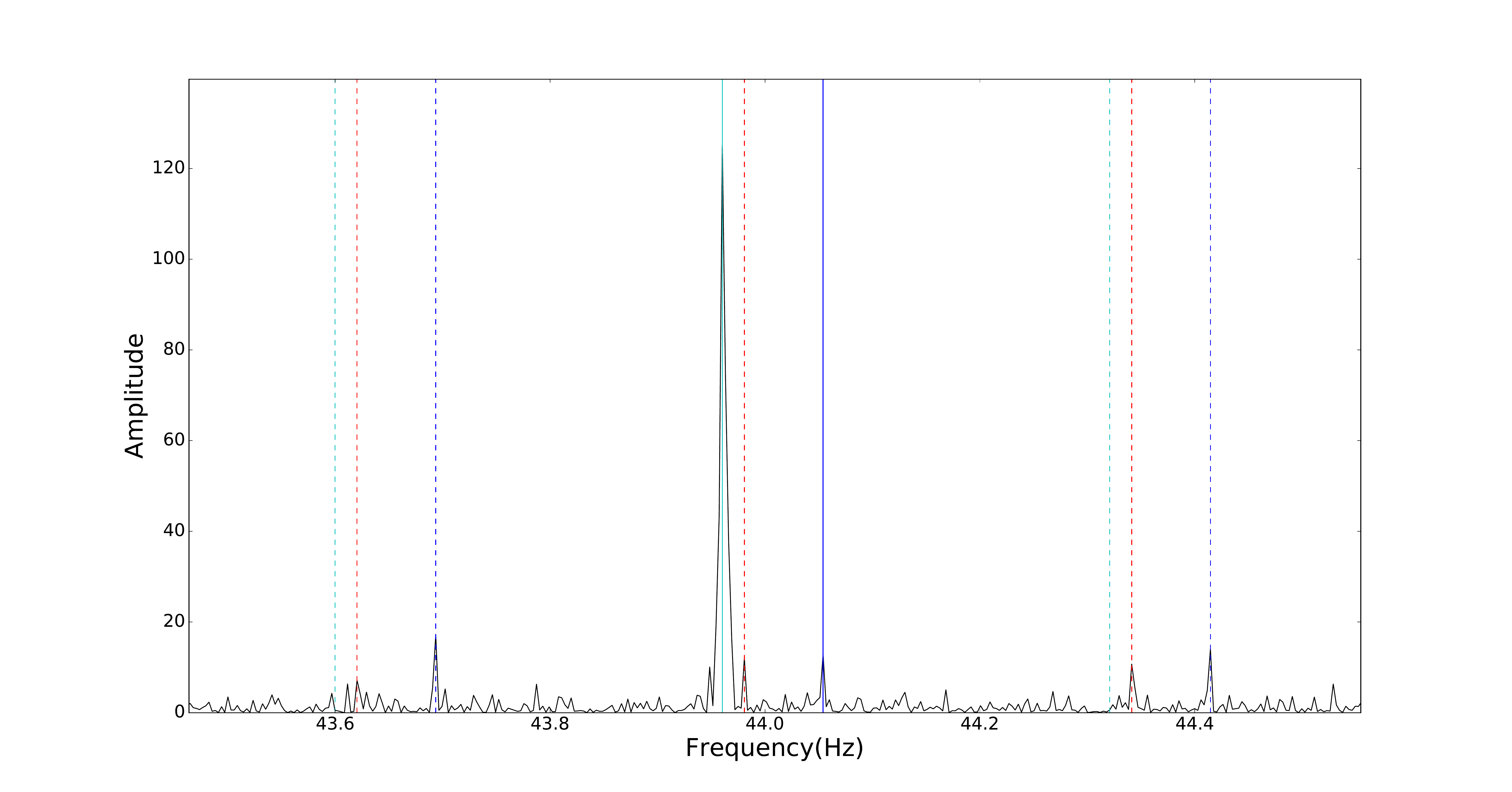}
		\caption[Fourier power spectrum for $s = 0$.]{Fourier power spectrum for $s = 0$. Vertical dashed red lines mark the harmonics from B's intrinsic signal. The fundamental frequency of the modulation signal $f_{A,0}$ is shown by a vertical solid blue line and vertical dashed blue lines mark the sidebands of the modulation signal. The emission from A's intrinsic signal is visible as the prominent peak close to the fundamental frequency of the modulation signal, marked by a solid vertical cyan line, which is not at $f_{A,0}$ due to the transformations applied above (see text). Similar to the sidebands of the modulation signal, we mark the positions of the sidebands for A's intrinsic signal.There is no power in the sidebands for A's intrinsic signal, indicating that we have successfully separated A's intrinsic signal from the modulation signal.}
		\label{harmonics_s0}
	\end{figure*}
	
	Since emission from A is stimulating emission in B, we can think of A as the ``carrier" signal which modulates the magnetosphere of B. This interpretation implies we would see a signal at the fundamental frequency of the carrier (in this case $f_{A,0}$) and sidebands of this signal separated by the modulation frequency (in this case $f_{B,0}$). Thus, we would expect to see a signal at frequencies $f_{A,0} \pm m \times f_{B,0}$ where $m = 0, 1, 2, ...$. We see this structure in the Fourier power spectra for the three different cases of $s$, with the power spectrum for $s = 0$ shown in Fig.~\ref{harmonics_s0} for reference. There is a peak at the fundamental frequency $f_{A,0}$ (marked by a blue solid vertical line) and its sidebands (marked by vertical dashed blue lines). We also mark the harmonics from B's signal (fundamental frequency of $f_{B,0} = 0.3605$ Hz). They are visible as distinct peaks in the power spectrum indicating different origins for the two signals.
	
	In addition to these signals, we see a strong signal close to $f_{A,0}$, marked in Fig.~\ref{harmonics_s0} by a solid vertical cyan line. This is the relic of the signal from A's emission, but shifted away from its native frequency of $f_{A,0}$ and reduced in amplitude by the transformations that we have applied. This is a key part of the analysis which allows us to distinguish the signal generated by A's intrinsic emission and the modulation signal. In Fig.~\ref{harmonics_s0}, we also do not detect any power in the sidebands associated with A's intrinsic signal. The presence of A's intrinsic signal, without the presence of sidebands, serves as evidence that the signal we see at $f_{A,0}$ after applying the transformations is from the modulation feature rather than from A itself, and that we have successfully separated the modulation signal from A's intrinsic emission.
	
	\begin{figure*}
		\includegraphics[width=\textwidth]{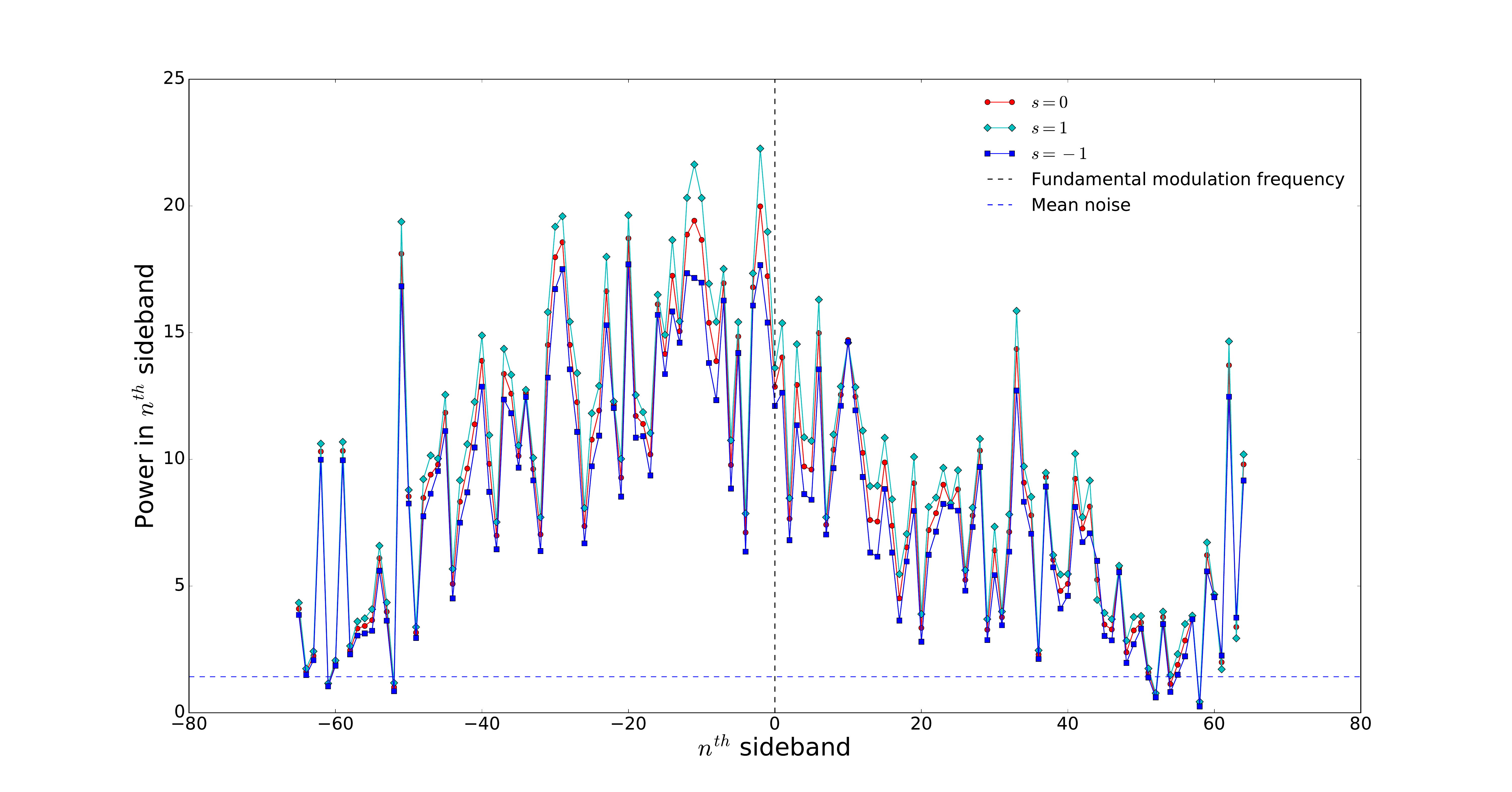}
		\caption[Fourier power in the modulation signal's fundamental frequency and sidebands for the first BP I on MJD 52997.]{Fourier power in the modulation signal's fundamental frequency and sidebands for the first BP I on MJD 52997. The power for all three cases of $s = 1,0, -1$ are plotted together for comparison. The location of the fundamental frequency is shown by a vertical dashed line, while the mean noise level is shown by a horizontal dashed line. The case of $s = 1$ has consistently high power over all components which indicates this is the true sense of rotation of A.}
		\label{BP1_1_52997}
	\end{figure*}
	
	\begin{figure*}
		\includegraphics[width = \textwidth]{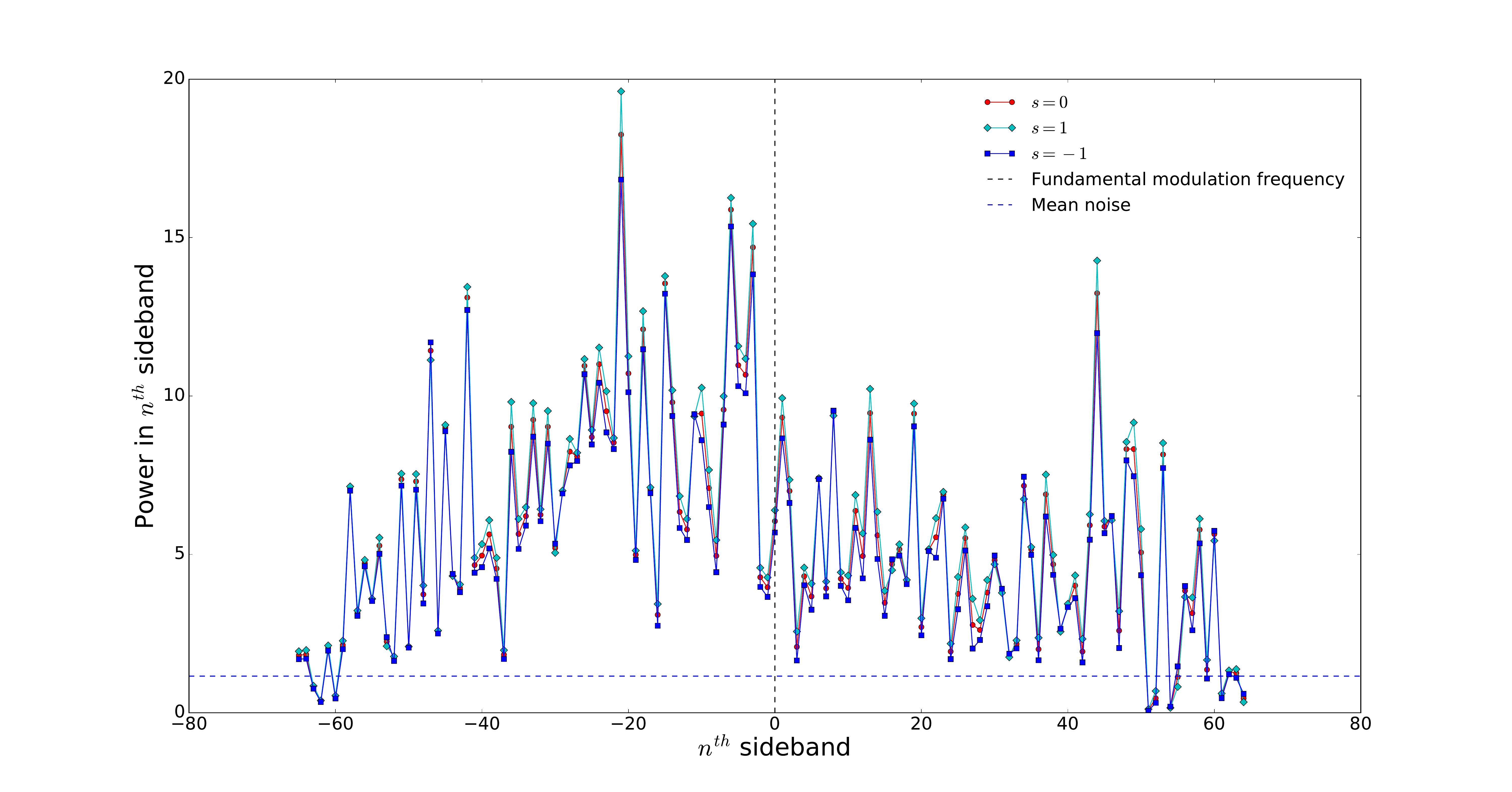}
		\caption[Fourier power in the modulation signal's fundamental frequency and sidebands for the second BP I on MJD 52997.]{Fourier power in the modulation signal's fundamental frequency and sidebands for the second BP I on MJD 52997. The power for all three cases of $s = 1,0, -1$ are plotted together for comparison. The location of the fundamental frequency is shown by a vertical dashed line, while the mean noise level is shown by a horizontal dashed line. The power in $s = 1$ is consistently higher than other values of $s$.}
		\label{BP1_2_52997}
	\end{figure*}
	
	\begin{figure*}
		\includegraphics[width = \textwidth]{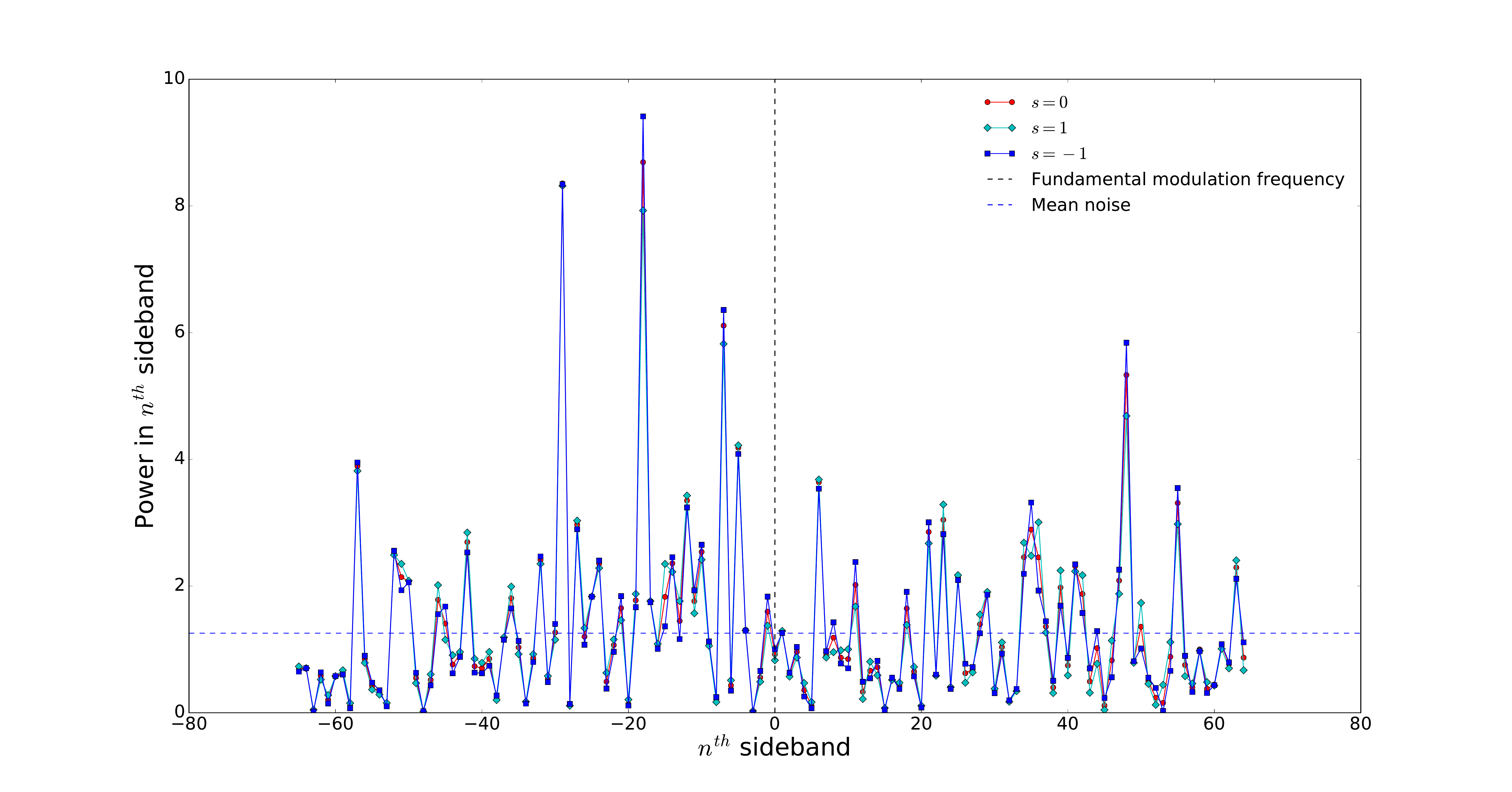}
		\caption[Fourier power in the modulation signal's fundamental frequency and sidebands for the first BP II on MJD 52997.]{Fourier power in the modulation signal's fundamental frequency and sidebands for the first BP II on MJD 52997. The power for all three cases of $s = 1,0, -1$ are plotted together for comparison. The location of the fundamental frequency is shown by a vertical dashed line, while the mean noise level is shown by a horizontal dashed line. In this orbital phase range, the modulation signal is not visible. Hence, we do not see any significant power for any value of $s$ at any of the sidebands of the modulation signal.}
		\label{BP2_1_52997}
	\end{figure*}
	
	\begin{figure*}
		\includegraphics[width = \textwidth]{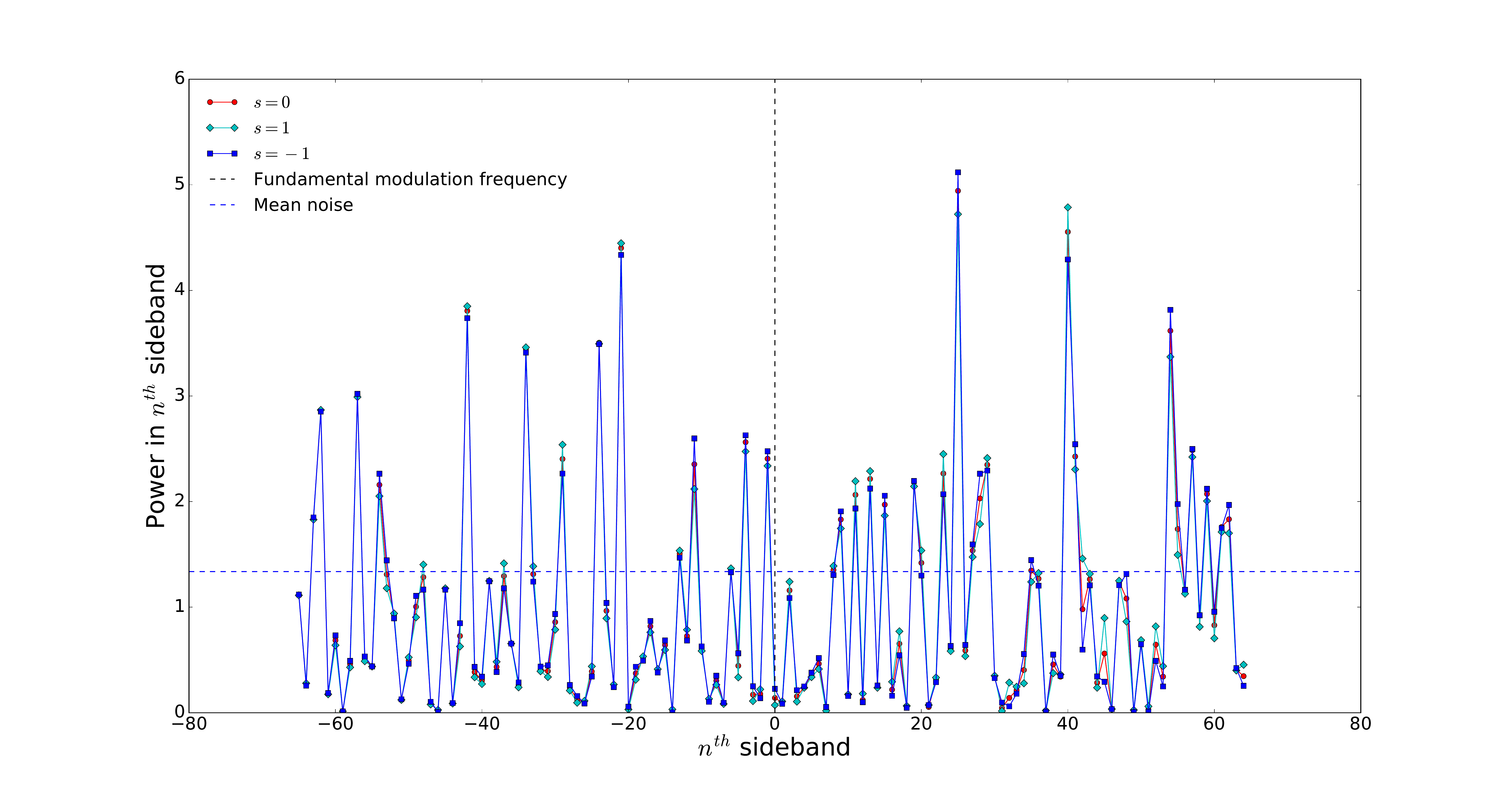}
		\caption[Fourier power in the modulation signal's fundamental frequency and sidebands for a weak phase ($40^{\circ}$ to $52^{\circ}$) on MJD 52997.]{Fourier power in the modulation signal's fundamental frequency and sidebands for a weak phase ($40^{\circ}$ to $52^{\circ}$) on MJD 52997. The power for all three cases of $s = 1,0, -1$ are plotted together for comparison. The location of the fundamental frequency is shown by a vertical dashed line, while the mean noise level is shown by a horizontal dashed line. In this orbital phase range, neither B nor the modulation signal is visible. Hence, we do not see any significant power for any value of $s$ at any of the sidebands of the modulation signal.}
		\label{WP_2_52997}
	\end{figure*}
	
    Finally, we compare the power in the signals at frequencies $f_{A,0} \pm n \times f_{B,0}$ for all three cases of $s$, with the value of $s$ with the highest power indicating the direction of rotation of A with respect to its orbit. Note that, as described in Sec.~\ref{obs}, we have observed two complete orbits of the Double Pulsar on MJD 52997. We plot the power at the fundamental frequency and its sidebands for the BP I of the first and second orbit in Figs.~\ref{BP1_1_52997}~and~\ref{BP1_2_52997}, respectively, and plot the BP II (where B is visible, but the modulation features are not visible) for the first orbit in Fig.~\ref{BP2_1_52997}. For completeness, we also plot the power at the fundamental frequency and its sidebands for some randomly selected weak phase ($40^{\circ}$ to $52^{\circ}$, where weak or no emission is observed in B's spectrum and no modulation signal is visible) for the first orbit in Fig.~\ref{WP_2_52997}.

    In both the power spectra for BP I (see Figs.~\ref{BP1_1_52997}~and~\ref{BP1_2_52997}), the fundamental frequency and its sidebands have consistently higher power in $s = 1$ than for $s = 0, -1$. The power at these frequencies is also significantly higher than the mean noise floor of the respective power spectra. By comparison, there is very little to no power in these frequencies for the other orbital phase ranges of B's orbit (see Figs.~\ref{BP2_1_52997}~and~\ref{WP_2_52997}). This is consistent with observations of the Double Pulsar where the modulation driftbands are not visible in any other orbital phase range apart from BP I \citep{Maura_mod_2004}. 
     
	 \begin{figure}
	     \centering
	     \includegraphics[width = \columnwidth]{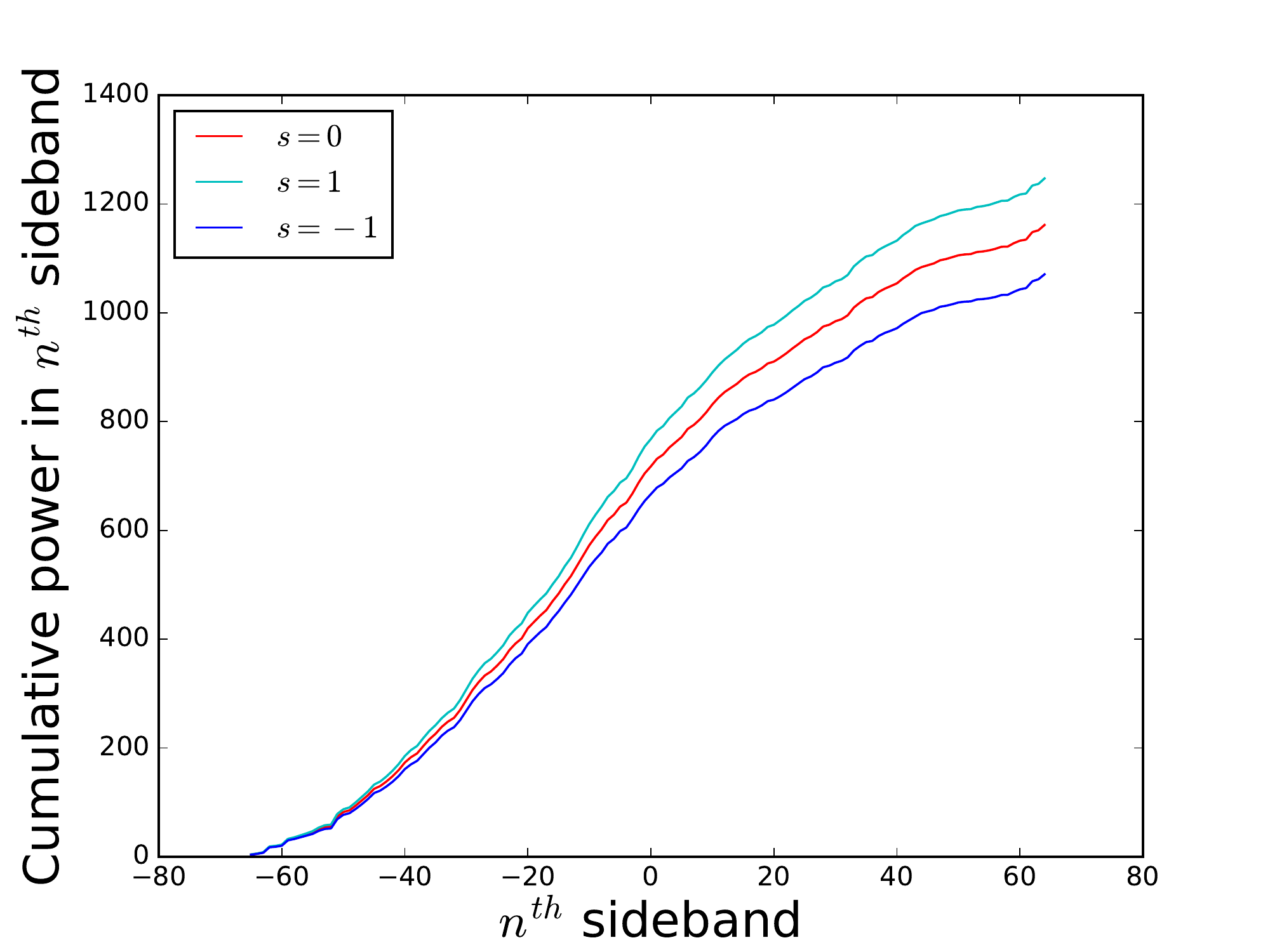}
	     \caption[Cumulative power across all sidebands for the first BP I on MJD 52997.]{Cumulative power across all sidebands for the first BP I on MJD 52997. The power in $s = 1$ is consistently higher than the power in the other values of $s$.}
	     \label{cum_bp1_1}
	 \end{figure}
	 
	 \begin{figure}
	     \centering
	     \includegraphics[width = \columnwidth]{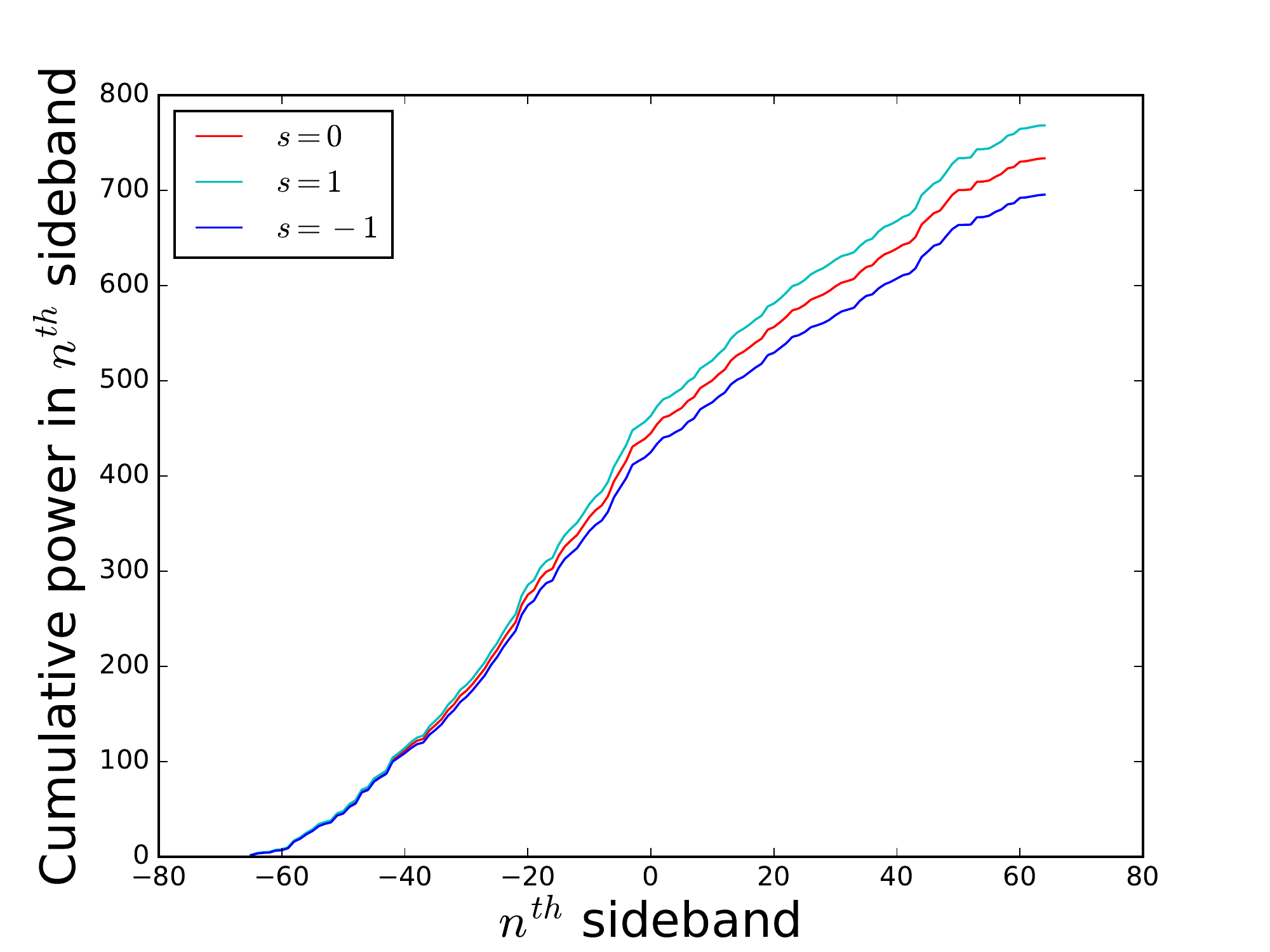}
	     \caption[Cumulative power across all sidebands for the second BP I on MJD 52997.]{Cumulative power across all sidebands for the second BP I on MJD 52997. Similar to the first BP I, the power in $s = 1$ is consistently higher than the power in the other values of $s$.}
	     \label{cum_bp1_2}
	 \end{figure}
	 
	 \begin{figure}
	     \centering
	     \includegraphics[width = \columnwidth]{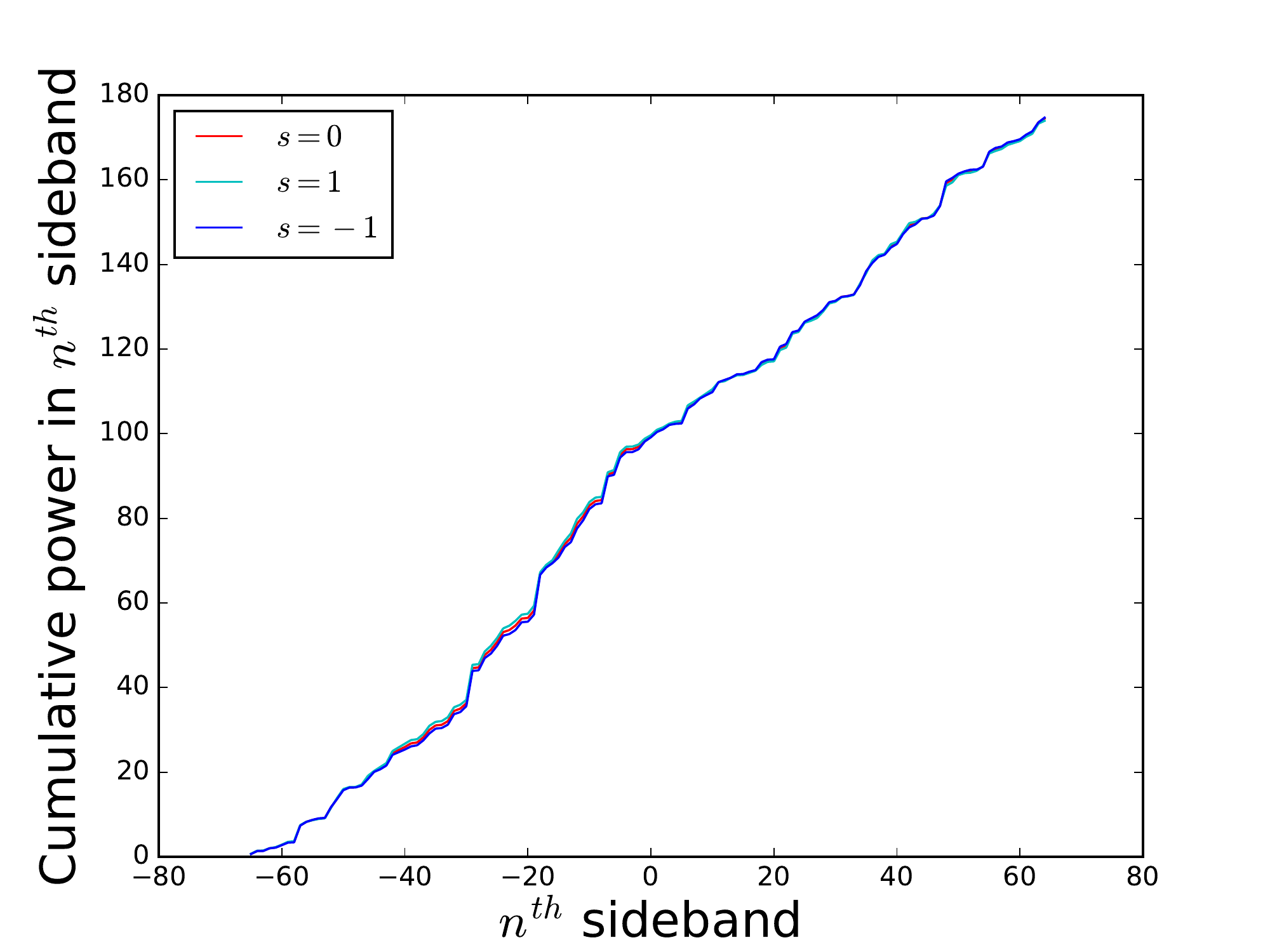}
	     \caption[Cumulative power across all sidebands for the first BP II on MJD 52997.]{Cumulative power across all sidebands for the first BP II on MJD 52997. No value of $s$ dominates over the other values in terms of total power. This is consistent with the conclusion drawn from Fig.~\ref{BP2_1_52997}.}
	     \label{cum_bp2_1}
	 \end{figure}
	 
	 \begin{figure}
	     \centering
	     \includegraphics[width = \columnwidth]{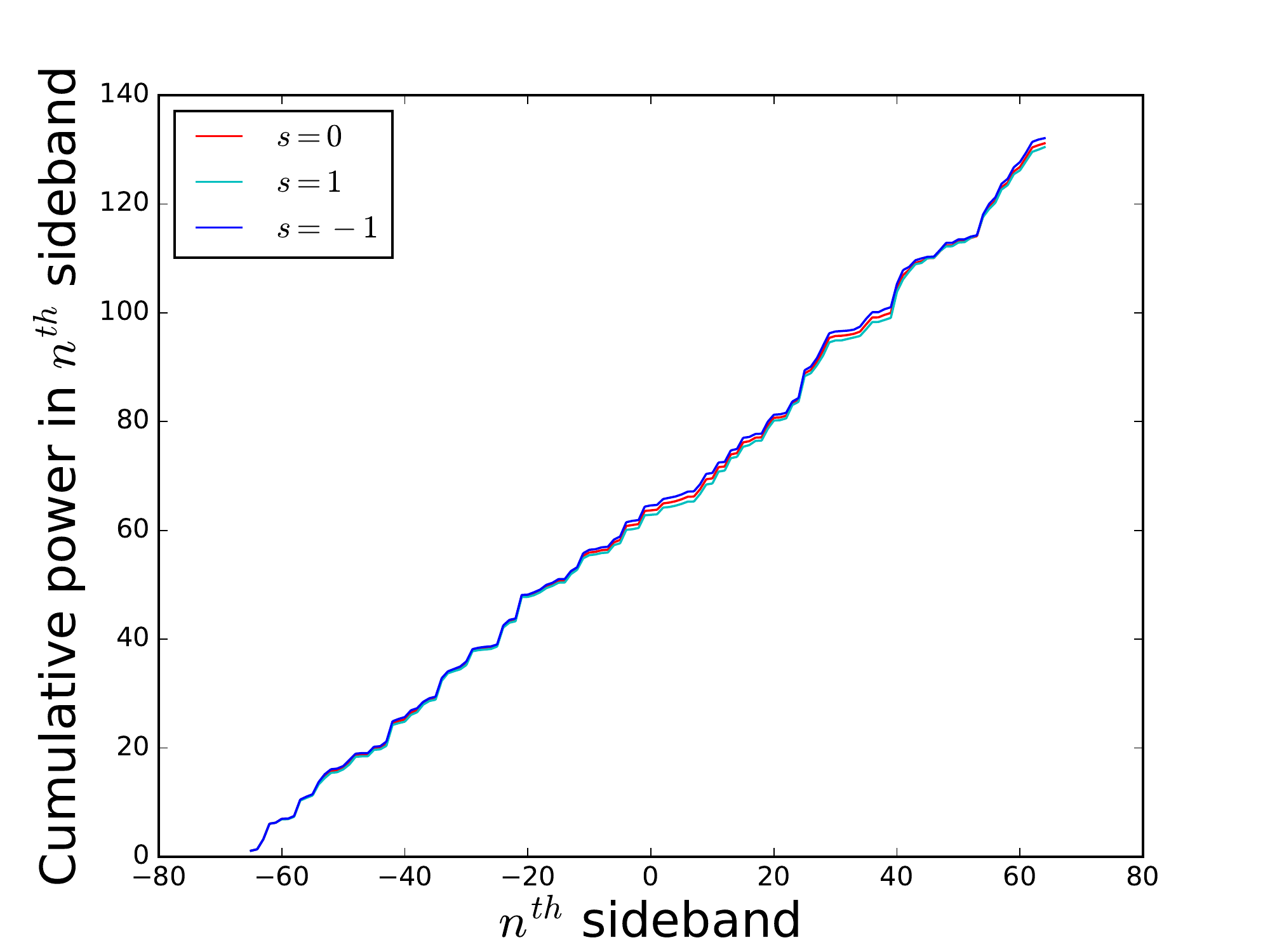}
	     \caption[Cumulative power across all sidebands for the weak phase on MJD 52997.]{Cumulative power across all sidebands for the weak phase on MJD 52997. No value of $s$ dominates over the other values in terms of total power. This is consistent with the conclusion drawn from Fig.~\ref{WP_2_52997}.}
	     \label{cum_nbp_2}
	 \end{figure}
	 
	 To illustrate the consistently higher power in BP I, we plot the cumulative power across all sidebands of the modulation signal in Figs.~\ref{cum_bp1_1}~and~\ref{cum_bp1_2}. In these plots, we begin with the power in the $-64^{th}$ sideband in Fig.~\ref{BP1_1_52997} and add it to the power in the next sideband and so on until we reach the $64^{th}$ sideband. These plots clearly indicate that the power in $s = 1$ is always greater than that in other values of $s$. For comparison, we plot in Figs.~\ref{cum_bp2_1}~and~\ref{cum_nbp_2} the cumulative power across all sidebands for BP II and the weak phase respectively. None of the values of $s$ dominate over the other values in terms of their total power.
	 
	 To test the significance of this result, we take the data from the two BP I sections and scrambled them such that we had a time series that resembled noise. Next, we pass this scrambled time series through the same analysis described above for the real data. This process of scrambling the data and passing it through the analysis pipeline is repeated 1000 times so that we have a collection of values for the total power in different cases of $s$. We compare the total power across all sidebands (including the fundamental frequency) for $s = 1, 0, -1$ in the scrambled time series with the total power in $s = 1, 0, -1$ obtained from the real time series, respectively, to obtain the standard deviation
	 \begin{equation}
	     \displaystyle \sigma = A \times \frac{B}{C},
	     \label{sigma_eqn}
	 \end{equation}
	 where A is the average standard deviation of total power in the scrambled data, B is the mean of the total power for $s = 1,0,-1$ for the real data, and C is the mean of the total power for $s = 1,0,-1$ for the scrambled data.
	 
	 Using the above $\sigma$, we compute the difference in total power for $s = 1,0,-1$ for the real data. For the first BP I, we find that $\sigma = 7.4$ and the total power in $s = 1$ is $11.6\sigma$ above $s = 0$ and $23.9\sigma$ above $s = -1$. Similarly, for the second BP I, we have $\sigma = 4.6$ and total power in $s = 1$ is $7.6 \sigma$ above $s = 0$ and $16.0 \sigma$ above $s = -1$. We perform a similar analysis for the BP II time series and find that none of the differences in total powers exceeds $1.5\sigma$. The observation of consistently higher power in $s = 1$ over $s = 0,-1$ along with the high significance of the $s = 1$ signal in two BP I time series leads us to the conclusion that $s = 1$ represents the true direction of rotation of A with respect to its orbit.

\section{Discussion and Conclusion} \label{Discussion}
	
	Based on this analysis, we conclude that A is rotating in a prograde direction with respect to its orbit. This is the first time, in 50 years of pulsar studies, that such a direct confirmation of the sense of rotation of a pulsar has been obtained. This is additional empirical evidence for the rotating lighthouse model \citep[earlier evidence was presented by][using special relativistic aberration of the revolving pulsar beam due to orbital motion in the B1534+12 system]{stairs_1534_aberration}. This model describes pulsars as rapidly rotating neutron stars emitting magnetic-dipole radiation from their polar cap region. This rapid rotation of the pulsar results in the periodic pulses of light that are characteristic of pulsar emission. This work provides direct confirmation of this model.
	
	This result will help constrain evolutionary theories of binary systems \citep{Alpar_recycle_1982} as well as improve constraints on B's supernova kick. \citet[][]{Ferdman_snevidence_2013} computed a mean 95\% upper limit on the misalignment angle between the spin and orbital angular momentum axes of A to be $3.2^{\circ}$ and concluded that A's spin angular momentum vector is closely aligned with the orbital angular momentum. Our result confirms that and earlier hypotheses \citep{Willems_0737formation_2006, Stairs_0737formation_2006, Ferdman_snevidence_2013, Tauris_DNS_formation} that the kick produced by B's supernova was small.
	
	Furthermore, knowing the direction of spin angular momentum of A will allow us to compute the sign of the relativistic spin-orbit coupling contribution to the post-Keplerian parameter $\dot{\omega}$, which in turn will allow us to determine A's moment of inertia \citep{Kramer_testofgravity_2009}. The moment of inertia of A, along with the well-determined mass of A will provide us with a radius, which will introduce fundamental constraints on the equation of state for dense matter \citep{Lattimer_EOS_2005}.
	
	This measurement of the sense of rotation of A was made using the frequency of the modulation signal and the rotational frequencies of A and B. An alternative way to measure the same effect is using times of arrivals of the pulses from the modulation signal and A's radio emission. \citet{Freire_model_2009} constructed a geometric model for the double pulsar system and used it to exploit the times of arrivals to measure the sense of rotation of A along with determining the height in B's magnetosphere at which the modulation signal originates. Their model will also provide another measurement of the mass ratio of A and B which will affect the precision of some of the tests of general relativity carried out in this binary system. Our preliminary results implementing the \citet{Freire_model_2009} model also indicate prograde rotation for A, and will be published in a future work. 

\newpage
\renewcommand{\thechapter}{5}

\chapter{Conclusion}
\label{chap:conc}

Using pulsar population synthesis analysis with the {\sc psrpoppy} software package, we analyze the population of DNS systems in the Galaxy. We find that given the current known DNS systems, the scale height of this population is consistent with the scale height of the canonical pulsar population. We also calculate a DNS merger rate of $\mathcal{R}_{\rm MW} = 37^{+24}_{-11}$~Myr$^{-1}$, where the errors represent 90\% confidence intervals. This DNS merger rate implies a LIGO DNS merger detection rate of $\mathcal{R} = 1.9^{+1.2}_{-0.6} \times (D_{\rm r} / 100 \, {\rm Mpc})^3$~yr$^{-1}$, where $D_{\rm r}$ is the range distance.

We also developed a neural network implementation to calculate the Doppler smearing due to a pulsar's orbital motion and integrated it into the {\sc psrpoppy} software package. Using this implementation, we investigated the population of ultra-compact ($1.5 \, {\rm min} \leq P_{\rm b} \leq 15\,\rm min$) NS--WD and DNS systems in the Galaxy. We place a 95\% confidence upper limit of $\sim$850 and $\sim$1100 ultra-compact NS--WD and DNS systems in the Galaxy, respectively. We also show that among the current radio pulsar surveys, the radio pulsar surveys at the Arecibo radio telescope have a $\sim$50\% chance of detecting one of these systems. Finally, we show that an integration time of $t_{\rm int} \sim 1$~min will maximize the the S/N ratio as well as the probability of detection of these systems.

These two results show that current ground-based GW observatories like LIGO and Virgo as well as planned space-based observatories like LISA have excellent prospects of detecting BNS systems. For LIGO-Virgo, this means that we can expect more events like GW170817 which in turn will help expand on the results obtained from this merger event. Our results show that there is a high probability for detecting at least one ultra-compact binary system before LISA is scheduled to launch in the early 2030s. These ultra-compact binaries will also allow, for the first time, simultaneous observations using GW and electromagnetic observatories and produce ground-breaking new science.

Finally, we study the Double Pulsar system and show that pulsar A rotates prograde with respect to its orbit. This is the first ever direct measurement of the sense of rotation of a pulsar and is another confirmation of the rotating lighthouse model for pulsars. This also confirms that the spin angular momentum is closely aligned with the orbital angular momentum and that the supernova that produced the second, younger, pulsar B had a small kick associated with it. This result will eventually aid in the first-ever measurement of the moment of inertia of pulsar A by fixing the sign of the spin-orbit coupling and thereby increasing the confidence in the measurement of the moment-of-inertia.
\newpage
\bibliographystyle{aasjournal.bst}
\bibliography{thesis} 
\renewcommand{\baselinestretch}{2}
\small\normalsize

\end{document}